\documentclass[longauth]{aa}
\usepackage{graphicx}
\usepackage[section]{placeins}
\usepackage{sidecap}
\usepackage[english]{babel}
\usepackage{txfonts}
\usepackage{multirow}
\usepackage{lscape}
\usepackage{longtable}
\usepackage{natbib}
\usepackage{color}
\usepackage{url}
\bibpunct{(}{)}{;}{a}{}{,} 

\newcommand{\OII}{$\left[\mathrm{O\textrm{\textsc{ii}}}\right]\,$}

\newcommand{\OIIll}{$\left[\mathrm{O\textrm{\textsc{ii}}}\right]\,(\lambda\lambda 3729,3726)$}
\newcommand{\OIII}{$\left[\mathrm{O\textrm{\textsc{iii}}}\right]\,$}
\newcommand{\OIIIl}{$\left[\mathrm{O\textrm{\textsc{iii}}}\right]\,(\lambda\lambda 4959,5007)$}

\newcommand{\Ha}{H${\alpha}\,$}
\newcommand{\Hb}{H${\beta}\,$}

\newcommand{\uflux}[0]{${\rm erg \cdot s^{-1} \cdot cm^{-2}}$}

\newcommand{\HII}{$\mathrm{H\textrm{\textsc{ii}}}\,$}

\begin{document}

   \title{The $0.1<z<1.65$ evolution of the bright end \\ of the \OII luminosity function}
\author{Johan Comparat\inst{1,2} \and Johan Richard\inst{4} \and Jean-Paul Kneib\inst{5} \and Olivier Ilbert\inst{3}  \and V.\,Gonzalez-Perez\inst{6,7} \and Laurence Tresse\inst{3} \and Julien Zoubian\inst{6} \and Stephane Arnouts\inst{3} \and Joel R. Brownstein\inst{12} \and Carlton Baugh\inst{7} \and Timothee Delubac\inst{5} \and Anne Ealet\inst{6} \and Stephanie Escoffier\inst{6} \and Jian Ge\inst{13} \and Eric Jullo\inst{3}   \and Cedric Lacey\inst{7} \and Nicholas P. Ross\inst{8} \and David Schlegel\inst{9} \and Donald P. Schneider\inst{14,15} \and Oliver Steele\inst{10}  \and Lidia Tasca\inst{3} \and Christophe Yeche\inst{11} \and Michael Lesser\inst{17} \and Zhaoji Jiang\inst{16} \and Yipeng Jing\inst{18} \and Zhou Fan\inst{16} \and Xiaohui Fan\inst{17} \and  Jun Ma\inst{16} \and Jundan Nie\inst{16} \and Jiali Wang\inst{16} \and Zhenyu Wu\inst{16} \and Tianmeng Zhang\inst{16} \and Xu Zhou\inst{16} \and Zhimin Zhou\inst{16} \and Hu Zou\inst{16}}


\institute{
Departamento de Fisica Teorica, Universidad Autonoma de Madrid, Spain\\ \and
Instituto de Fisica Teorica  UAM/CSIC, Spain\\ \and
Laboratoire d'Astrophysique de Marseille - LAM, Universit\'e d'Aix-Marseille \& CNRS, UMR7326, 38 rue F. Joliot-Curie, 13388 Marseille Cedex 13, France\\ \and
CRAL, Observatoire de Lyon, Universit\'e Lyon 1, 9 Avenue Ch. Andr\'e, 69561 Saint Genis Laval Cedex, France\\ \and
Laboratoire d'astrophysique, \'Ecole Polytechnique F\'ed\'erale de Lausanne (EPFL), Observatoire de Sauverny, 1290 Versoix, Switzerland\\ \and
CPPM, Universit\'e d'Aix-Marseille, CNRS/IN2P3, Marseille, France\\ \and
Institute for Computational Cosmology, Department of Physics, University of Durham, South Road, Durham, DH1 3LE, UK \\ \and
Department of Physics, Drexel University, 3141 Chestnut Street, Philadelphia, PA 19104, USA \\ \and
Lawrence Berkeley National Laboratory, 1 Cyclotron Road, Berkeley, CA 94720, USA \\ \and
Institute of Cosmology and Gravitation (ICG), Dennis Sciama Building, Burnaby Road, Univ. of Portsmouth, Portsmouth, PO1 3FX, UK \\ \and
CEA, Centre de Saclay, IRFU, F-91191 Gif-sur-Yvette, France \\ \and 
Department of Physics and Astronomy, University of Utah, 115 S 1400 E, Salt Lake City, UT 84112, USA\\ \and
Astronomy Department, University of Florida, 211 Bryant Space Science Center, Gainesville, FL\\ \and
Department of Astronomy and Astrophysics, The Pennsylvania State University, University Park, PA 16802 \\ \and 
Institute for Gravitation and the Cosmos, The Pennsylvania State University, University Park, PA 16802\\ \and
Key laboratory of Optical Astronomy, National Astronomical Observatories, Chinese Academy of Sciences, Beijing, 100012, China\\ \and
Steward Observatory, University of Arizona, Tucson, AZ 85721\\ \and
Center for Astronomy and Astrophysics, Department of Physics and Astronomy, Shangai Jiao Tong University, Shangai 200240,China}

\date{\today}

  \abstract{
We present the \OIIll$ $ luminosity function measured in the redshift range $0.1<z<1.65$ with unprecedented depth and accuracy. Our measurements are based on medium resolution flux-calibrated spectra of emission line galaxies with the FORS2 instrument at VLT and with the SDSS-III/BOSS spectrograph. The FORS2 spectra and the corresponding catalog containing redshifts and line fluxes are released along with this paper.

In this work we use a novel method to combine the aforementioned surveys with GAMA, zCOSMOS and VVDS, which have different target selection, producing a consistent weighting scheme to derive the \OII luminosity function.

The measured luminosity function is in good agreement with previous independent estimates. The comparison with two state-of-the-art semi-analytical models is good, which is encouraging for the production of mock catalogs of \OII flux limited surveys. We observe the bright end evolution over 8.5 Gyr: we measure the decrease of $\log L_*$ from $42.4$ erg/s at redshift $1.44$ to $41.2$ at redshift $0.165$ and we find that the faint end slope flattens when redshift decreases.

This measurement confirms the feasibility of the target selection of future baryonic acoustic oscillation surveys aiming at observing \OII flux limited samples. 
}


\titlerunning{\OII LF in $0.1<z<1.65$}

\maketitle

\section{Introduction}
\label{introduction}
In the current $\Lambda$CDM paradigm the matter dominated Universe at redshift $1.65$ becomes driven by dark energy at $z=0.2$ \citep{2013arXiv1303.5076P}. 
This is one of the reasons proving the great interest in understanding the evolution of the Universe during this time span.

To understand the structural evolution of the Universe during this epoch, we need the largest possible map. Measuring rapidly accurate galaxy positions (redshifts) is key to build precise maps. The measurement of the emission-line-based redshifts in the optical domain with ground-based optical spectrographs is the least telescope time-consuming observing mode to build such maps. Luckily, narrow spectroscopic signatures in emission are abundant and enable a precise redshift measurement. The strongest emission line in an optical galaxy spectra is the \Ha $\lambda \, 6562\AA$ emission line. It allows constructing galaxy maps to redshift $0<z\lesssim0.53$ (for example, the CFR Survey \citep{1995ApJ...455...50L,1998ApJ...495..691T}, or the GAMA survey \citep{2011MNRAS.413..971D,2013MNRAS.433.2764G}. The second strongest set of emission lines is \OIIIl$ $ and \Hb, which allows making maps to redshift $0<z\lesssim1.1$ (for example the VVDS wide survey \citep{2013A&A...559A..14L} or the zCOSMOS survey \citep{Lilly_2009}. Though, to map the complete range $0<z<1.65$ with these lines, it is necessary to observe spectra in infra-red where the atmosphere is less transparent, i.e. longer exposure times. The third strongest set of lines is the \OIIll$ $ emission line doublet that allows an accurate redshift estimate throughout the redshift range $0<z<1.7$. The DEEP2 survey \citep{2013ApJS..208....5N} measured the redshifts in the range $0.7<z<1.4$ with the resolved \OII doublet.

The \OII luminosity function (LF hereafter) and its evolution is therefore pivotal for the planning of future spectroscopic surveys, that will target the most luminous \OII galaxies until reaching the required density to address the fundamental question of the nature of dark energy.

The \OII luminosity functions have been previously derived by \citet{2002ApJ...570L...1G,2007ApJ...657..738L,2009A&A...495..759A,2009ApJ...701...86Z,2010MNRAS.405.2594G,2013MNRAS.428.1128S,2013ApJ...769...83C,2013MNRAS.433..796D}. 
In these analysis (spectroscopy or narrow band photometry), the bright end of the \OII luminosity function is however not well constrained, partly due to the fact that either the survey areas are small or the redshift selection is very narrow. With our study we aim to get a better constrain of the bright end of the \OII LF by using new deep spectroscopic measurements.

In this study, we gather an \OII emission line sample from publicly available data to which we add newly acquired spectroscopy. In section \ref{sec:DATA}, we describe current publicly available \OII data and the new flux-calibrated spectroscopic data acquired by ESO VLT/FORS2 and by SDSS-III/BOSS spectrograph.
With this combined sample we measure the \OII luminosity function, in section \ref{sec:LF}. We also project the luminosity function to inform the planning of future surveys. Finally, we compare our measurement with semi-analytical models in section \ref{sec:SAMS}. 
Although this new sample is suited for such an analysis, we do not perform a new calibration of the relation \OII -- SFR -- dust, and leave it for future studies.

Throughout the paper, we quote magnitudes in the AB system \citep{1982STIN...8311000O} and we provide the measurements in Planck cosmology $h=0.673$, $\Omega_m=0.315$, $\Omega_\Lambda=0.685$, $w_{\rm Dark \, Energy}=-1$ \citep[see][]{2013arXiv1303.5076P}.

\section{\OII spectroscopic data}
\label{sec:DATA}
To measure the LF, we collected publicly available \OII flux-calibrated spectroscopy from which no \OII LF was previously derived.
 
\subsection{GAMA}
The Galaxy and Mass Assembly (GAMA) survey released its spectroscopic data and corresponding catalogs \citep{2011MNRAS.413..971D,2014MNRAS.441.2440B}. They provide a magnitude-limited sample ($r<19.8$) with spectroscopic redshifts (it extends to redshift $\sim0.4$) and \OII flux measurements corrected from the aperture on one of their fields of the stripe 82 (48 deg$^2$ near $\alpha_{\rm J2000}\sim217^\circ$ and $\delta_{\rm J2000}\sim0$ ) \citep{2013MNRAS.430.2047H}. We matched this sample to the stripe 82 deep co-add \citep{2011arXiv1111.6619A} to obtain the $u,g,r,i,z$ optical counter part of each \OII emitter.

\subsection{VVDS}
The VIMOS VLT Deep Survey final data release \citep{2013A&A...559A..14L} provides catalogs and spectra of all observations. We use a restricted set of catalogs where spectroscopic redshifts and the fits of the emission lines on the spectra are provided : the `deep' and `ultra deep' observations of the 2h field and the  `wide' observations on the 22h field. Spectral features were measured with the same pipeline as in  \citep{2009A&A...495...53L}. The $u,g,r,i,z$ optical magnitudes for these samples are taken from the CFHT-LS deep and wide observations \citep{Ilbert_06,Coupon_2009,2012A&A...545A..23B}.

To derive the integrated line flux, we convert the measured equivalent width (EW) into a flux density using 
\begin{equation}
f_{\lambda}^{\rm total}=-EW_{\rm measured}\; 10^{-(m+48.6)/2.5} \frac{c}{\lambda^2_{\left[\mathrm{O\textrm{\textsc{ii}}}\right]}} {\rm erg\; cm^{-2}\;s^{-1}\; \AA^{-1}}
\end{equation}
where $m$ is the broad-band magnitude of the CFHTLS filter containing \OII. Table \ref{color:weight:scheme:tab} gives the magnitude used as a function of the redshift of the galaxy. We compare this flux density with the one measured in the slit to make sure the discrepancy is of the order of magnitude of an aperture correction and use $f^{\rm total}$ in the LF.

\subsection{zCOSMOS}
We use the public zCOSMOS 10k bright spectroscopic sample on COSMOS field \citep{2007ApJS..172....1S,Lilly_2009}; which provides spectroscopic redshifts and fits of the emission lines in the spectra. The corresponding optical photometry is taken from \citet{scoville2007,Ilbert_2009}. The zCOSMOS survey provides the correction of the aperture correction for 1 arc second slits along the dispersion axis.

\subsection{New data from ESO/VLT on the COSMOS field}
\label{sec:data:vlt}
The data described in this subsection is thoroughly documented and publicly available here\footnote{\url{http://eboss.ft.uam.es/~comparat/website/ELG_VLT/}}. 

We have constructed an optimized color-box using the extremely rich ground-based photometry of the COSMOS/HST-ACS field \citep{2007ApJS..172...99C,scoville2007,Ilbert_2009} in order to target galaxies with strong emission lines that are expected in the redshift range $0.9<z<1.7$. 

We observed 2265 targets with the VLT/FORS-2 instrument equipped with the 600z+23 holographic grating, which is the unique multi-object spectroscopic ESO instrument that reaches out to $1\mu m$ allowing one to probe galaxies with redshift $z\lesssim$1.7 using the \OII emission line. The spectral range 737 nm - 1070 nm is sampled with a resolution 
$\lambda/\Delta \lambda=1390$ at 900 nm.
We made two short exposures of 309 seconds each and we observed emission lines with a flux $> 10^{-16}$ \uflux at a signal-to-noise ratio $>7$ at redshift $z>1.3$ and with a better SNR at lower redshift.

\begin{figure}
\begin{center}
\includegraphics[height=78mm]{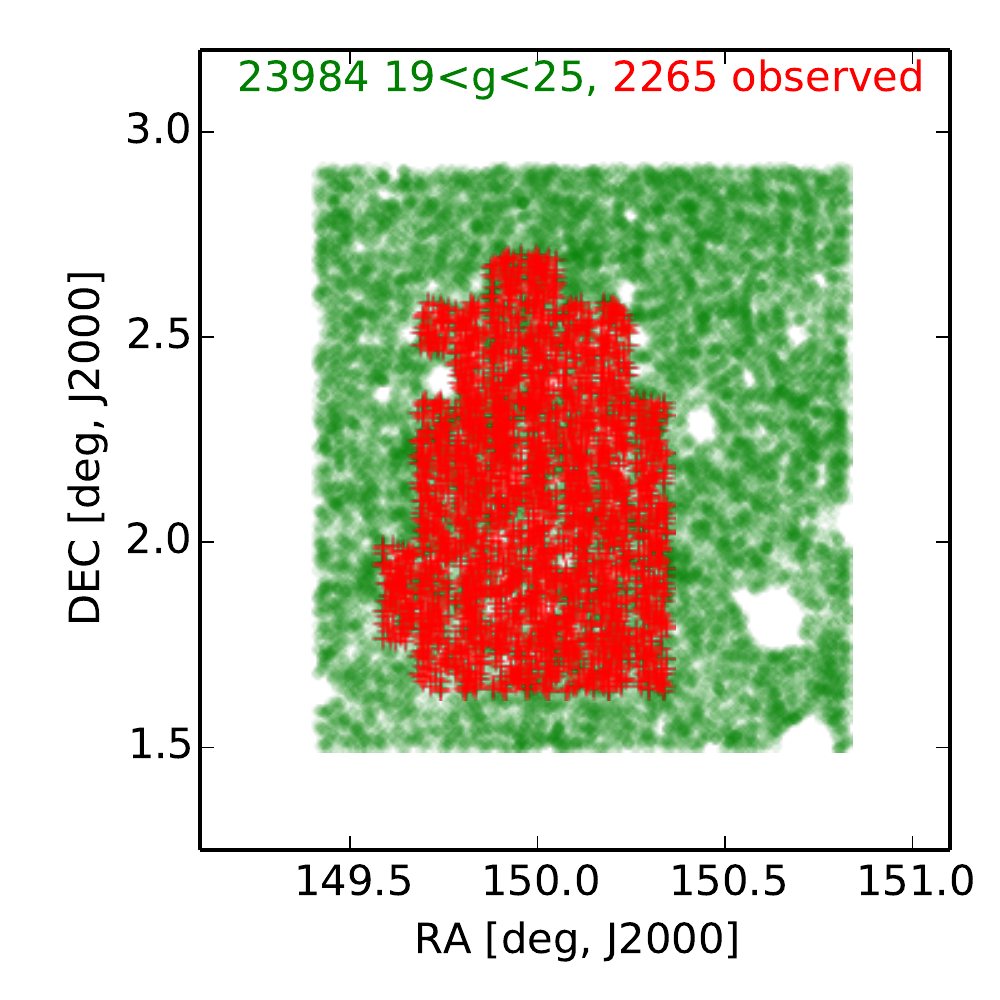}
\caption{Location on the sky of the galaxies observed with the VLT in this paper (red crosses). All COSMOS photometric redshift catalog detections are in green dots.}
\label{fig:location:targets}
\end{center}
\end{figure}

The targets are on the COSMOS field centered at $RA_\mathrm{J2000}=150^\circ$ and $DEC_\mathrm{J2000}=2.2^\circ$; see Fig. \ref{fig:location:targets}. 
The field covered is not perfectly continuous. Indeed, the area where slits can be placed is smaller than the complete mask area and we designed the pointings without considering this effect. Therefore there are vertical empty stripes of 36 arcseconds between each row of observation. The same effect applies horizontally, but is smaller $<3$ arcseconds.

We perform the target selection on the COSMOS photoZ catalog from \citet{Ilbert_2009} that contains detections over 1.73 deg$^2$ (effective area).

\subsubsection{Selection}
\begin{figure}
\begin{center}
\includegraphics[width=40mm]{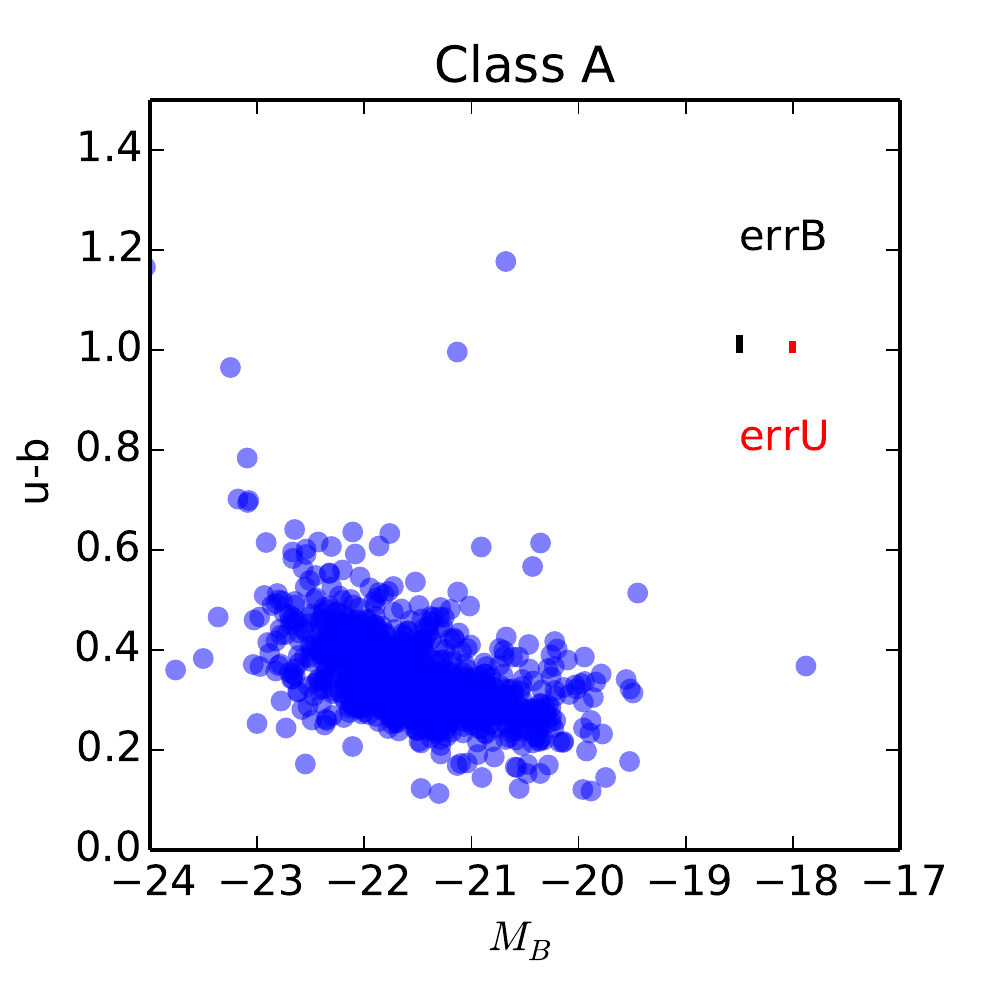}
\includegraphics[width=40mm]{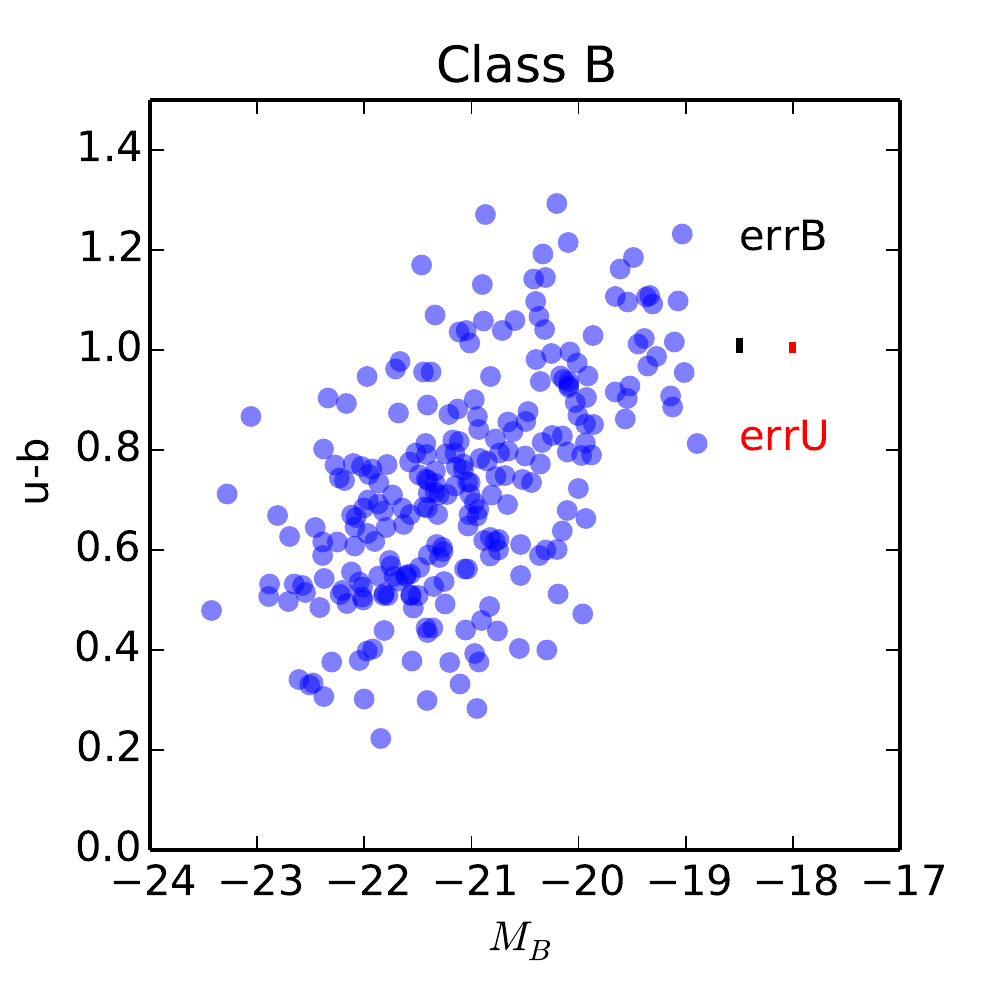}\\
\includegraphics[width=40mm]{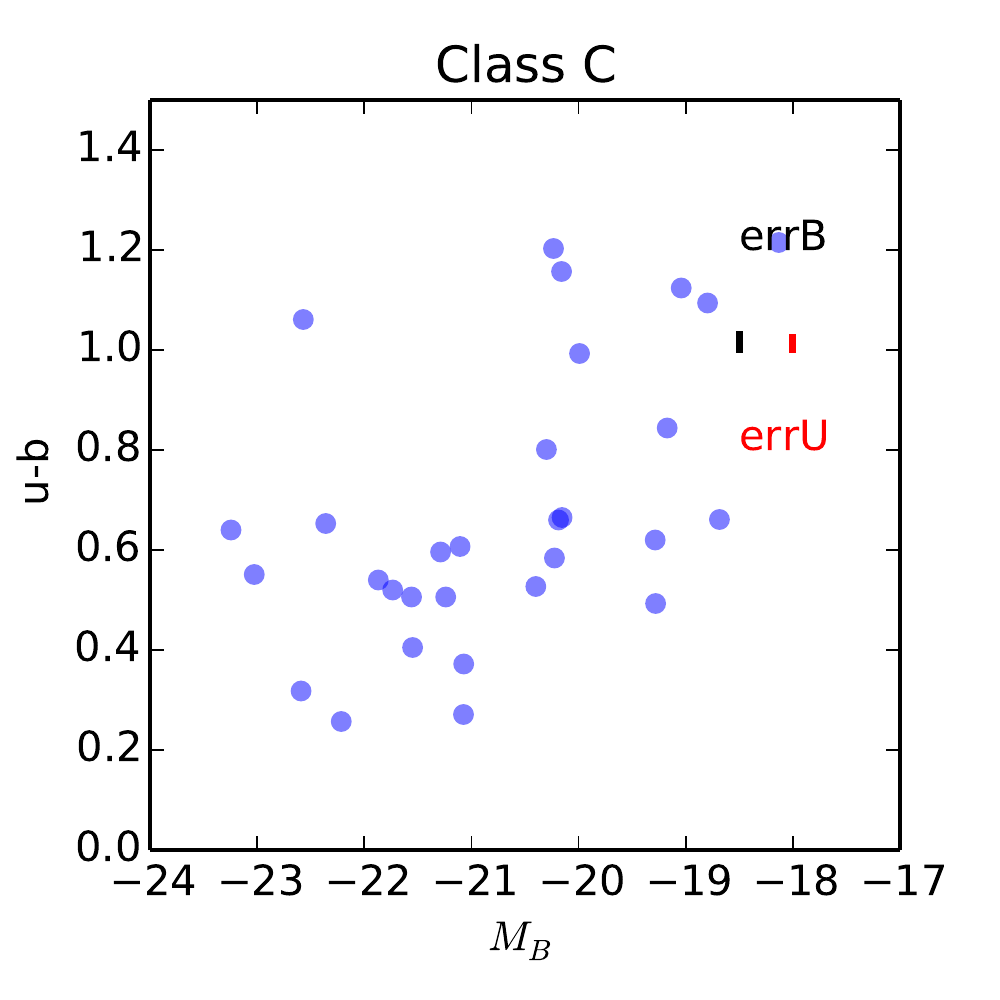}
\includegraphics[width=40mm]{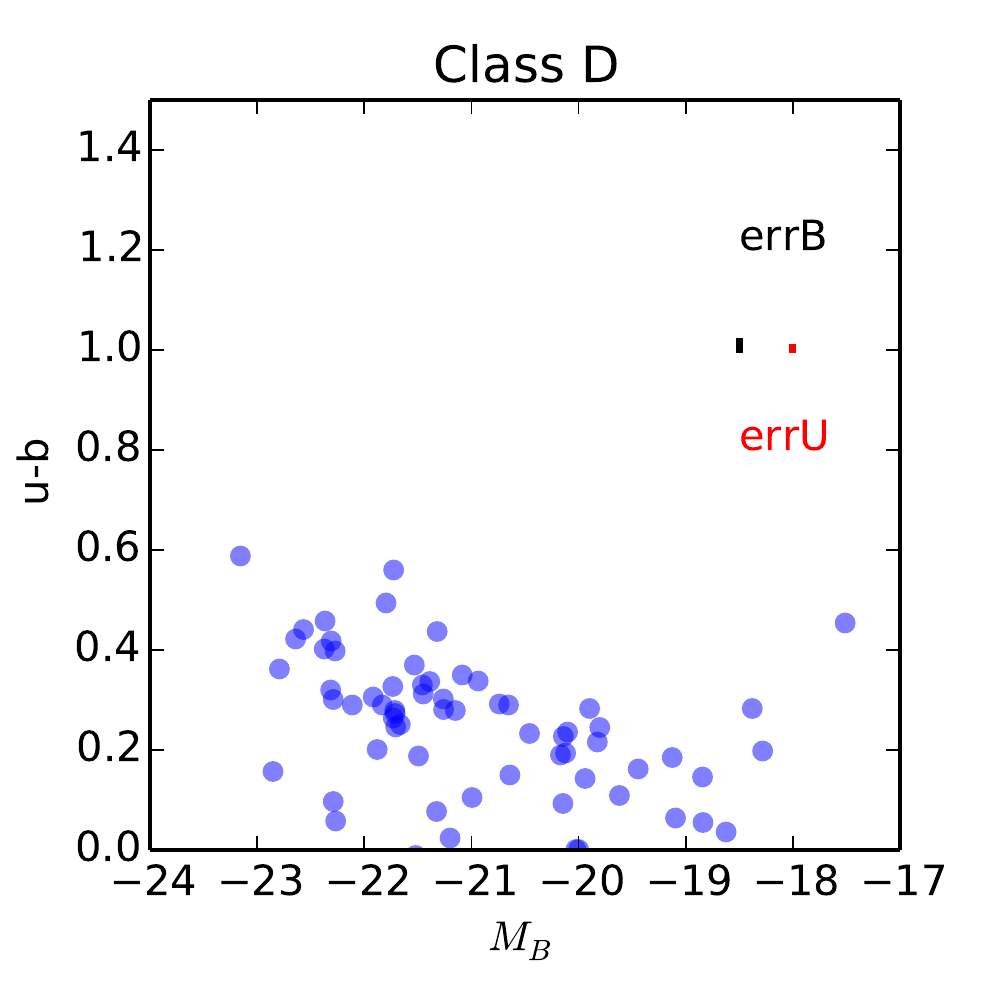}\\
\includegraphics[width=40mm]{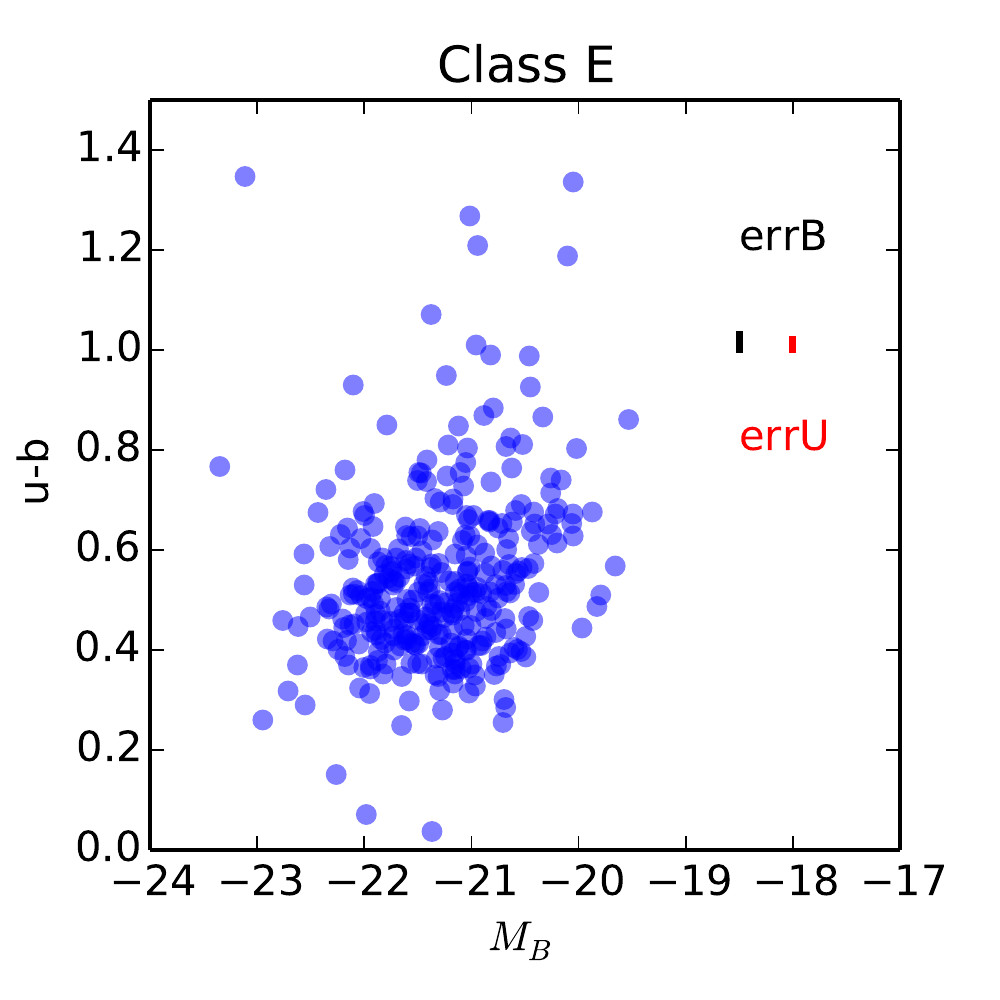}
\includegraphics[width=40mm]{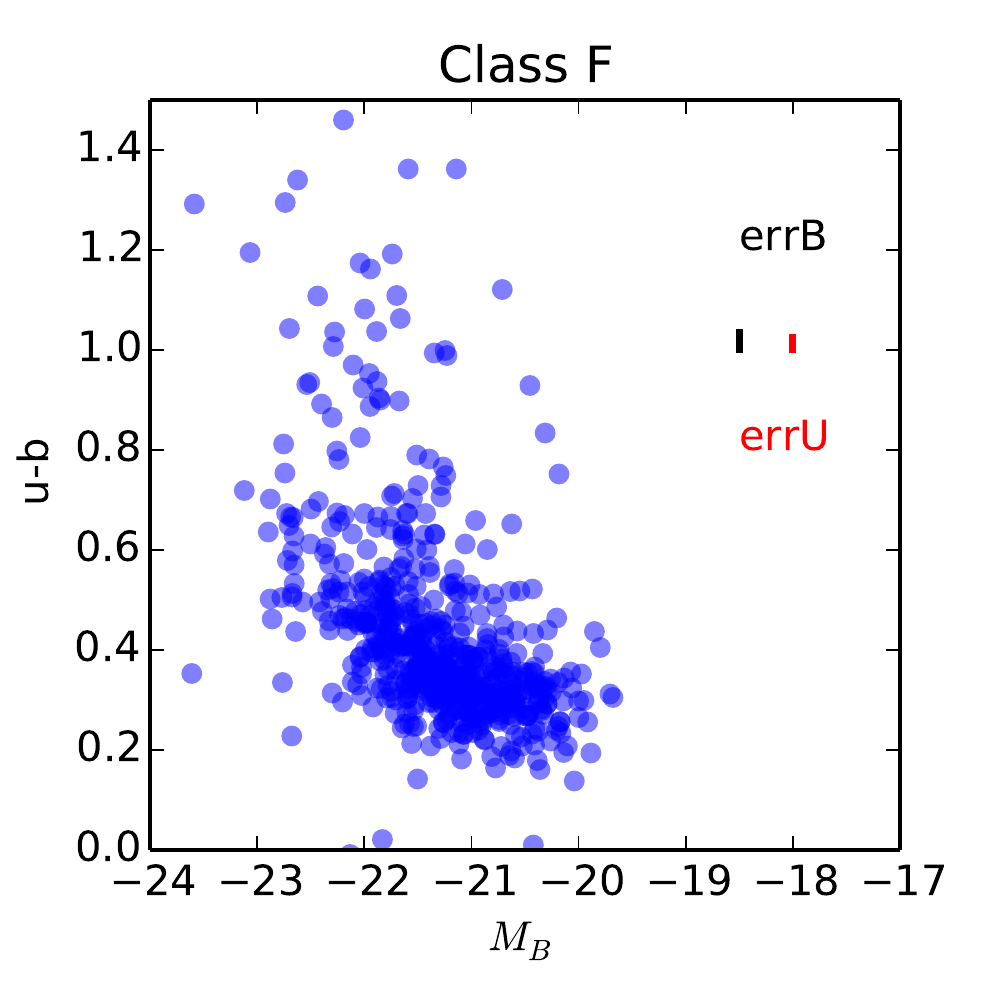}
\caption{Selection. $u-b$ color vs. $M_B$ absolute magnitude for each selection with $M_B$ derived using the photometric redshift. The vertical bars show the mean $1\sigma$ error on the $u$ and $b$ bands for each selection. Class A objects are bright and blue.}
\label{fig:selection:targets}
\end{center}
\end{figure}

To fill completely the slit masks, we used six different selection schemes using the COSMOS catalog magnitudes (we use \textsc{MAG AUTO} not corrected from galactic extinction): 
\begin{itemize}
\item $Class$ A: a $griz+$3.6$\mu m$ selection with $20<g<24$ and $i-z>(g-i)/2-0.1$ and $r-z>(i-mag_{3.6\mu m})/3$ and $r-z>0.8(g-i)+0.1$; these criteria select strong \OII emitters that are bright and blue. It was designed using the Cosmos Mock Catalog \citep{2009A&A...504..359J}; 
\item $Class$ B: Herschel detected galaxies at $5\sigma$ \citep{2011A&A...532A..90L};
\item $Class$ C: MIPS detected galaxies at $5\sigma$ \citep{2009ApJ...703..222L};
\item $Class$ D: $ugr$ selection from \citet{2013MNRAS.428.1498C} ($20<g<23.5$, $-0.5<u-g<1$, $-1<g-r<1$, $-0.5<u-r<0.5$);
\item $Class$ E: $gri$ selection from \citet{2013MNRAS.428.1498C} ($-0.1<g-r<1.1$, $0.8<r-i<1.4$, $20.5<i<23.5$);
\item $Class$ F: a photometric redshift selection $1<z_{phot}<1.7$ and $20<i<24$ to fill the remaining empty area of the masks.
\end{itemize}
$Class$ A is the only sample on which one can perform a standalone statistical analysis. In fact, the slits placed on the other selections were constrained by the slits placed on the $Class$ A targets, therefore the obtained samples are not random sub-samples of their parent selection. Table \ref{tab:prio:quality} describes the quantity of each class observed and the quality of the redshift obtained (see next subsection). The location of these samples in the $u-b$ versus $M_B$ absolute magnitude band is presented in Fig. \ref{fig:selection:targets}. It shows that the $Class$ A and C targets are bright and blue. A comparison of this selection with DEEP 2 observations \citep{2013ApJ...767...89M} demonstrates that $Class$ A galaxies have stellar masses between $10^{10}M_\odot$ and $10^{11}M_\odot$ and a star formation rate $SFR>10^{1.2} M_\odot$ yr$^{-1}$, and therefore possess strong \OII emission lines. The redshift efficiency for the $Class$ A selection is 701/947=74.0\% (692/947=73\% have $0.6<z<1.7$ and 552/947=58\% have $1<z<1.7$). 
The selections may seem complicated on first sight, though they were useful to minimize telescope time and measure a large amount of \OII emitters. The Class A, D and E selections try to mimic an absolute magnitude MB selection using optical bands. Class A is successful as it is 16\% more efficient than the magnitude limited selection represented by Class F.

\begin{table}
\caption{Number of spectra observed. $Class$ is the selection scheme used. Q is the redshift quality of the spectrum observed. 'eff' is the efficiency defined as the ratio (in per cent) between the number of spectra with a quality of 3 or 4 and the total.}
\begin{center}
\begin{tabular}{c c c c c c c c }
\hline \hline
Q&\multicolumn{6}{c}{Class} & \\
&A&B&C&D&E&F&total \\
\hline
0&149&43&11&39&93&168&503\\
1&24& 5& 2& 1&13&18&63\\
2&73&23& 4&10&60&79&249\\
3&307&62& 8&15&116&262&770\\
4&394 &117&10&19&47&93&680\\
\hline
total&947&250 & 35&  84& 329& 620&2265\\
\hline
eff &74.0 &  71.6 &  52.9 & 41.5 & 50.3 & 58.1& 64.7\\
\hline
\end{tabular}
\end{center}
\label{tab:prio:quality}
\end{table}%

\subsubsection{Data reduction, spectroscopic redshifts}
\begin{figure}
\begin{center}
\includegraphics[width=80mm]{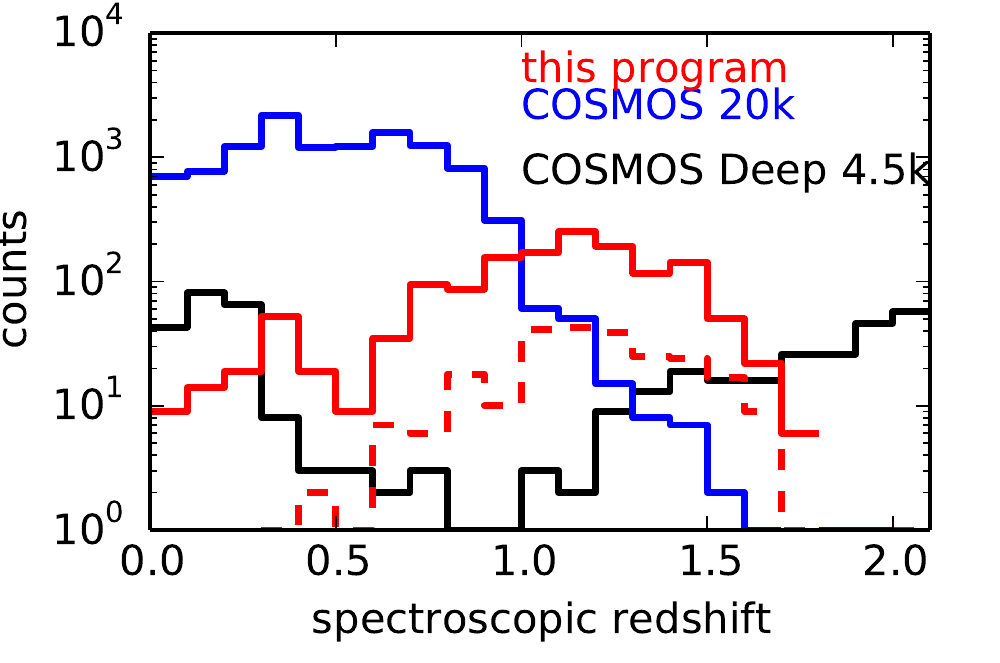}\\
\includegraphics[width=80mm]{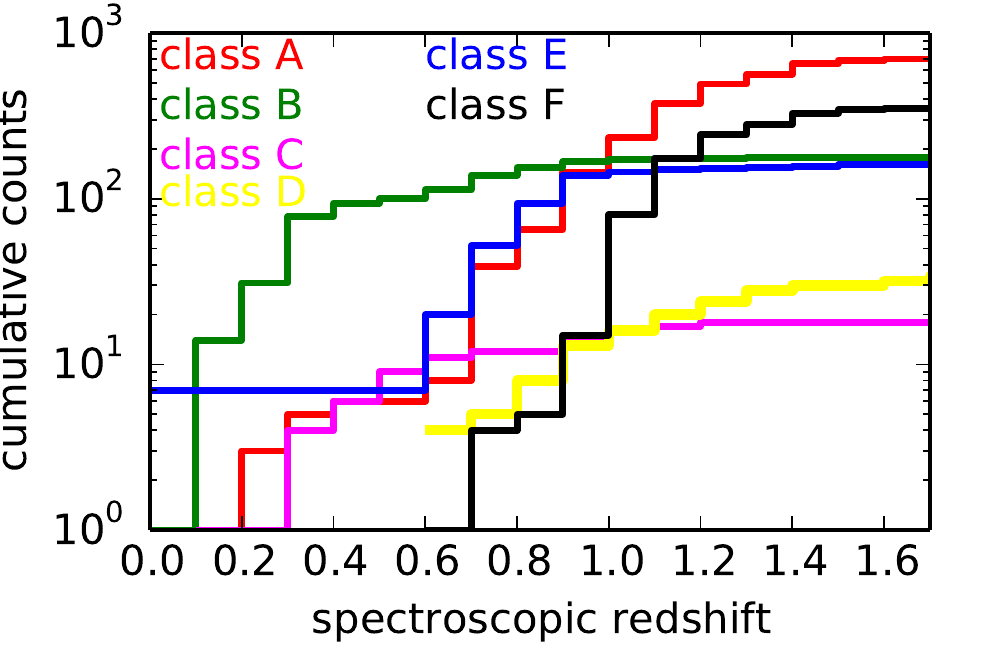}
\caption{Redshift distributions. {\it Top.} The distribution of our sample (red solid line for $Q=3$ or 4, dashed line for $Q=2$) compared to previous spectroscopic programs on COSMOS: the COSMOS deep 4.5k (black) and COSMOS 20k (blue). {\it Bottom.} Quality 3 or 4 redshift cumulative distributions for each class of target in our sample.}
\label{fig:redshift:hist}
\end{center}
\end{figure}

The data processing pipeline performs an extraction of the spectrum that allows the estimation of the flux in the \OII emission line. This procedure has two steps.
\begin{enumerate}
\item We apply the scripts and procedures from the ESO pipeline document\footnote{ftp://ftp.eso.org/pub/dfs/pipelines/fors/fors-pipeline-manual-4.4.pdf}. We use \textsc{fors\_bias} for the master bias creation, \textsc{fors\_calib} for the master calibration creation, and \textsc{fors\_science} to apply the calibrations and subtract the sky. Figures 8.2 and 8.3 from that document show an example of the data reduction cascade. We use the `unmapped' result obtained in the middle of the \textsc{fors\_science} part of the reduction, and the matching frames containing the wavelength values. Both `average' and `minimum' combinations of the 2 frames are obtained on each mask.
\item Emission lines are visually identified on both average and minimum reductions. The visual inspection is more efficient than an automated measurement because of the presence of cosmic ray residuals when combining the two exposures, and the strong variations of signal-to-noise across the 2d spectra (both spatially and spectrally).
A quality flag $Q$ is assigned on each object, from 4 (secure redshift identification), 3 (clear single line redshift identification), 2 (possible line, not 100\% convincing), 1 (a rough estimate), 0 (strong defect preventing redshift
measurement / line identification). We consider $Q=3$ or 4 as reliable redshifts; see Table \ref{tab:prio:quality}. Given the wavelength coverage of the spectrum, there are only a few spectra with a single line that could be mis-identified. In those particular cases, they were attributed a quality $Q=2$.
\end{enumerate}
As shown in Fig. \ref{fig:redshift:hist}
 the redshifts measured for the galaxies in our study fill the gap between `COSMOS 20k' and `COSMOS Deep 4.5k' \citep{2007ApJS..172....1S,Lilly_2009}. The highest redshift in our sample is $z_{max}=1.73$. The classes B and C mainly contain galaxies at redshifts below $z<0.8$. 

\begin{figure}
\begin{center}
\includegraphics[width=40mm]{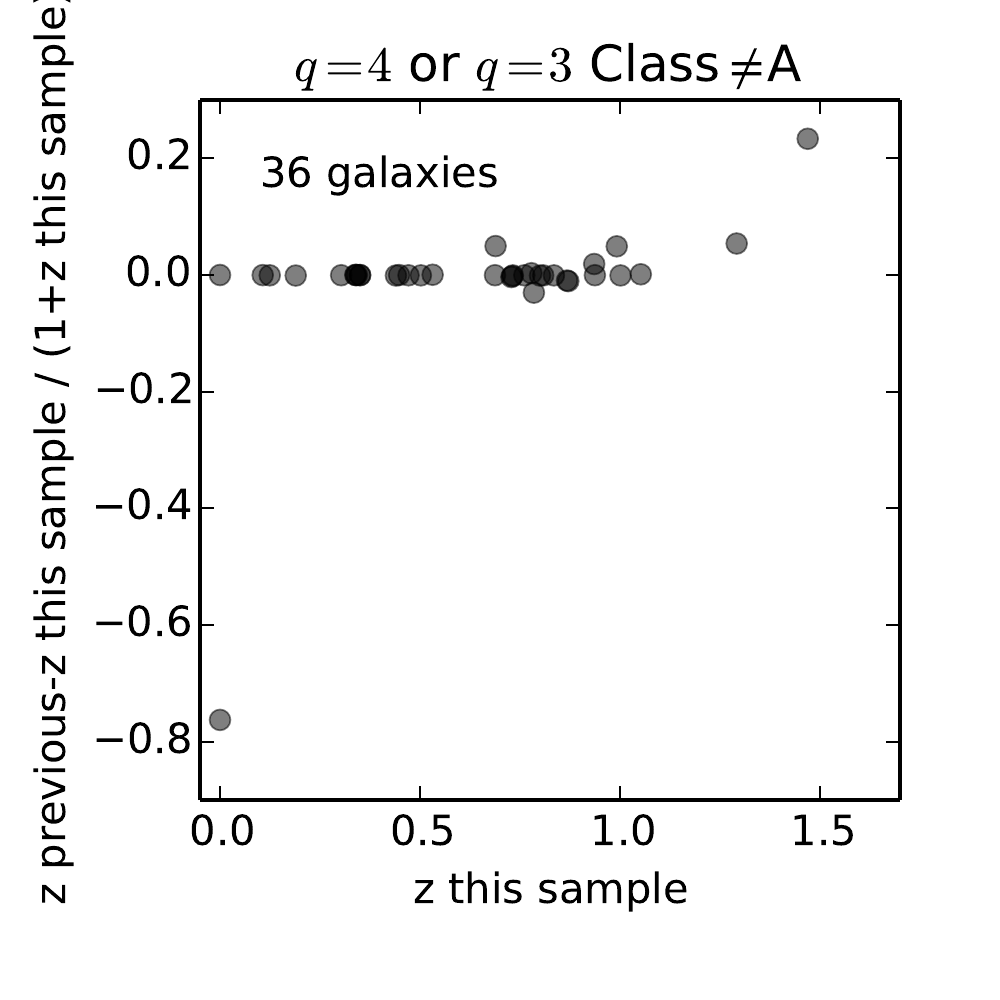}
\includegraphics[width=40mm]{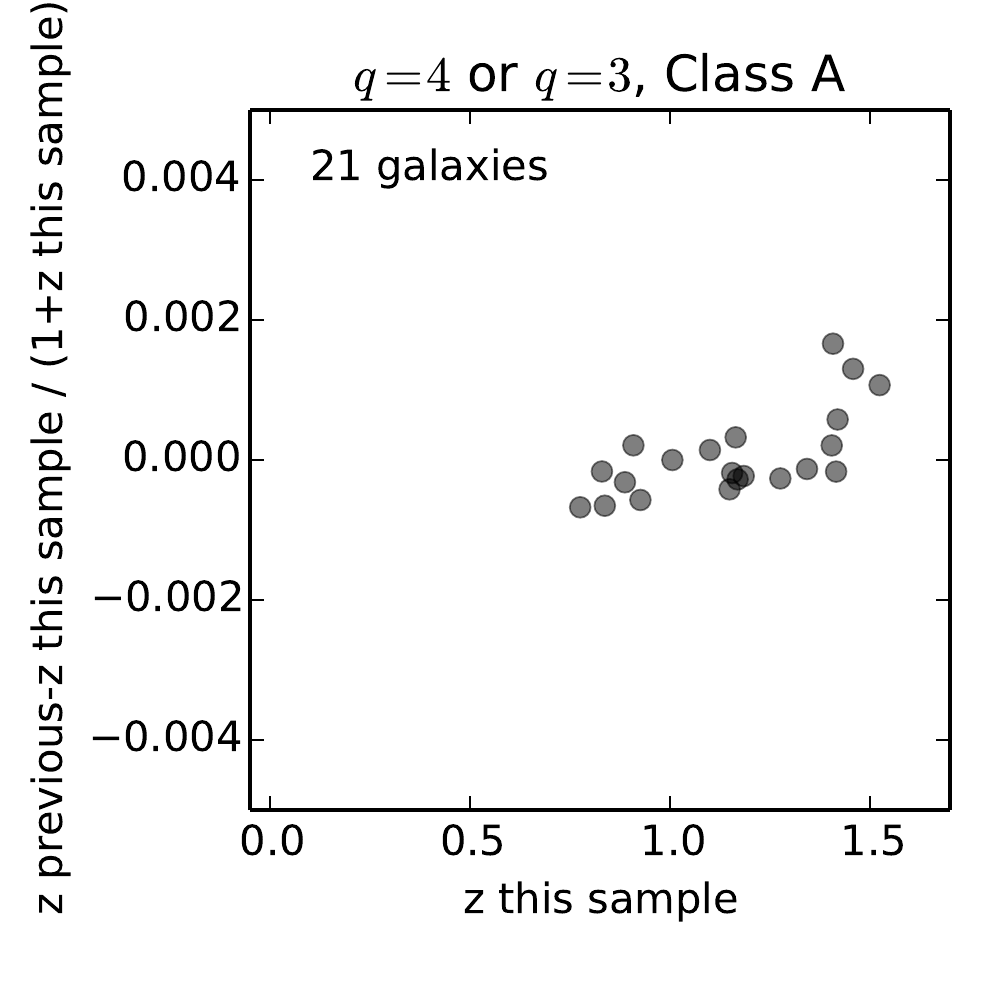}
\caption{Comparison between the spectroscopic redshift and the spectroscopic redshift from previous surveys.}
\label{fig:redshiftComparison}
\end{center}
\end{figure}

There is a small overlap between the sample presented here and previous COSMOS spectroscopic samples; this allows us to compare the redshift of objects observed twice (the positions on the sky match within 0.1 arc second). For the set of objects with $Class$ A, 7 galaxies with $Q=3$ and 26 with $Q=4$ have a counterpart with a high quality flag. For the set of objects with $Class\neq A$, 15 galaxies with $Q=3$ and 21 with $Q=4$ have a counterpart with a high quality flag. Among these matches, only two galaxies have a spectroscopic redshift that do not agree at the 10\% level ($dz>0.1(1+z)$). After a second inspection of those redshifts we found that the redshifts we obtained are correct; see Fig. \ref{fig:redshiftComparison}. 

\begin{figure}
\begin{center}
\includegraphics[width=40mm]{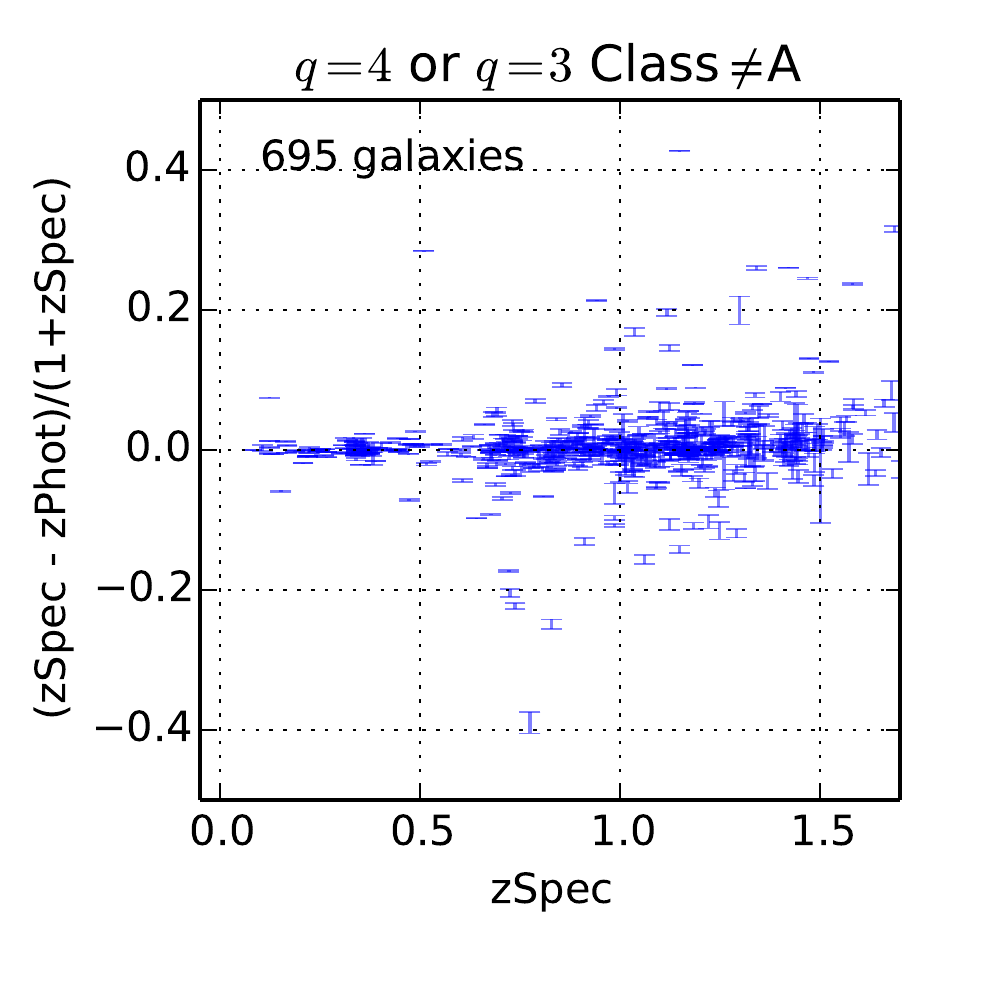}
\includegraphics[width=40mm]{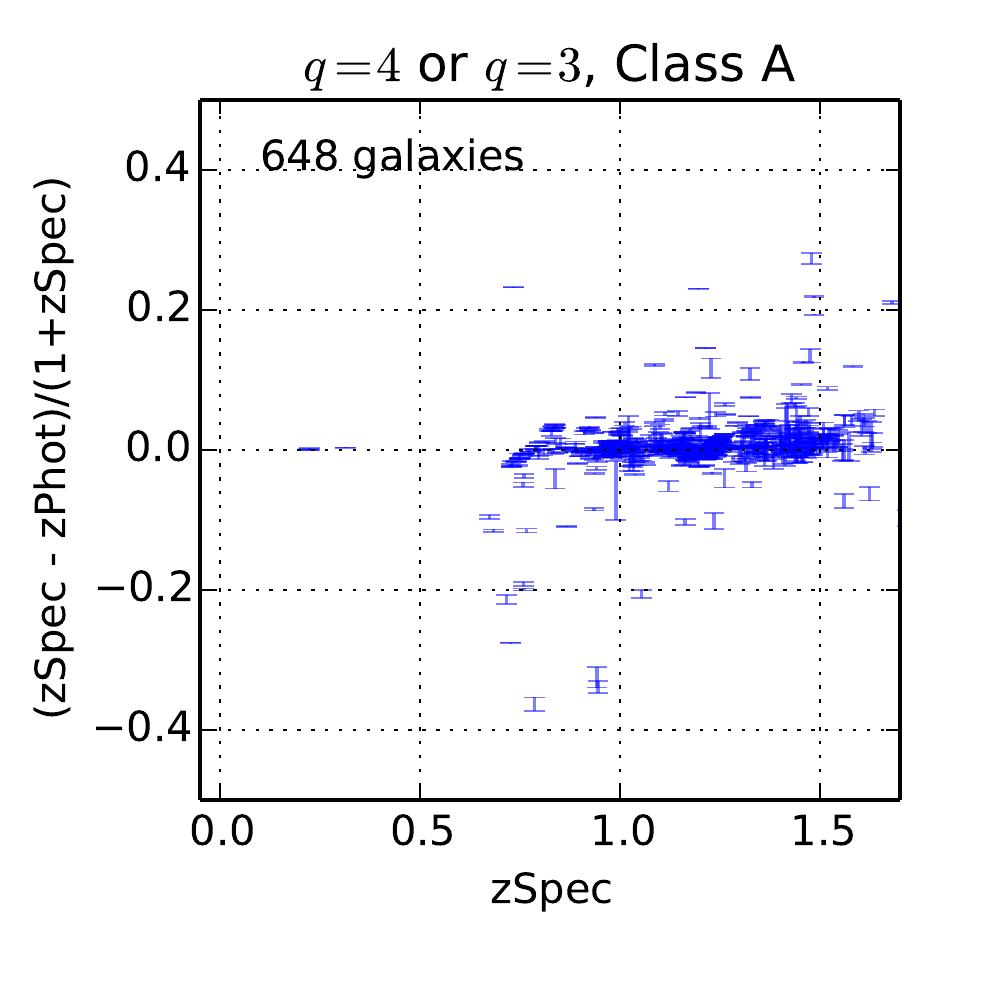}
\caption{Comparison between the spectroscopic redshift and photometric redshift.}
\label{fig:redshiftComparison2}
\end{center}
\end{figure}

We also compare our spectroscopic redshifts to the photometric redshifts from \citet{Ilbert_2009,2013A&A...556A..55I}. 
We consider 1344 objects with both good spectroscopic redshift and photometric redshift. A total of 97.3\% (1306) of the photometric redshifts are in agreement with the spectroscopic redshift within a 15\% error, and 89.9\% (1207) within a 5\% error (see Fig. \ref{fig:redshiftComparison2}). The agreement is excellent.

\begin{figure}
\begin{center}
\includegraphics[width=80mm]{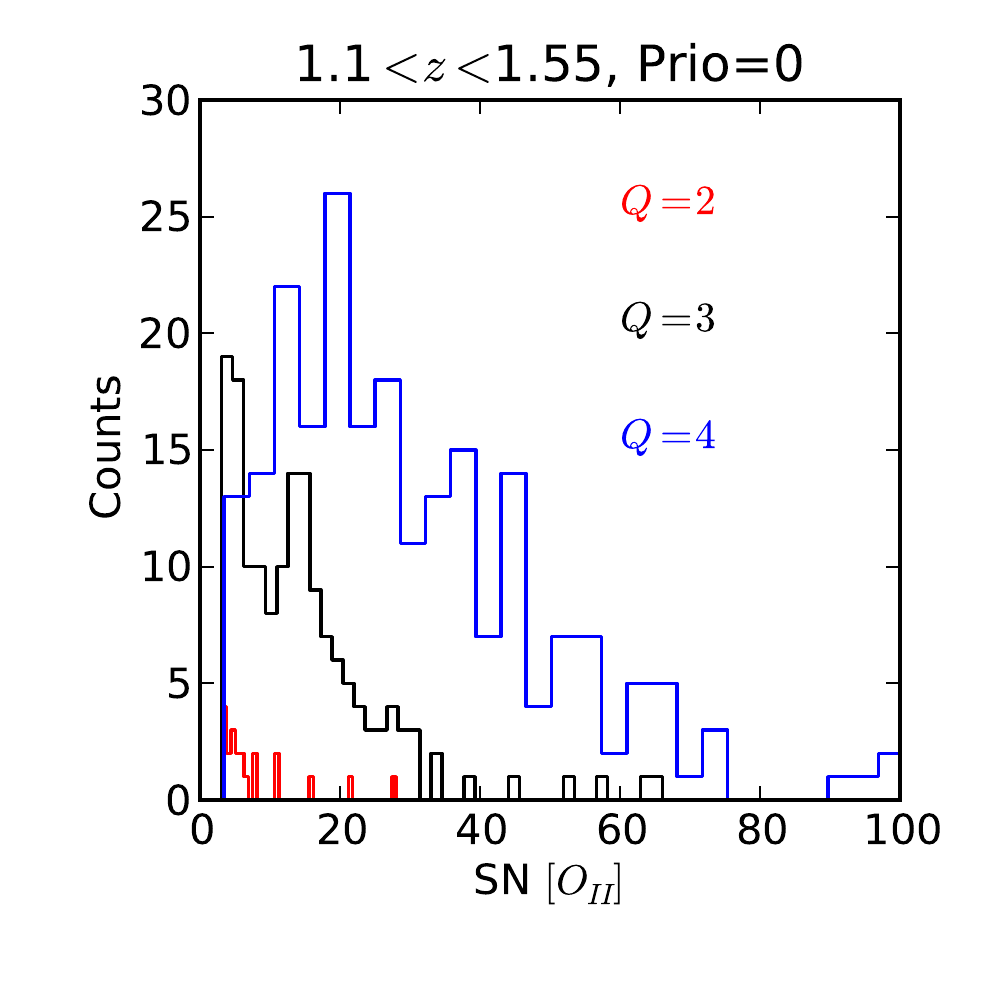}
\caption{SNR in the \OII lines for different redshift quality flags. The galaxies with $Q=3$ or 4 are used to determine the luminosity function.}
\label{fig:SNOII:hist}
\end{center}
\end{figure}

Finally, we fit the emission lines detected in the spectra using a simple Gaussian model for every line. 
Given the resolution of the spectrograph, one cannot detect the difference between the fit of a single Gaussian and the fit of a doublet for the \OII line.
From this model we determine the emission line flux and the signal-to-noise ratio (SNR) of the detection. \OIII is detected in the redshift range $0.45<z<1.05$ and \OII in the range $0.9<z<1.75$. The SNR distribution is correlated to the quality flag of the redshifts; see Fig. \ref{fig:SNOII:hist}. The exposure times were too short to measure the continuum, we therefore estimate the equivalent widths and the levels of the continuum using the broadband magnitude that contains the emission line.

\begin{figure}
\begin{center}
\includegraphics[width=80mm]{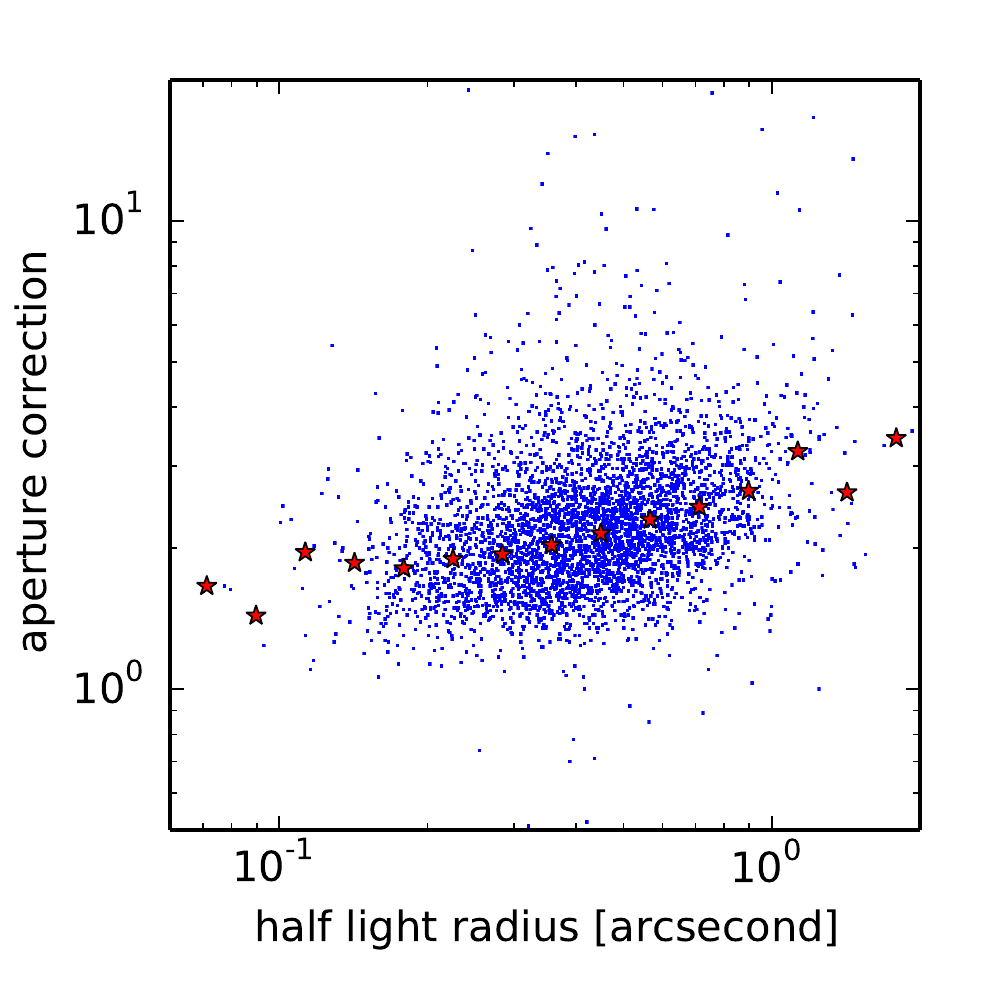}
\caption{Aperture correction factor vs. half light radius on Cosmos. The red stars represent the median correction for bins of half light radius.}
\label{cosmos:app:corr}
\end{center}
\end{figure}

We use an extrapolation of the median aperture correction as a function of half light radius, based on the zCOSMOS corrections, to correct the fluxes from the aperture; see Fig. \ref{cosmos:app:corr}.

\begin{figure}
\begin{center}
\includegraphics[width=8cm]{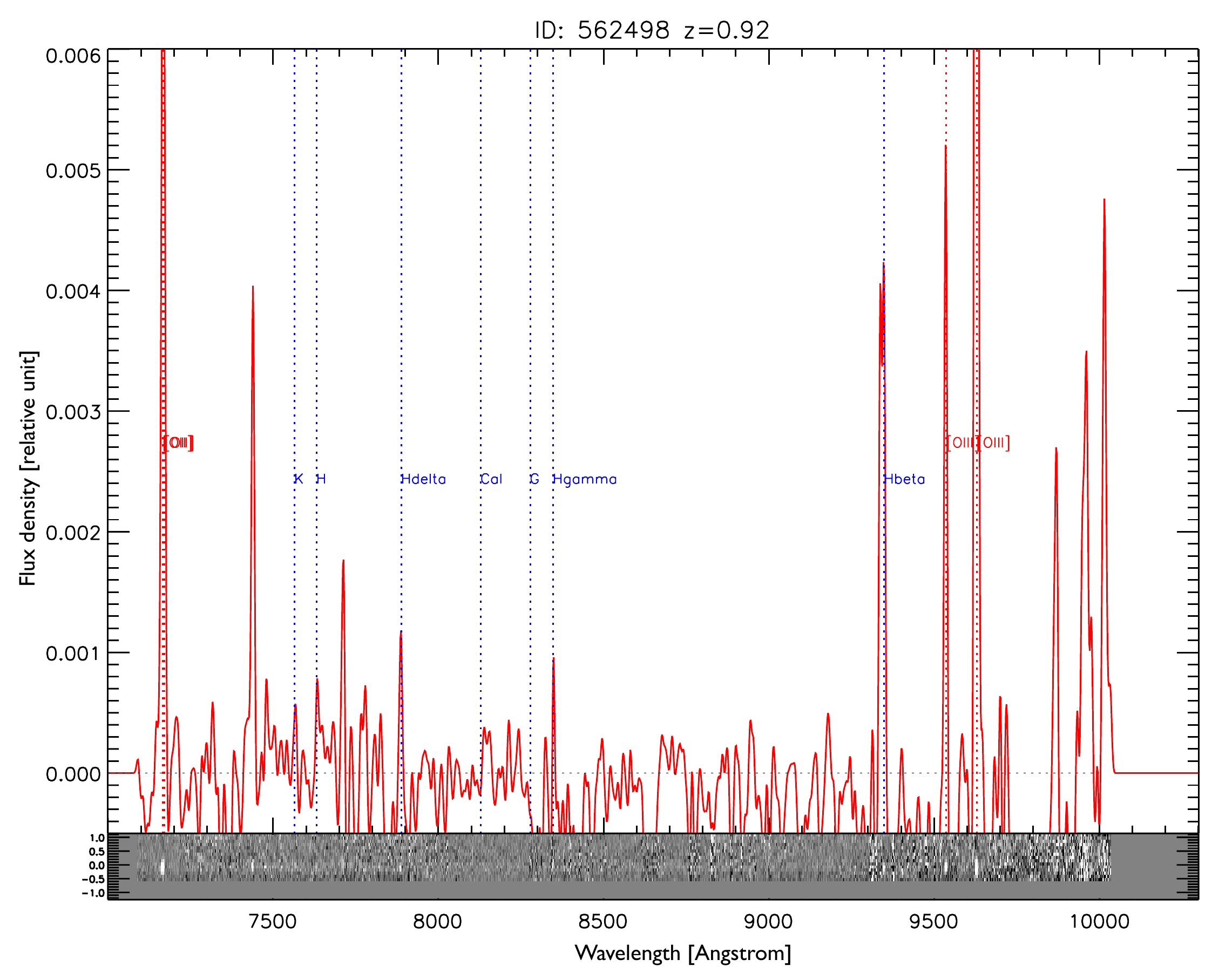}
\includegraphics[width=8cm]{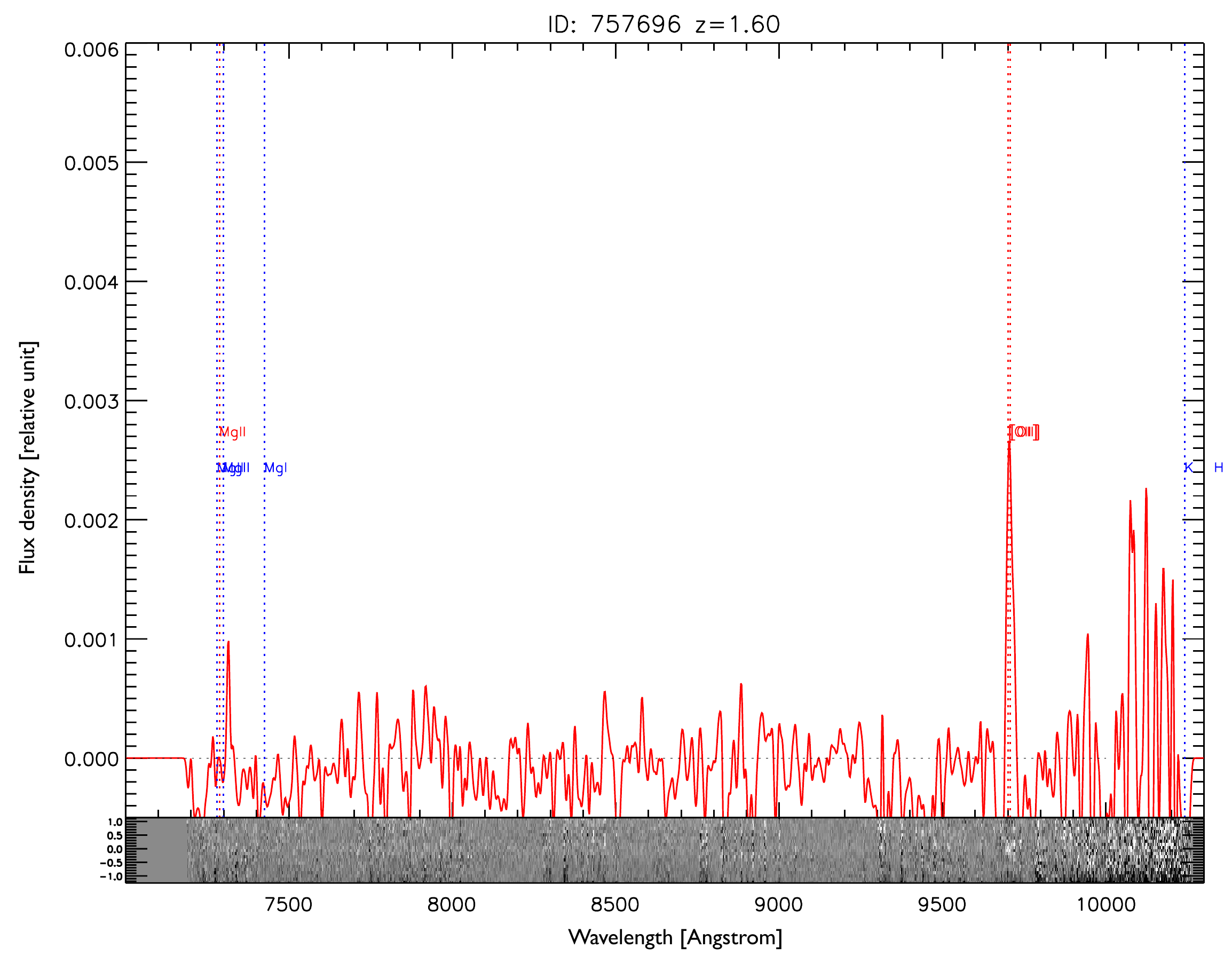}
\caption{Two spectra from the ESO VLT/FORS2 data set at redshift 0.92 and 1.6. The one dimensional spectra is on top of the two dimensional spectra.}
\label{fig:example:spectra}
\end{center}
\end{figure}

Two examples of spectra are displayed on Fig. \ref{fig:example:spectra}.

\subsubsection{Galaxies with a flux \OII $>10^{-15}$ \uflux}

There are three galaxies with \OII fluxes greater than $10^{-15}$ \uflux. Two are compact and have broad emission lines and are probably AGNs. For the luminosity function analysis, we removed the broad-line AGNs in order to be able to easily compare our results with other studies. The third galaxy appears disturbed and might be ongoing a merging process.

\subsection{Data from SDSS-III/BOSS ELG ancillary program}
\label{sec:data:bossw3}

Within the Sloan Digital Sky Survey III collaboration (SDSS \citealt{2000AJ....120.1579Y,2011AJ....142...72E,2013AJ....145...10D}), were observed galaxies with strong emission lines (ELG) in the redshift range $0.4<z<1.6$ to test the target selection of emission line galaxies on two different photometric systems for the new SDSS-IV/eBOSS survey. These observations are part of the SDSS-III/BOSS ancillary program and are flagged `ELG' or `SEQUELS\_ELG'. These spectra will be part of the SDSS Data Release 12 (DR12).

\subsubsection{CFHT-LS $ugri$ selection}
During a first observation run, a total of 2292 fibers were allocated over 7.1 deg$^2$ (the area of an SDSS-III plate). The ELG were observed during 2 hours with the spectrograph \citep{2013AJ....146...32S} of the 2.5m telescope located at Apache Point Observatory, New Mexico, USA, \citep{2006AJ....131.2332G}. The targets were selected from the Canada-France-Hawaii Telescope Legacy Survey (CFHT-LS) Wide W3 field catalogue with $ugri$ bands. The data and cataloguing methods are described in \citet{Ilbert_06,Coupon_2009}, and the T0007 release document\footnote{http://www.cfht.hawaii.edu/Science/CFHTLS/}.

The color selection used is 

$-0.5<u-r<0.7\cdot(g-i)+0.1$ and $20<g<22.8$. 

\begin{figure*}
\begin{center}
\includegraphics[width=7cm]{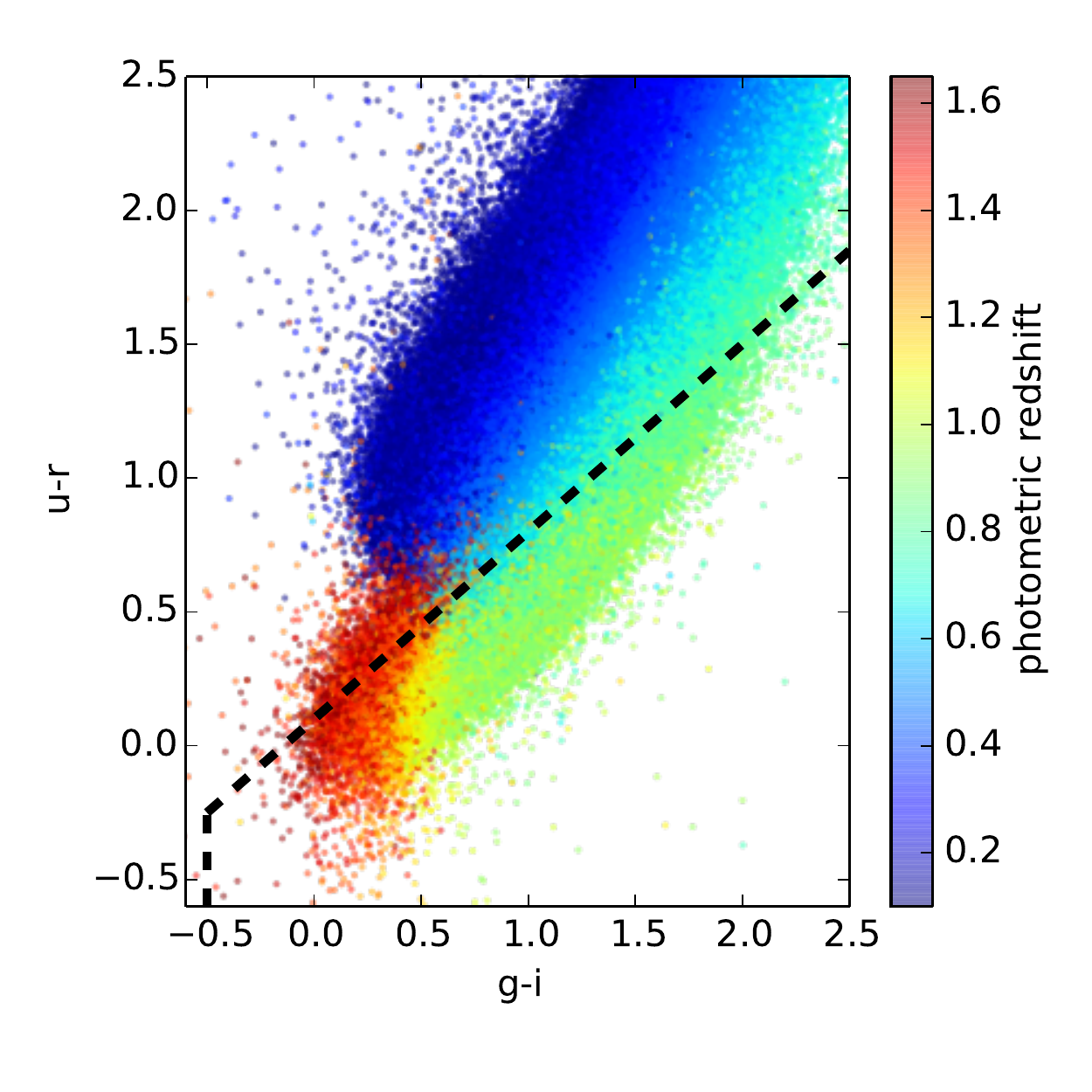}
\includegraphics[width=7cm]{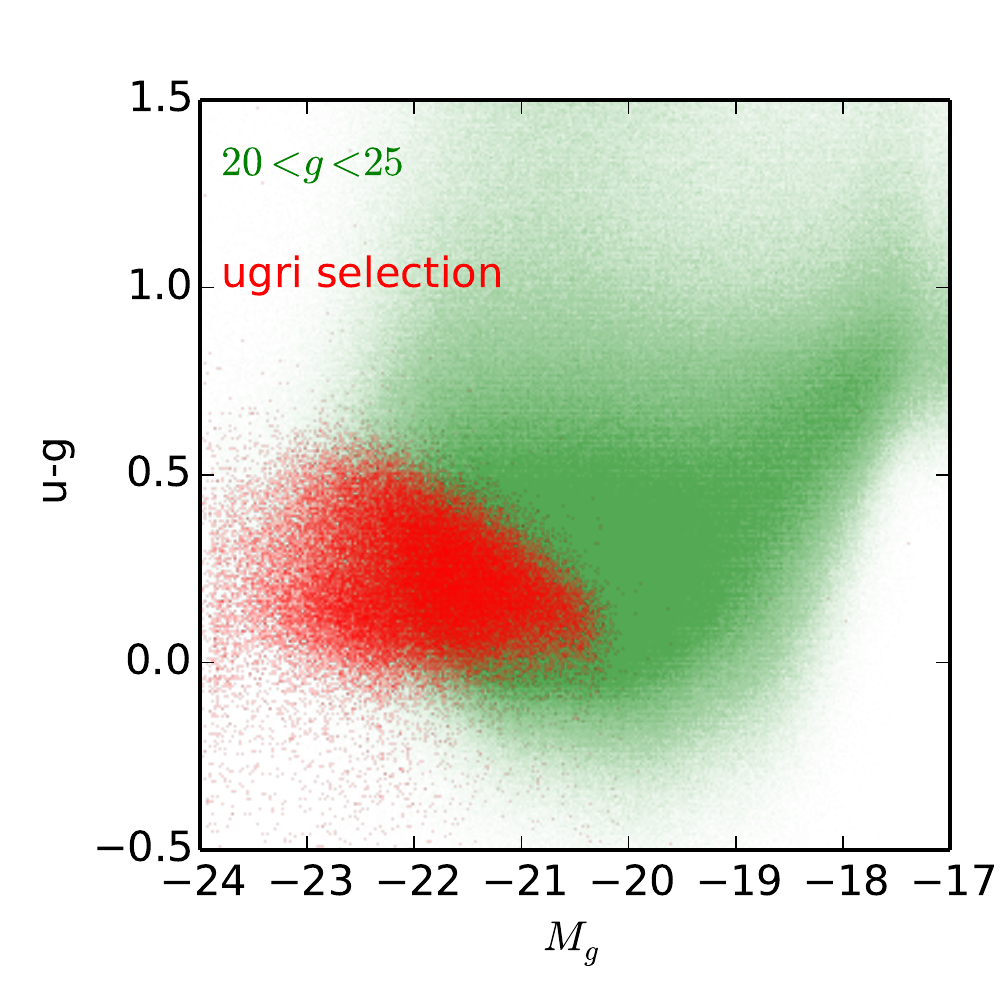} \\
\includegraphics[width=7cm]{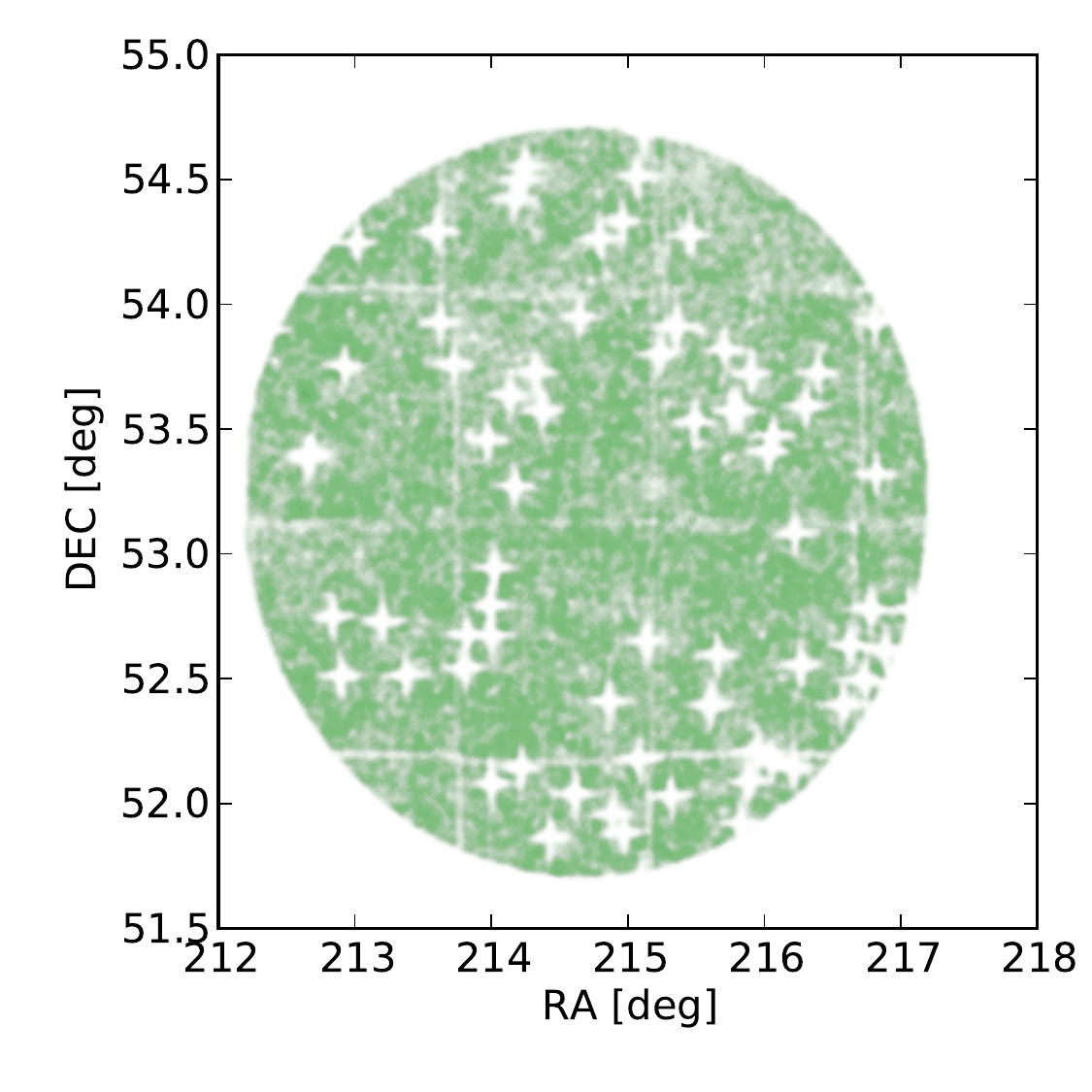}
\includegraphics[width=7cm]{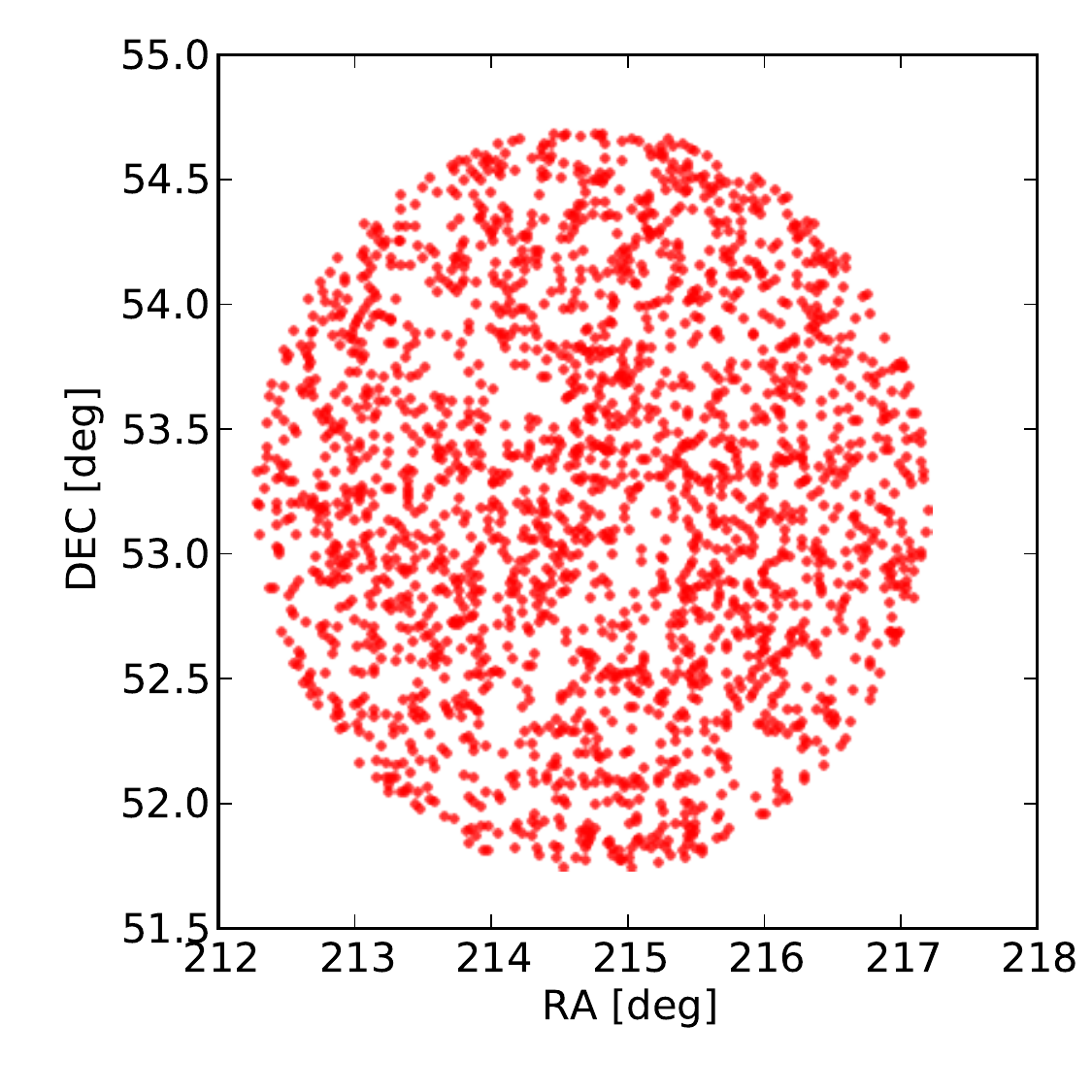}\\
\includegraphics[width=7cm]{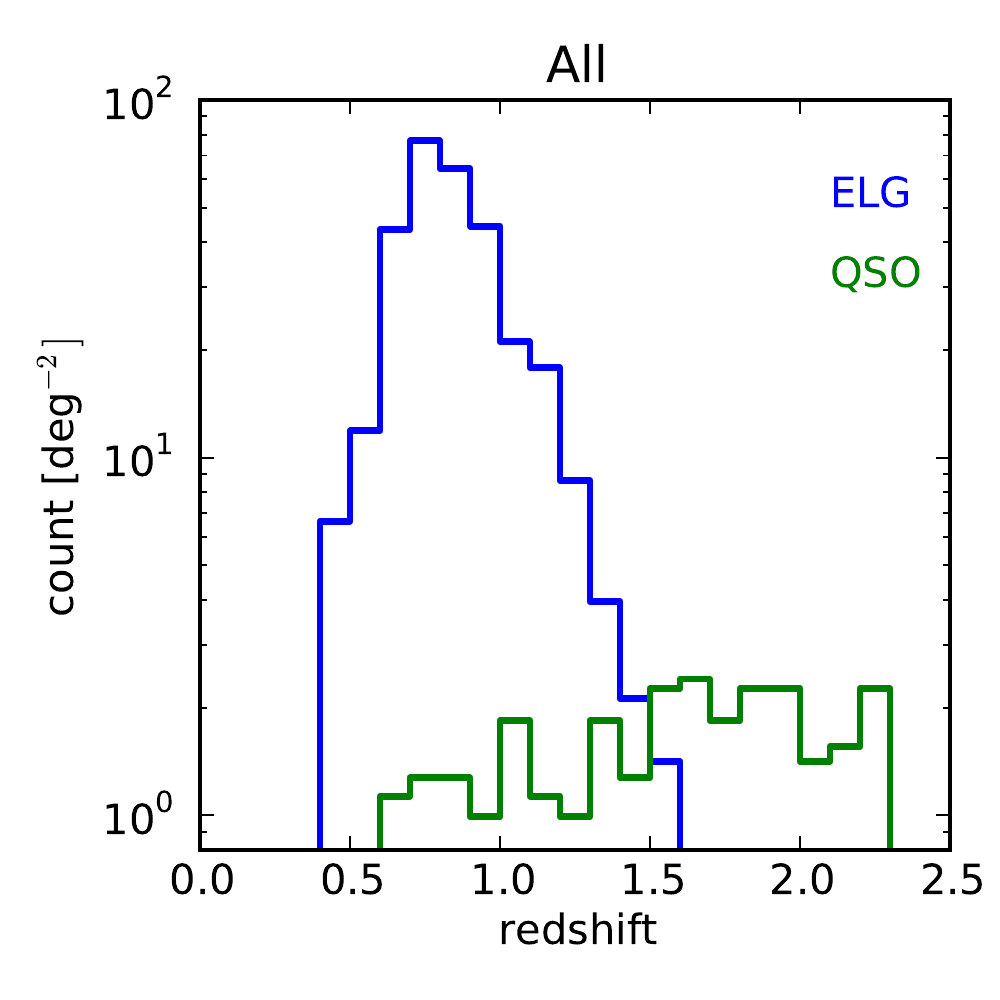}
\includegraphics[width=7cm]{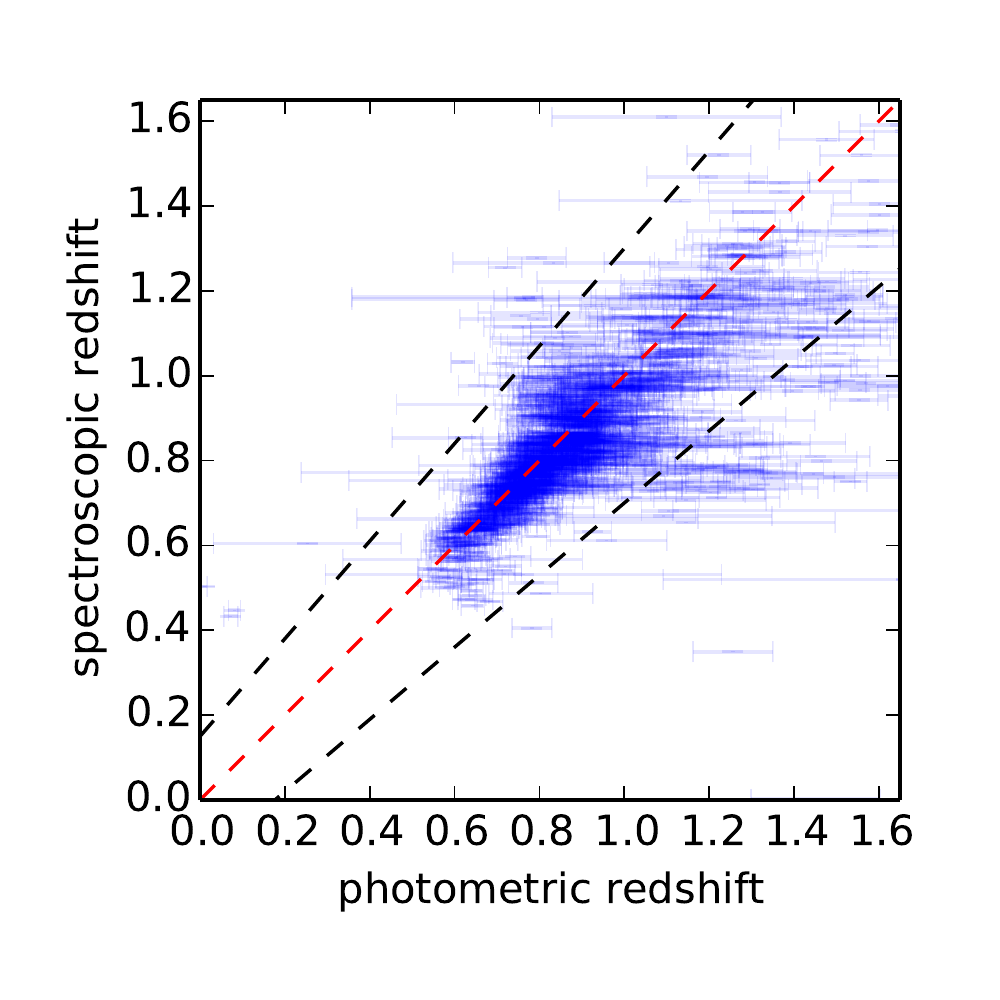}
\caption{Observations with SDSS telescope based on the CFHTLS photometry. {\bf Top row}. Left. u-r vs. g-i colored according to the photometric redshift. The selection applied is the dashed box. Right. Color selection projected in the u-g vs MG plane which shows our selection aims for the brightest bluest objects. 
{\bf Middle row}. RA, DEC in degree (J2000). Left. $g<25$ objects from the CFHT-LS W3 photometric redshift catalog. Right. Targets observed spectroscopically.
{\bf Bottom row}. Left. Distribution of the redshifts observed. It is very efficient for selecting redshift 0.8 ELGs. Right. Spectroscopic redshift and photometric redshift. dashed lines represent the error contours at $dz=0.15(1+z)$. The agreement is good.}
\label{color:selection:w3}
\end{center}
\end{figure*}

The selection function focuses on the brightest and bluest galaxy population; see Fig. \ref{color:selection:w3}. The selection provides 3784 targets, we observed 2292 of them. 
The observations of this sample were obtained on the three SDSS plates number 6931, 6932, 6933. 

The reduction and the fit of the redshift is fully automated and performed by the version `v5\_6\_elg' of the BOSS pipeline \citep{2012AJ....144..144B}. A total 88.7\% of the spectra observed have sufficient signal to be assigned a reliable redshift; see Table \ref{table:obs:general:w3}. Of this sample 82.3\% are emission line galaxies (ELGs), 5.6\% are quasars (QSOs) and 0.7\% are stars. A total of 11.3\% of the spectra have insufficient signal to noise ratio to obtain a reliable spectroscopic redshift. 
The spectroscopic redshift distribution obtained is presented in Fig. \ref{color:selection:w3}. 
The photometric redshifts from T0007 on CFHT-LS W3 perform well. Of the 1609 galaxies with a photometric redshift and a good spectroscopic redshift, 92.3\% (1486) are within a 15\% error and 71.9\% (1157) in a 5\% error. 
The bluer objects tend to have larger uncertainties on the photometric redshifts.
Improving CFHT-LS photometric redshifts for such population is of great interest but beyond the scope of this paper.  
\begin{table}
\caption{Observations on the CFHT-LS Wide 3 field. The observed objects are a random subsample of the selection. As three plates were super-imposed, the fiber collision has a negligible effect.}
\begin{center}
\begin{tabular}{l r r r r r c}
\hline \hline
\multicolumn{2}{l}{class} &  N & percent & z success rate & N deg$^{-2}$\\ \hline
\multicolumn{2}{l}{$ugri$ selection} 	& 3784 &&&535.3\\
\multicolumn{2}{l}{observed} 		& 2292 & 60.5& 100\%&324.5	\\
\multicolumn{2}{l}{redshift measured} & 2032 	&53.7&88.7\%&287.5\\
	&ELG 					& 1888	&&	82.3\% &267.1\\
	&QSO					& 128	&&	5.6\%  &18.1\\
	&stars					& 16		&&	0.7\%  &2.3	\\
\multicolumn{2}{l}{bad data}		& 260	&&	11.3\% & 36.8	\\
\hline
\hline
\end{tabular}
\end{center}
\label{table:obs:general:w3}
\end{table}%

\subsubsection{SCUSS + SDSS $ugri$ selection}
A similar target selection was applied to a combination of SCUSS $u$-band survey\footnote{\url{http://batc.bao.ac.cn/Uband/survey.html}} (Xu Zhou et al., 2014, in preparation, Hu Zou et al., 2014, in preparation) and SDSS $g,r,i$ photometry on a region of the sky of 25.7 deg$^2$ around $\alpha_{\rm J2000}\sim23^\circ$ and $\delta_{\rm J2000}\sim20^\circ$. 
This observation run, with a total of 8099 fibers allocated, measured ELG spectra with exposure times of 1h30 and covering 25.7 deg$^2$. 

The color selection used is similar as before but with a $u$ magnitude limit instead of a $g$ limit:

$-0.5<u-r<0.7\cdot(g-i)+0.1$ and $20<u<22.5$.

We also had a low priority ELG selection criterion (LOWP) to fill the remaining fibers. It is the same criterion stretched in magnitude and color to investigate the properties of the galaxies around the selection:
[ $20<u<22.7$ and $-0.9<u-r$ ] and [ $u-r<0.7\cdot(g-i)+0.2$ or $u-r<0.7$ ].

\begin{figure*}
\begin{center}
\includegraphics[width=7cm]{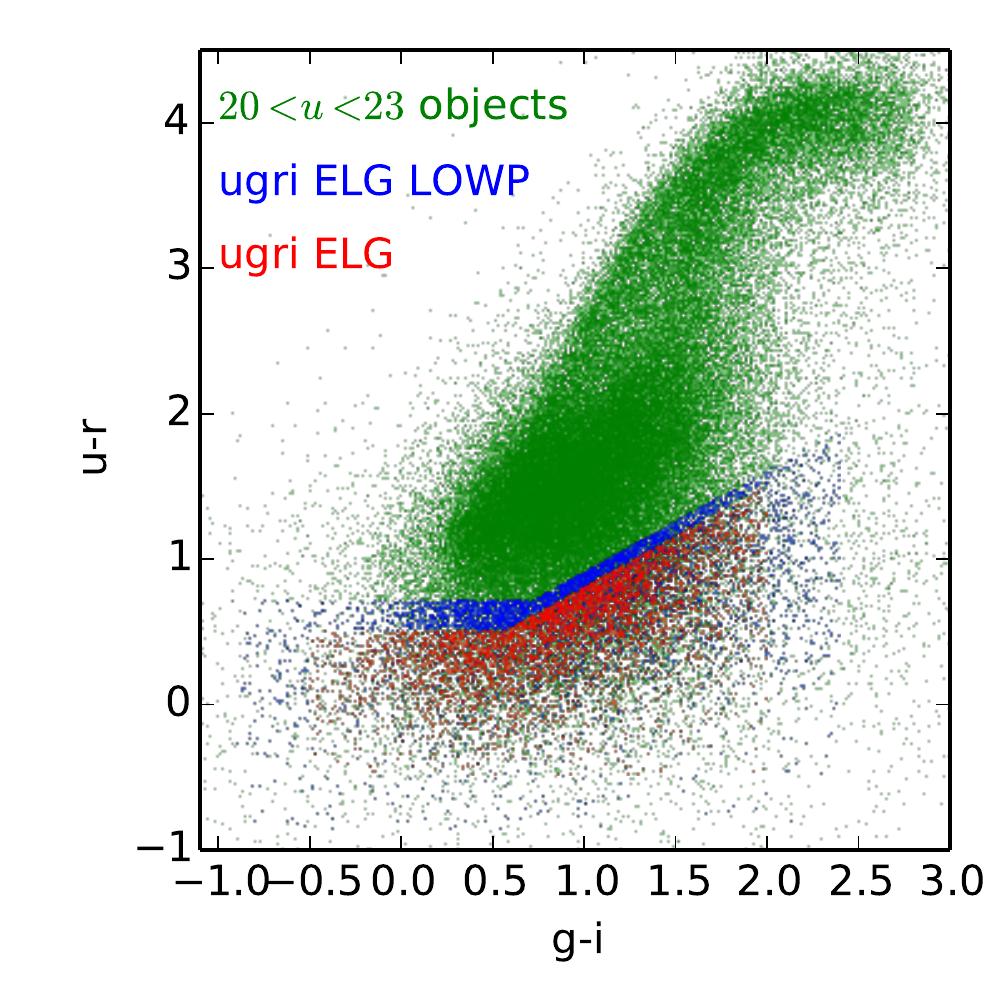}
\includegraphics[width=7cm]{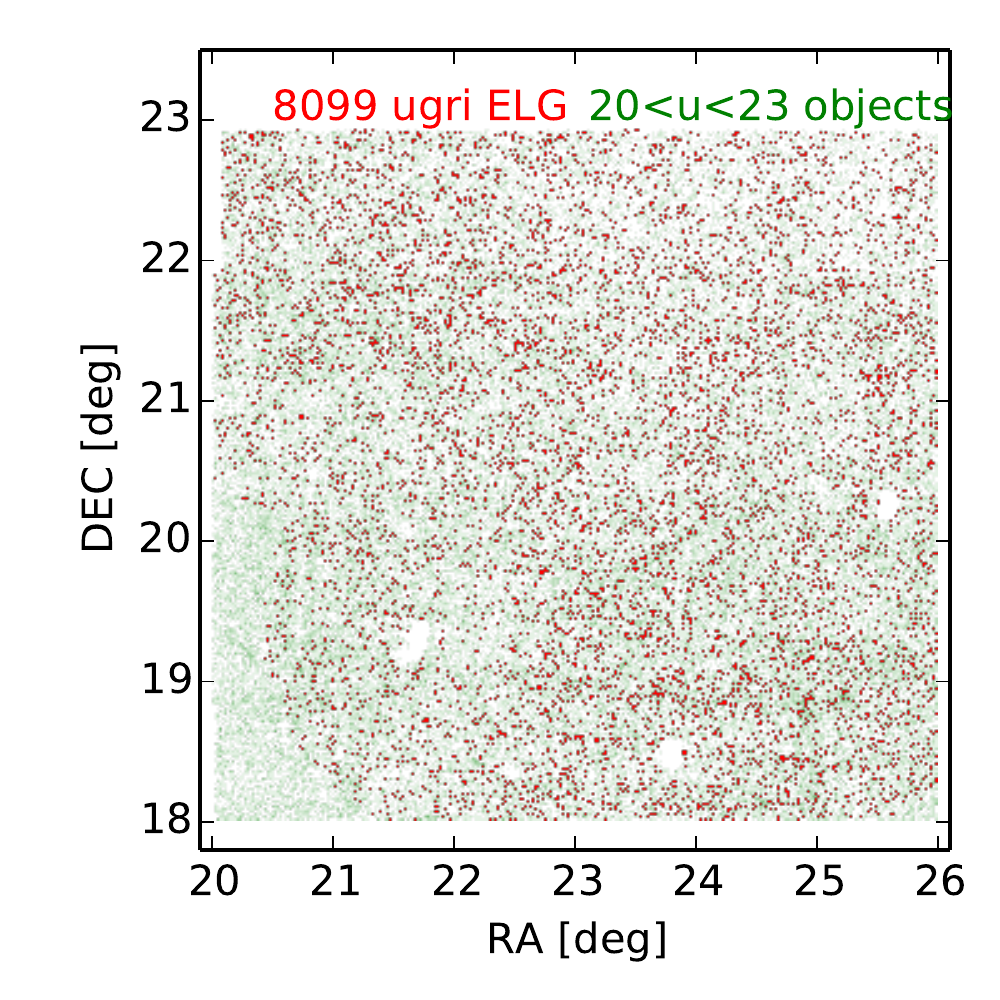} \\
\includegraphics[width=12cm]{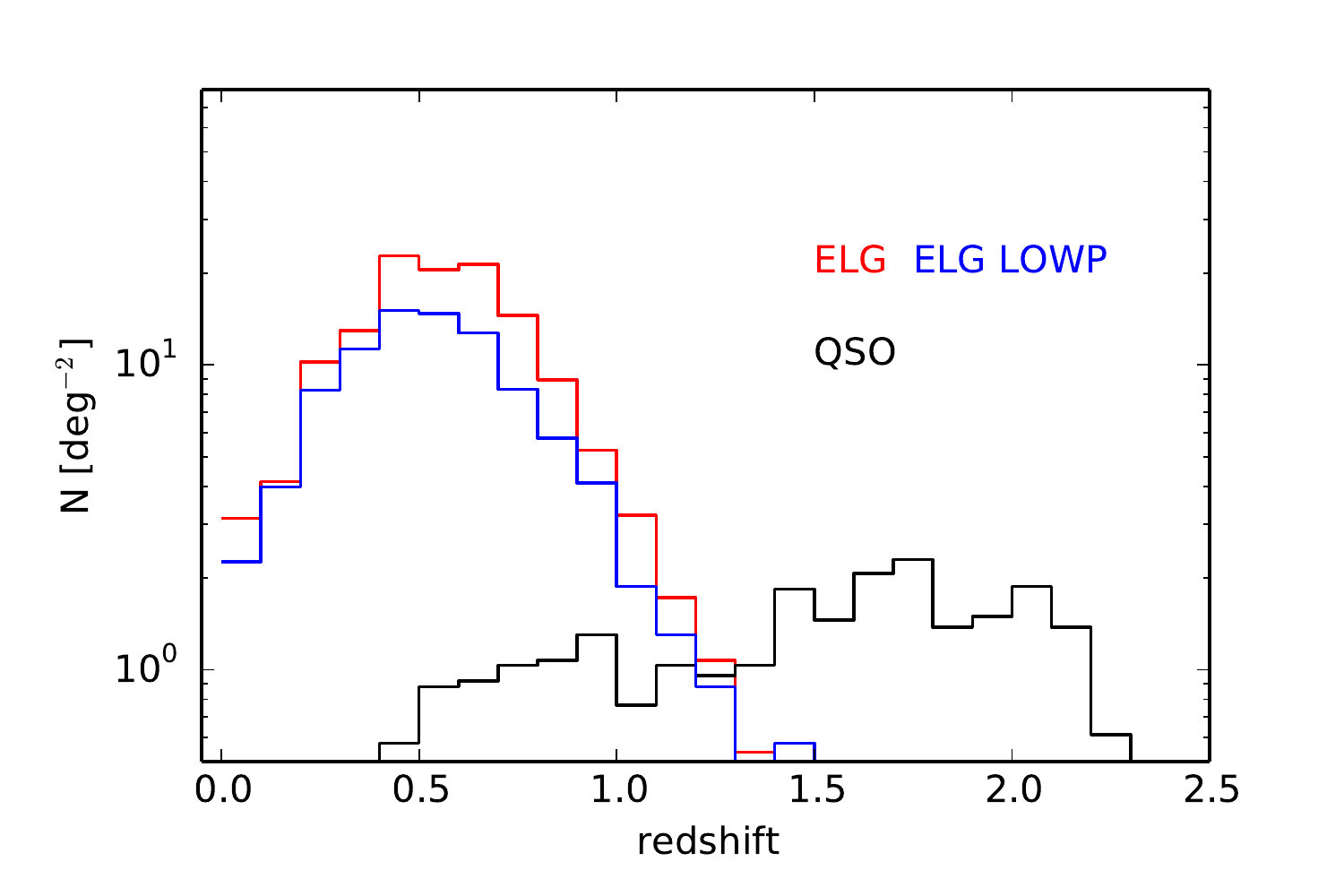}
\caption{Observations with SDSS telescope based on the SCUSS photometry. {\bf Top row}. Left panel. Color selection applied to the SDSS + SCUSS photometry. Right panel. 
RA, DEC in degree (J2000). All the $20<u<23$ objects (green) and the spectroscopic targets (red). 
{\bf Bottom}.  Observed redshift distributions per square degree per $\Delta z=0.1$ High priority ELG (red) are at slighter higher redshift than the low priority ELG (blue). This selection based on a shallower photometry than the CFHT-LS is not as efficient at selecting redshift 0.8 ELGs.}
\label{redshift:dist}
\end{center}
\end{figure*}

The BOSS pipeline was used to process the data and all the spectra were inspected to confirm the redshifts. The redshift distribution of this sample is shifted towards lower redshifts, see Fig. \ref{redshift:dist}, compared to the previous sample due to the shallower photometry (between 5 and 10 times shallower) from which the targets were drawn. This sample is complementary in terms of redshift and luminosity. Table \ref{table:observations} summarize the   the results of the observations. 

We retain the split of the two $ugri$ ELG samples because the parent photometry catalog are very different.

\begin{table*}
\caption{SCUSS -- SDSS-III/BOSS ELG observed as a function of spectral types.}
\begin{center}
\begin{tabular}{ c c c c c c c c c}
\hline \hline
Selection & type & \multicolumn{3}{c}{reliable redshift}  & \multicolumn{3}{c}{low confidence redshift}  \\ 
& & N & N [deg$^{-2}$]& percent &N & N [deg$^{-2}$]& percent & \\ \hline
\multirow{5}{*}{ELG} & All &  4914  &  188.26  & 100 &  \\ 
 & Galaxies &  3419  &  130.99  &  69.58  &  29  &  1.11  &  0.59   \\ 
 & Quasars &  676  &  25.9  &  13.76  &  85  &  3.26  &  1.73   \\ 
 & Stars &  129  &  4.94  &  2.63  &  3  &  0.11  &  0.06   \\ 
 & Lousy &  573  &  21.95  &  11.66   \\ \hline 
\multirow{5}{*}{ELG LOWP} & All &  3185  &  122.02 & 100 & \\ 
 & Galaxies &  2392  &  91.64  &  75.1  &  33  &  1.26  &  1.04   \\ 
 & Quasars &  206  &  7.89  &  6.47  &  22  &  0.84  &  0.69   \\ 
 & Stars &  71  &  2.72  &  2.23  &  0  &  0  &  0   \\ 
 & Lousy &  461  &  17.66  &  14.47   \\ \hline 
\end{tabular}
\end{center}
\label{table:observations}
\end{table*}%

\subsubsection{ Emission line flux measurement on BOSS spectra}

The flux in the emission line derived from BOSS spectra were estimated using two different pipelines, the redshift pipeline \citep{2012AJ....144..144B} and the Portsmouth pipeline\footnote{\url{http://www.sdss3.org/dr9/algorithms/galaxy_portsmouth.php#kinematics}} \citep{2013MNRAS.431.1383T}. These estimators produce consistent measurements.
The BOSS spectra were observed with fiber spectroscopy, and objects are typically larger than the area covered by the fibers, which have a diameter of 2 arcsecond. To correct this effect we compute the difference between the magnitude in a 2 arcsec aperture and the total model magnitude. Table \ref{color:weight:scheme:tab} gives the magnitude used as a function of the redshift of the galaxy. If the difference is within the error of the total magnitude, we do not correct the measured \OII flux. If the difference is greater than the error, then we rescale the \OII flux, $f_{\rm measured}$, using the magnitudes $m$ in which the \OII line is located. The correction used is:
\begin{eqnarray*}
& f_{\rm total}=f_{\rm measured} 10^{| m-m_{\rm fiber}|/2.5} \; {\rm if }\; |m-m_{fiber}|>{\rm err}_m \\
& f_{\rm total}=f_{\rm measured} \; {\rm if }\; |m-m_{fiber}| \leqslant {\rm err}_m 
\end{eqnarray*}

We cannot tell if the part of the galaxy located outside of the fiber actually emits more or less than the part measured within the fiber. 
The mean correction in flux is $\sim2.7$, {\it i.e.}, the fiber captured on average $\sim$40\% of the total flux.

\subsection{Galactic dust correction}
We correct the measured \OII fluxes from the extinction of our galaxy using the \citet{1994ApJ...429..582C} law 
$f_{\rm corrected}(\lambda)=f_{\rm observed}10^{0.4 \;  {\rm E(B-V)} \;}$ where E(B-V) is taken from the dust maps made by \citet{1998ApJ...500..525S}.



\subsection{Final sample}
 
We combine the data samples previously described, to measure the observed \OII LF in the redshift range $0.1<z<1.65$. We set 8 redshift bins of width $\sim$1 Gyr.

We select reliable redshifts that have well-defined photometry in the two or three of the $ugriz$ optical bands that we are using for the weighting scheme (see next paragraph). Moreover, we request a detection of the \OII lines with a signal to noise ratio greater than 5. In total, we use around 20 000 spectra. The total amount of spectra provided by each survey as a function of redshift is given in Table \ref{tab:data:available:vol}. 

This conjunction of surveys has a gap in redshift around redshift 0.45. We minimized the impact of this gap by shrinking to the minimum the redshift bin $0.4<z<0.5$ and removed it from the analysis. 

We tested the robustness of LF against the SNR limits between 3 and 10. We distinguished two regimes: for an SNR limit decreasing from 10 to 5, the uncertainty on the LF decreases and the luminosity limit decreases as the sample grows in size. For a SNR limit at 4 or 3, the number of detections increases, but the LF is not determined with more precision. In fact, for such low significance detections, the weights have a large uncertainty, which impacts the LF. The optimum results are obtained using an SNR limit of 5.

We do not use the DEEP2 \citep{2009ApJ...701...86Z,2013ApJS..208....5N}, HETDEX \citep{2013ApJ...769...83C}, and narrow band survey data from the Subaru Deep Field\citep{2007ApJ...657..738L}, from HiZELS \citep{2012MNRAS.420.1926S}, or the UKIDSS Ultra Deep Survey field \citep{2013MNRAS.433..796D} as an \OII LF measurement was already performed. Rather we compare their \OII LF measurements to ours. 

\begin{figure*}
\begin{center}
\includegraphics[height=52mm]{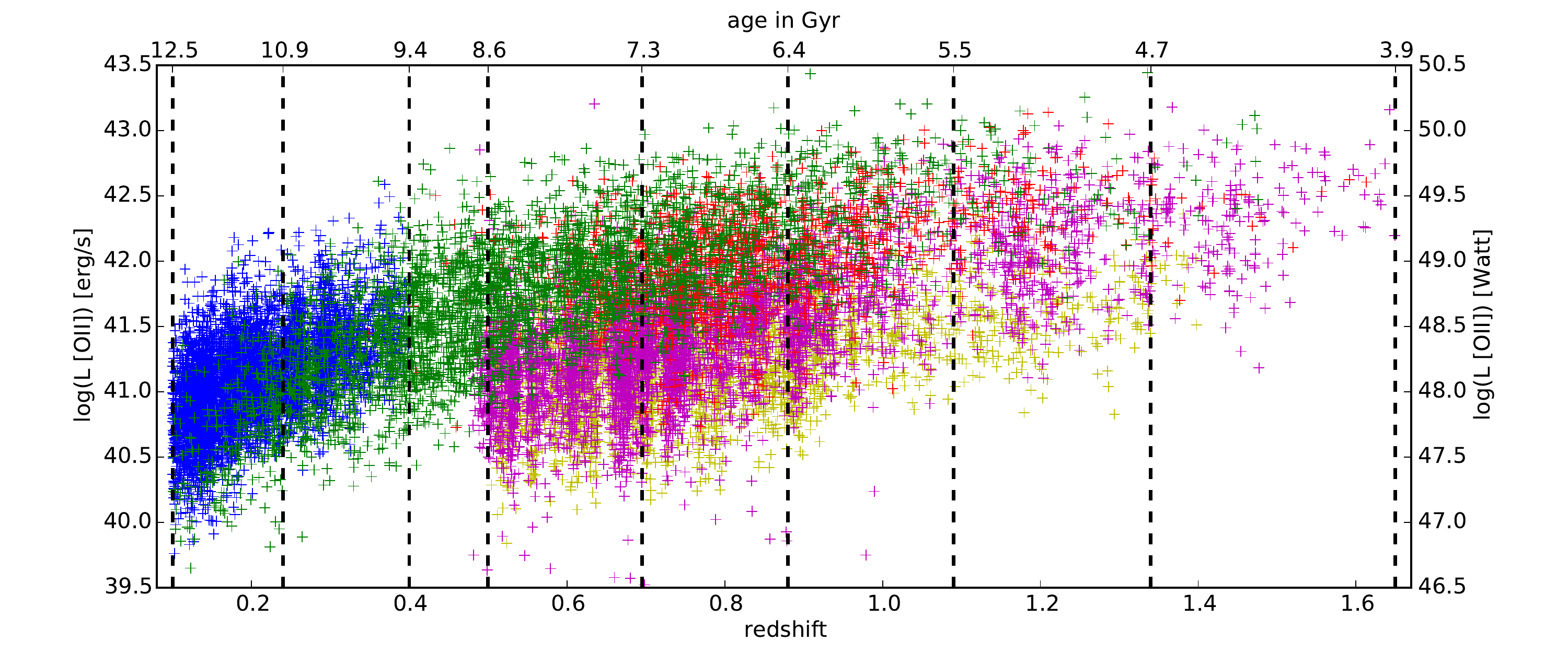}
\includegraphics[height=52mm]{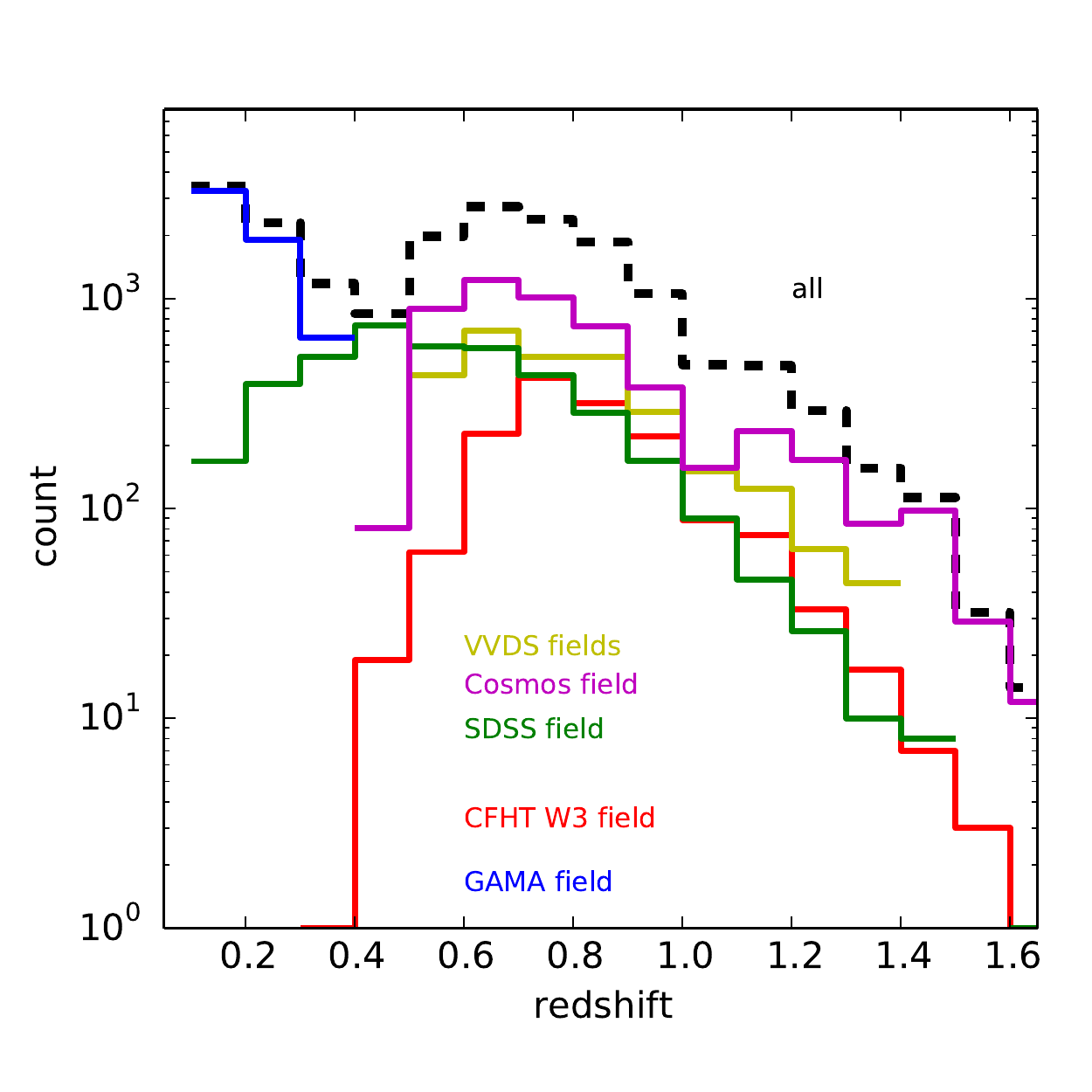}
\caption{\label{LF:weights:redshift} Left panel. Luminosity vs. redshift. In each redshift bin, the horizontal black lines show the $5\sigma$ completeness limit of the data presented, except in the $0.4<z<0.5$, where we show the high completeness limit (see subsection 4.1 for details). Right panel. Redshift distribution by survey, on the Cosmos field (purple), on the CFHT-LS W3 field (red), and on the SCUSS - SDSS field (green).}
\end{center}
\end{figure*}

\begin{landscape}
\begin{table}
\caption{Number of galaxies used in each redshift bin, by survey. The ELG VLT objects are at the same location of the sky that zCOSMOS and gathered in the same column. The data span over six different sky regions (VVDS has two).}
\begin{center}
\begin{tabular}{r r r r r r r r r r r r r r}
\hline \hline
$z_{min}$ & $z_{max}$ & GAMA & SEQUELS & BOSS W3 & VVDS & COSMOS & N total & 100$\sqrt{N}/N$ & volume & age at $z_{max}$  \\ 
 &   & & & & & & & & [$10^6$ Mpc${^3}$]& [Gyr] \\ 
\hline 
0.1  &  0.24  &  4136 & 314 & 0 & 0 & 0 &  4450  &  1.5  &  0.1  &    (12.5) 10.8  \\
0.24  &  0.4  &  1708 & 774 & 1 & 0 & 0 &  2483  &  2.0  &  0.32  &  9.4  \\
0.4  &  0.5  &  0 & 749 & 19 & 0 & 81 &  849  &  3.4  &  0.34  &  8.6  \\
0.5  &  0.695  &  0 & 1155 & 279 & 1113 & 2032 &  4579  &  1.5  &  0.97  &  7.3  \\
0.695  &  0.88  &  0 & 671 & 704 & 924 & 1652 &  3951  &  1.6  &  1.28  &  6.4  \\
0.88  &  1.09  &  0 & 315 & 342 & 579 & 711 &  1947  &  2.3  &  1.82  &  5.5  \\
1.09  &  1.34  &  0 & 84 & 126 & 231 & 465 &  906  &  3.3  &  2.55  &  4.7 \\
1.34  &  1.65  &  0 & 12 & 22 & 15 & 182 &  231  &  6.6  &  3.57  &  3.9  \\
\multicolumn{2}{c}{total in 0.1$<z<$1.65}   &  5844 & 4074 & 1493 & 2862 & 5123 &  19396  \\
\hline
\multicolumn{2}{c}{area [deg$^2$]} & 48.0 & 25.7 & 7.1 & 0.4 or 3.2 & 1.7    \\

\hline
\end{tabular}
\end{center}
\label{tab:data:available:vol}
\end{table}%
\end{landscape}

\section{\OII luminosity function}
\label{sec:LF}
Based on the \OII catalog constructed in the previous section, we measure the \OII luminosity function.

\subsection{Weighting scheme}
The weighting scheme relates the observed galaxies to their parent distribution, in this subsection, we describe a novel technique to compute the weights that allows combining different surveys.
\subsubsection{Principle}
In a dust-free theory, the \OII emitter population can be completely described by three parameters, the redshift, the continuum under the line (or the line equivalent width, EW) and the UV-slope that produced this emission. We denote $f$ the function that connects to a point in the three dimension space ($z$,EW,UV-slope) to a unique \OII flux. Observationally, these three parameters correspond to the emission line flux, the magnitude containing the emission line and the color preceding the emission line. 
In reality, the dust and orientation of each galaxy induces scatter in this parameter space introducing some scatter to the function f. The surroundings of each parameter of the relation should be considered so that $f$ can still be used to relate the observed distribution to the parent distribution. 

\begin{figure*}
\begin{center}
\includegraphics[width=160mm]{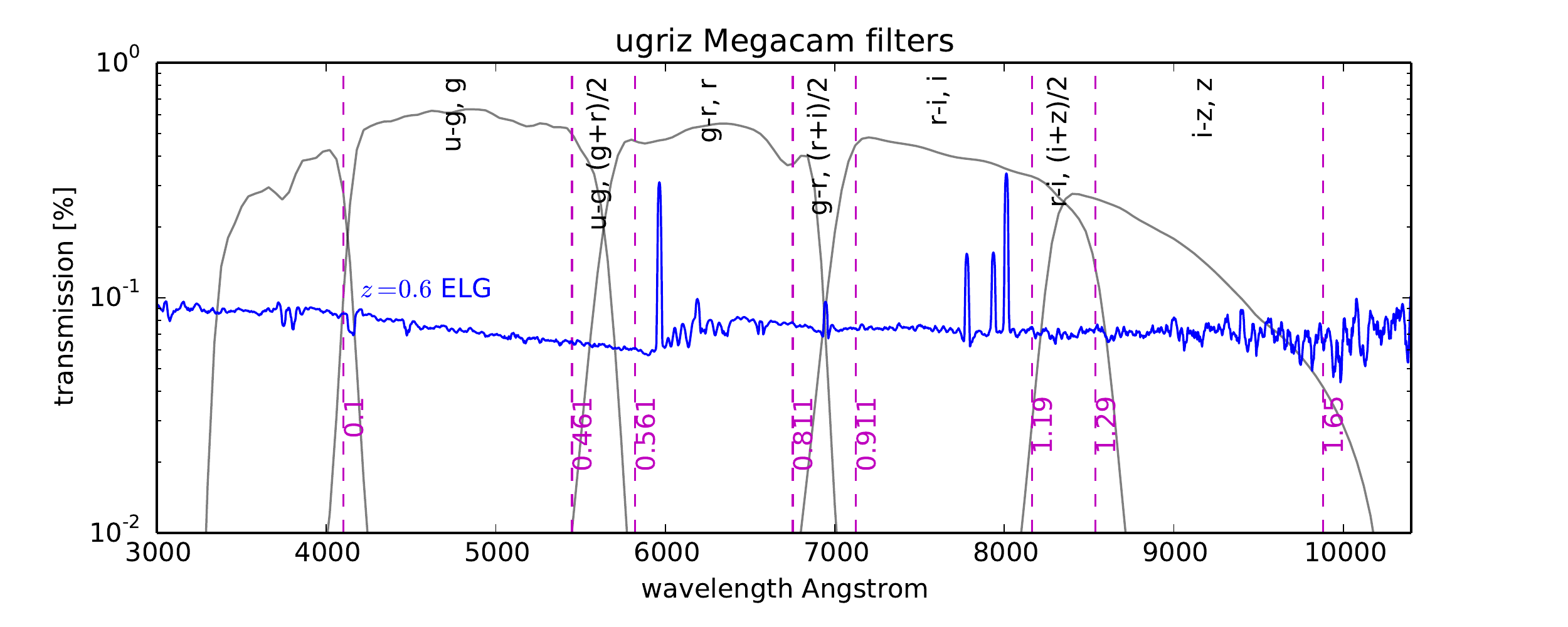}
\caption{Weighting scheme. Megacam transmission filters (grey lines) and the redshift bins (purple vertical lines) used to calculate the completeness weights. A rescaled spectrum of a typical emission line galaxy at redshift 0.6 is displayed in blue.}
\label{color:weight:scheme}
\end{center}
\end{figure*}

To implement this weighting scheme, we use the Megacam\footnote{\url{http://www.cfht.hawaii.edu/Instruments/Imaging/Megacam/specsinformation.html}} and SDSS photometric broad band $ugriz$ filters \citep{1996AJ....111.1748F,1998AJ....116.3040G} systems (the SCUSS $u$ filter is the same as the SDSS $u$ filter) to assign a magnitude and a color as a function of the redshift of each galaxy. For example, in the Megacam system, a galaxy with a redshift in $0.1<z<0.461$ will see its \OII line fall in the $g$ band, we thus use the $g$ magnitude and the $u-g$ color to compare this galaxy to the complete population. For \OII redshifts in $0.461<z<0.561$, we consider this zone as the overlap region between the $g$ and the $r$ filters. The boundaries are defined by the corresponding redshift of transition between the two bands $z_b$ broadened by 0.05, thus a transition of $z_b\pm0.05$. In this bin, we use $(g+r)/2$ as the magnitude and $u-g$ as the color. Table \ref{color:weight:scheme:tab} and Fig. \ref{color:weight:scheme} present the color and magnitude assigned for the weighing. This process allows a consistent weighting scheme for the complete data-set that has the same physical meaning when redshift varies.

\begin{table}
\caption{Weighting scheme as a function of redshift for the \OII lines. The redshift bins correspond to Megacam's or SDSS ugriz filter sets transitions.}
\begin{center}
\begin{tabular}{ c c c c r r r}
\hline \hline
CFHT $z$ range &SDSS $z$ range & magnitude & color  \\ \hline
 0.1 - 0.461 & 0.1 - 0.41 &  $g$		& $u-g$ 	\\
0.461 - 0.561 &0.41 - 0.51 & $(g+r)/2$	& $u-g$	\\
0.561 - 0.811 & 0.51 - 0.78 &$r$		& $g-r$	\\
 0.811 - 0.911 &0.78 - 0.88 & $(r+i)/2$	& $g-r$ 	\\
 0.911 - 1.19 &  0.88 - 1.17 &$i$		& $r-i$ 	\\
 1.19 - 1.29 &1.17 - 1.27 	 & $(i+z)/2$	&$r-i$ 	\\
 1.29 - 1.65 &1.27 - 1.65	 &	$z$		& $i-z$ 	\\
 \hline
\end{tabular}
\end{center}
\label{color:weight:scheme:tab}
\end{table}%

We compute the observed density deg$^{-2}$ of galaxies with a signal to noise ratio in the \OII emission line greater than 5 , $N_{\mathrm{observed \; with\; SNR} [O_{II}]>5}(z_{spec},m,c)$, as a function of spectroscopic redshift, magnitude $m$ and color $c$. This value is compared to the complete galaxy population $N_\mathrm{total}(z_{phot},m,c)$ to obtain a completeness weight,
\begin{equation}
W(z,m,c)=\frac{N_{\mathrm{observed \; with\; SNR} [O_{II}]>5}(z_{spec},m,c)}{N_\mathrm{total}(z_{phot},m,c)}.
\end{equation}

To obtain the information on the complete galaxy population $N\_\mathrm{total}(z_{phot},m,c)$, we use the photometric redshift catalogs from the CFHT-LS deep fields 1, 2, 3, 4 (WIRDS) \citep{Ilbert_06,2013A&A...556A..55I,2012A&A...545A..23B} and the Stripe 82 SDSS Coadd photometric redshift catalog \citep{2011arXiv1111.6619A,2012ApJ...747...59R} that span 3.19 deg$^2$ and 275 deg$^2$, respectively.

For redshifts $z<0.7$, we use the number densities observed on the Stripe 82 SDSS Coadd. For redshifts above, we the CFHT-LS deep photometric redshift catalog to obtain the best possible estimate of the parent density of galaxies (note that it is not necessary to have the parent photometry and the observed data on the same location of the sky).

We convert all the magnitudes to the CFHT Megacam system using the calibrations of \citet{Regnault_2009} to have a consistent weighting scheme among the various surveys.

\subsubsection{Implementation}
To implement the weights properly and avoid edge effects due to data binning, we adopt two 3D-tree\footnote{\url{http://eu.wiley.com/WileyCDA/WileyTitle/productCd-EHEP001657.html}} (one for the data and one for the parent sample) containing the redshift, the magnitude and the color normalized at their first and last deciles (D10 and D90), {\it i.e.}, we remap the three quantities so that the information is primarily contained in the interval 0 - 1
\begin{eqnarray}
{\rm z_{tree}=\frac{z - D10_{z}}{D90_{z}-D10_{z}}}, \\ {\rm m_{tree}=\frac{mag - D10_{mag}}{D90_{mag}-D10_{mag}}}, \\  {\rm c_{tree}=\frac{color - D10_{color}}{D90_{color}-D10_{color}}}.
\end{eqnarray}
This transformation allows the estimation of distance between two points in the trees without being biased by the distribution of each quantity. In this manner the distance between two points $i$, $j$,
\begin{equation}
\Delta_{i,j}^2=|z_{tree}^i - z_{tree}^j |^2 + |m_{tree}^i - m_{tree}^j |^2 + |c_{tree}^i - c_{tree}^j |^2
\end{equation}
represents more equally the three axes: color, magnitude and redshift.
We compute the number of galaxies around each galaxy $i$ in the observed sample tree and in the parent sample tree: $N(\Delta_\mathrm{i,obs}<0.15)$ and $N(\Delta_\mathrm{i,parent}<0.15)$. The ratio of the two numbers gives the individual weight for each observed galaxy.

We tested, using the jackknife method, the technique against different remapping schemes and tree distances and found the values mentioned above to be stable and reliable; see appendix \ref{sec:weights} for more details. This method is more reliable than constructing color and magnitude bins, as binning can be very sensitive to the fine tuning of each bin value, in particular at the edges of the distributions. 

We estimate the sample variance uncertainty on $N(\Delta_\mathrm{i,obs}<0.15)$ and $N(\Delta_\mathrm{i,parent}<0.15)$ in two ways: the Poisson error and the \citet{2011ApJ...731..113M} `cosmic variance' estimator. In areas of high density of the 3D-tree, the Poisson error is negligible compared to the cosmic variance estimation. In regions of small densities, this relation is reversed. To avoid underestimating the error, we consider the sample variance error as the maximum of the two estimators.

The uncertainty on the weight ($wErr$) is computed by varying the position of the galaxy in both trees in all directions of the redshift, magnitude and color space by its error in each dimension. This approach produces an upper and lower value for the weight. Note that the uncertainty in redshift is negligible compared to the error on the magnitude and color. 

The final uncertainty on the weight is the combination of the sample variance and of the galaxy weight error. The weight error dominates on the edges of the redshift - magnitude - color space, where densities are sparse. The sample variance error dominates in the densest zones of the redshift, magnitude, color space.

This method is very similar to that of \citet{2009ApJ...701...86Z} where they express the probability for \OII to be measured, denoted $f$, as a function of the object magnitude compared to the R-band cut, the B-R, R-I color cuts and of the probability to measure a good redshift with \OII in the spectra. From this probability they extract a completeness weight for each galaxy.

\subsection{Observed luminosity function}
We define the luminosity in the \OII lines by
\begin{equation}
L_{\left[\mathrm{O\textrm{\textsc{ii}}}\right]\,} \; [erg\, s^{-1}]=4 \pi \; \left( {\rm flux \; \left[\mathrm{O\textrm{\textsc{ii}}}\right] \; [erg\, s^{-1} cm^{-2}] } \right) \left[  \frac{{\rm d_{L(z)}}}{[{\rm cm}]}
 \right]^2,
\end{equation}
where ${\rm d_{L}}$ is the luminosity distance in cm expressed as:
\begin{equation}
{\rm d_{L(z)}}=(1+z) \frac{c}{H_0}\; \; \int_{u=0}^{u=z} \frac{du}{\sqrt{\Omega_\Lambda + \Omega_m (1+u)^3}}.
\end{equation}
The completeness limit is given by the completeness limit of the data sample with the highest sensitivity.

\begin{figure*}
\begin{center} 
\includegraphics[type=pdf,ext=.pdf,read=.pdf,width=6.cm]{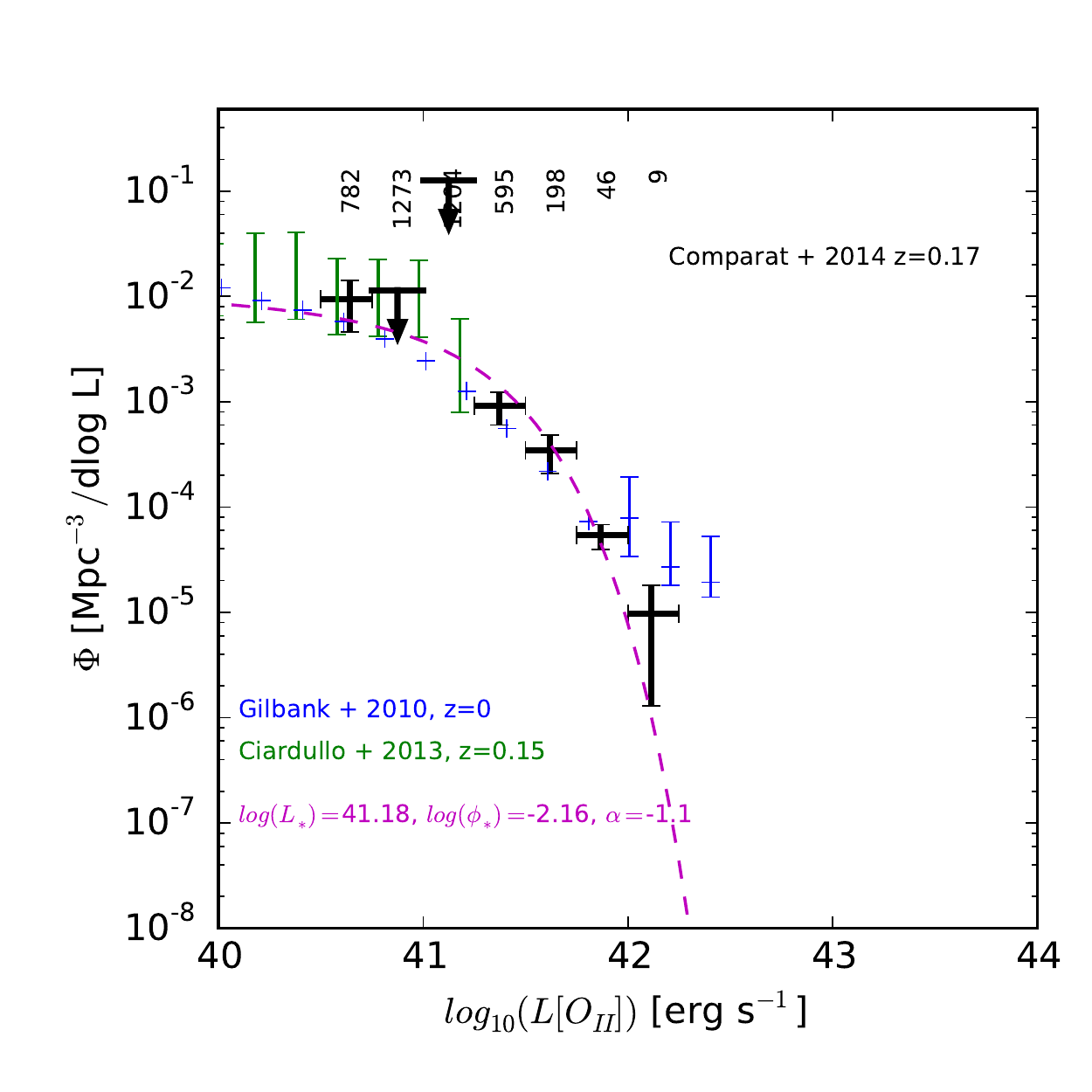}
\includegraphics[type=pdf,ext=.pdf,read=.pdf,width=6.cm]{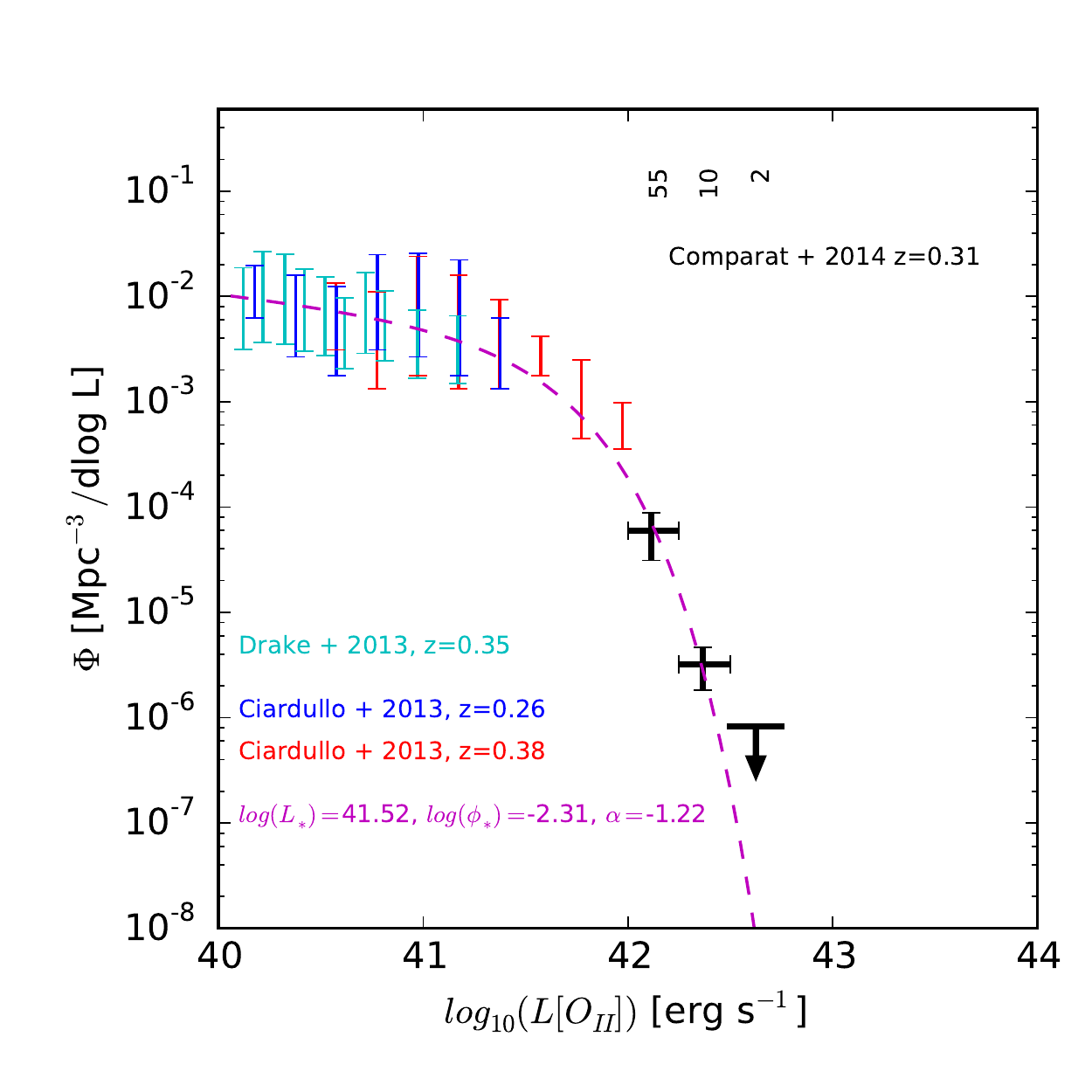}
\includegraphics[type=pdf,ext=.pdf,read=.pdf,width=6.cm]{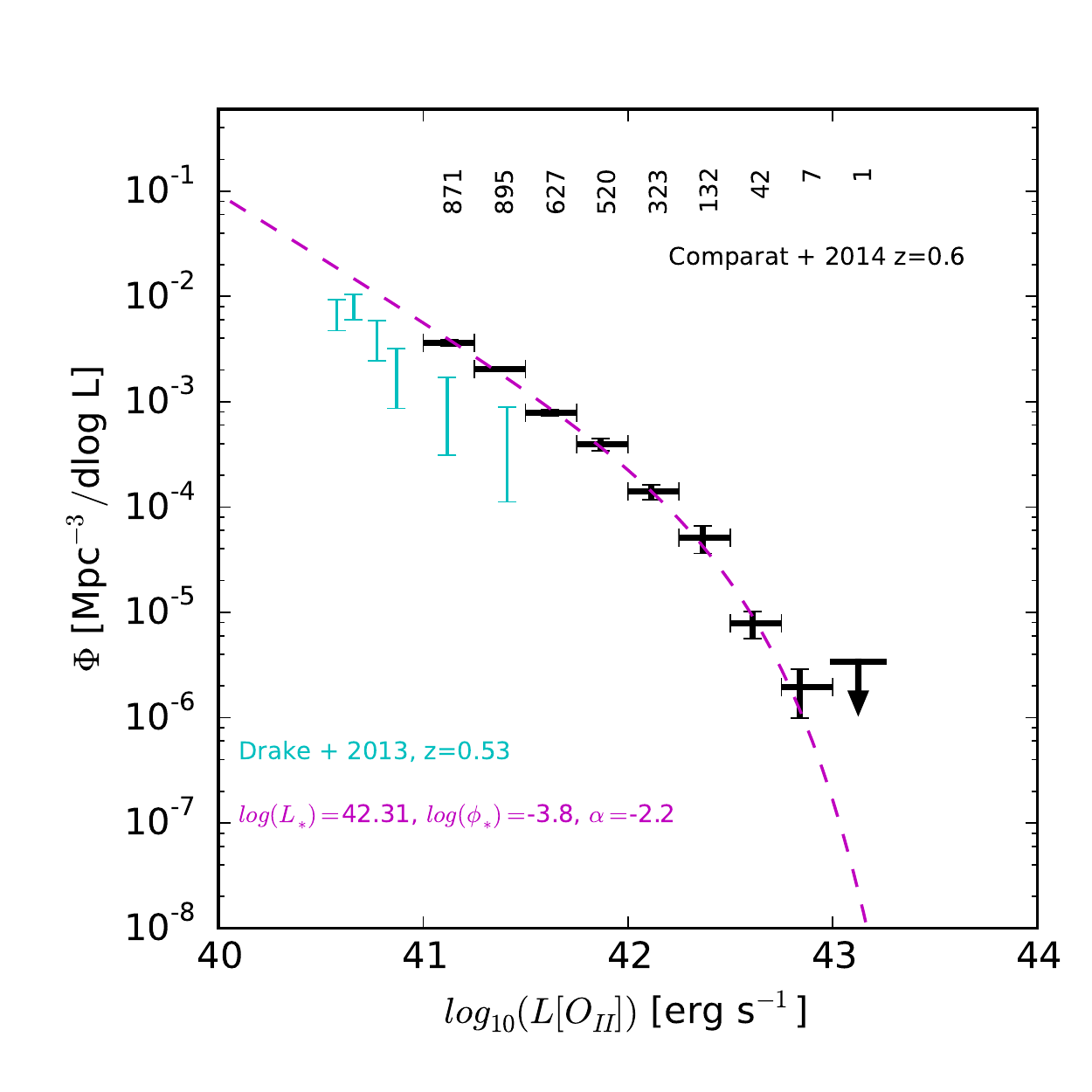}\\
\includegraphics[type=pdf,ext=.pdf,read=.pdf,width=6cm]{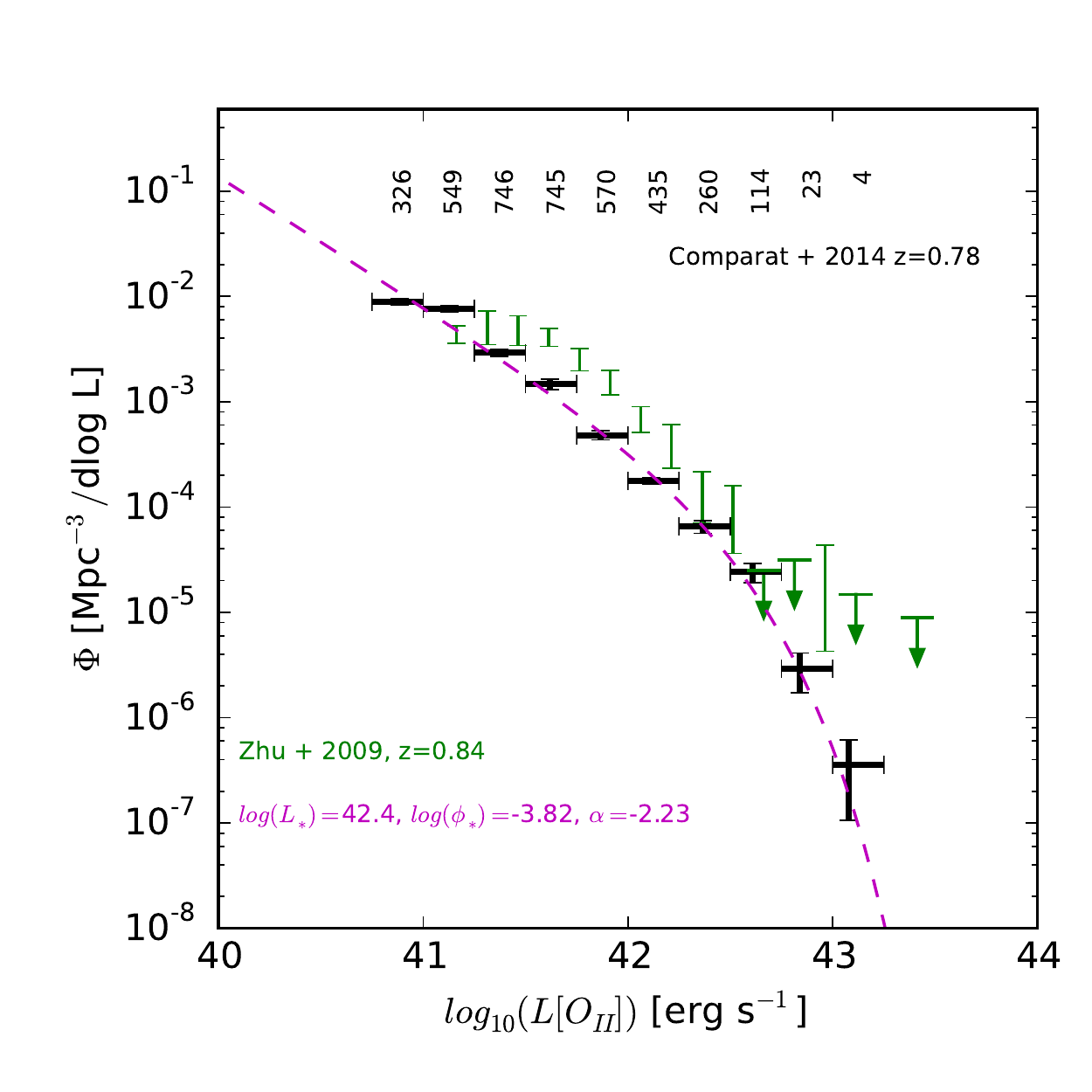}
\includegraphics[type=pdf,ext=.pdf,read=.pdf,width=6cm]{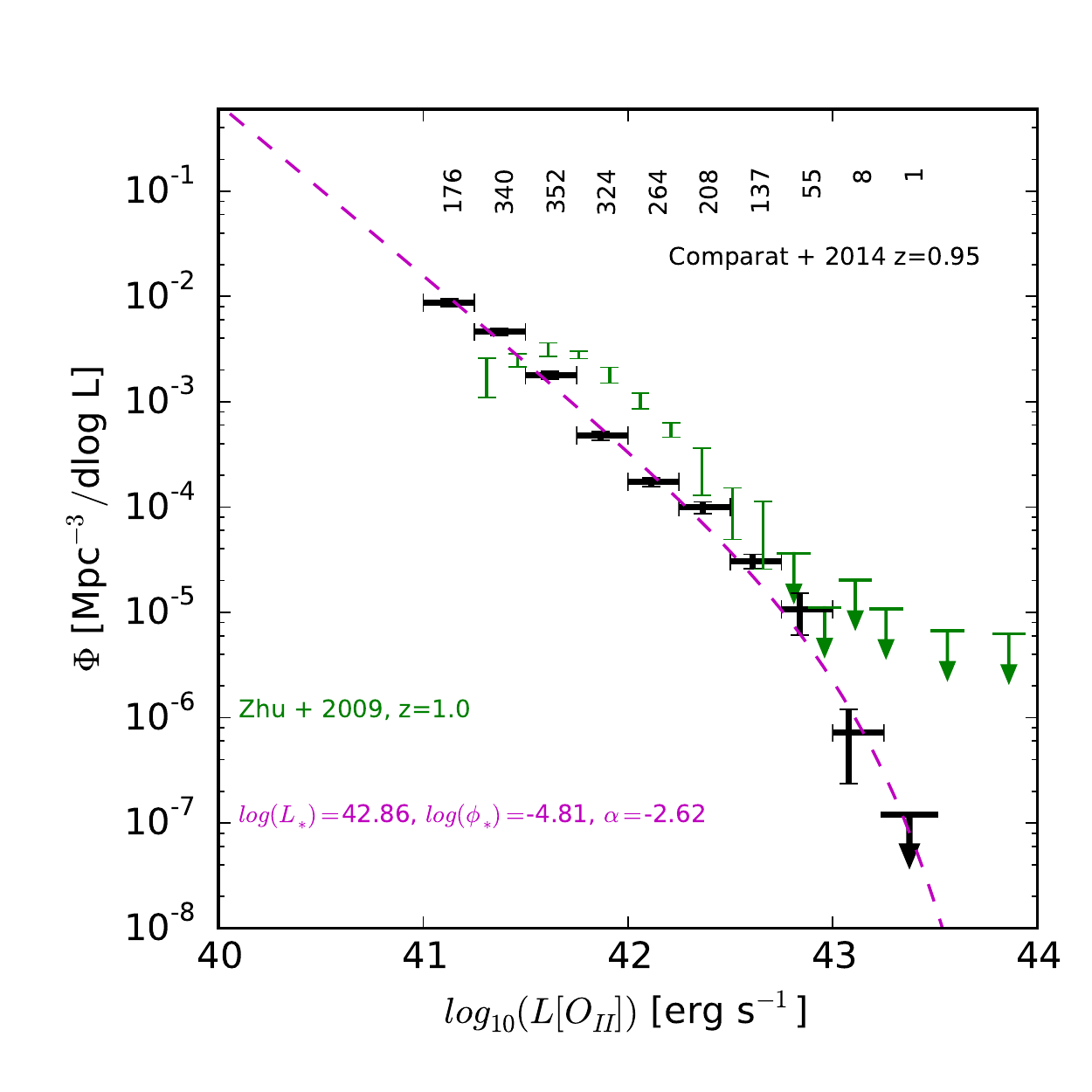}
\includegraphics[type=pdf,ext=.pdf,read=.pdf,width=6cm]{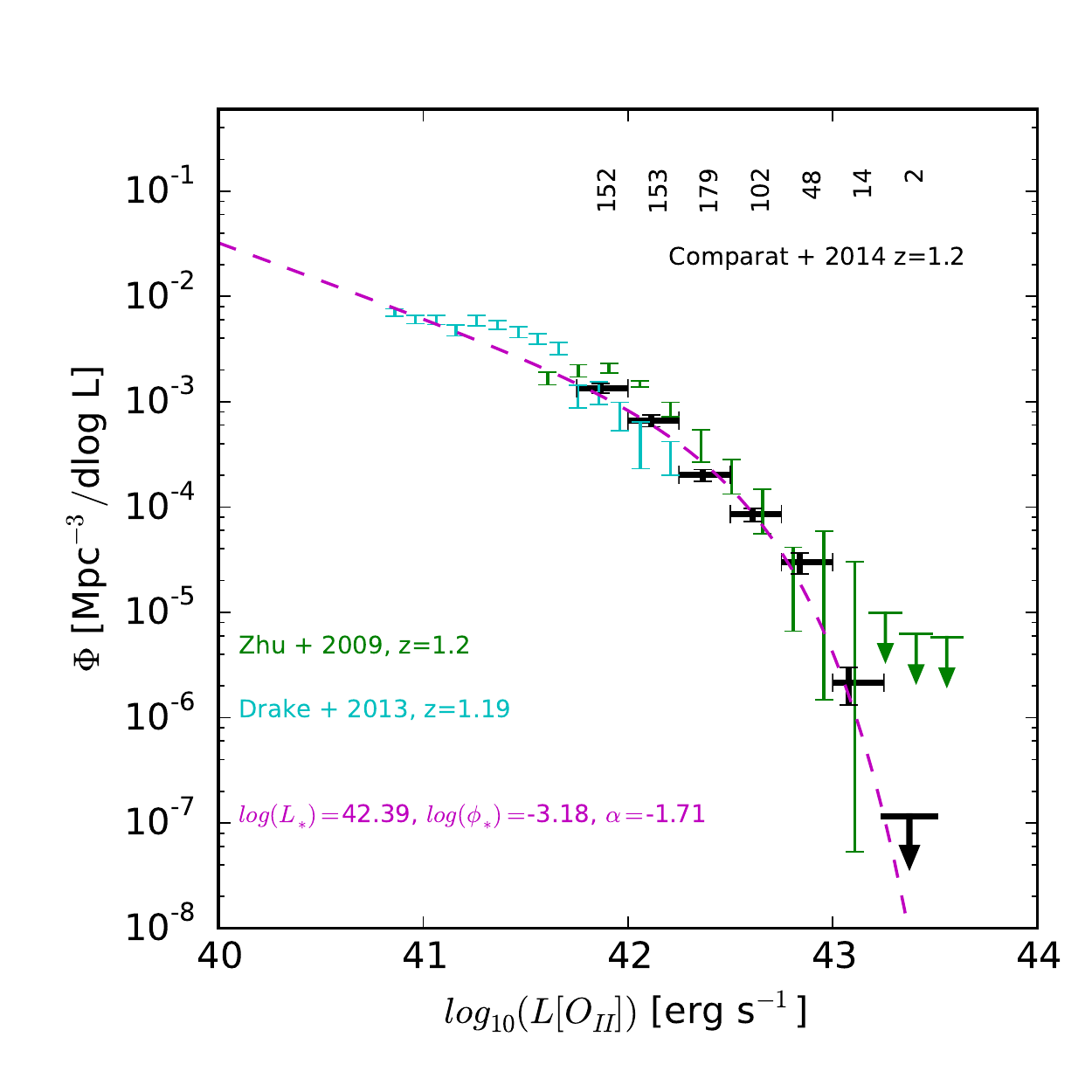}\\
\includegraphics[type=pdf,ext=.pdf,read=.pdf,width=6cm]{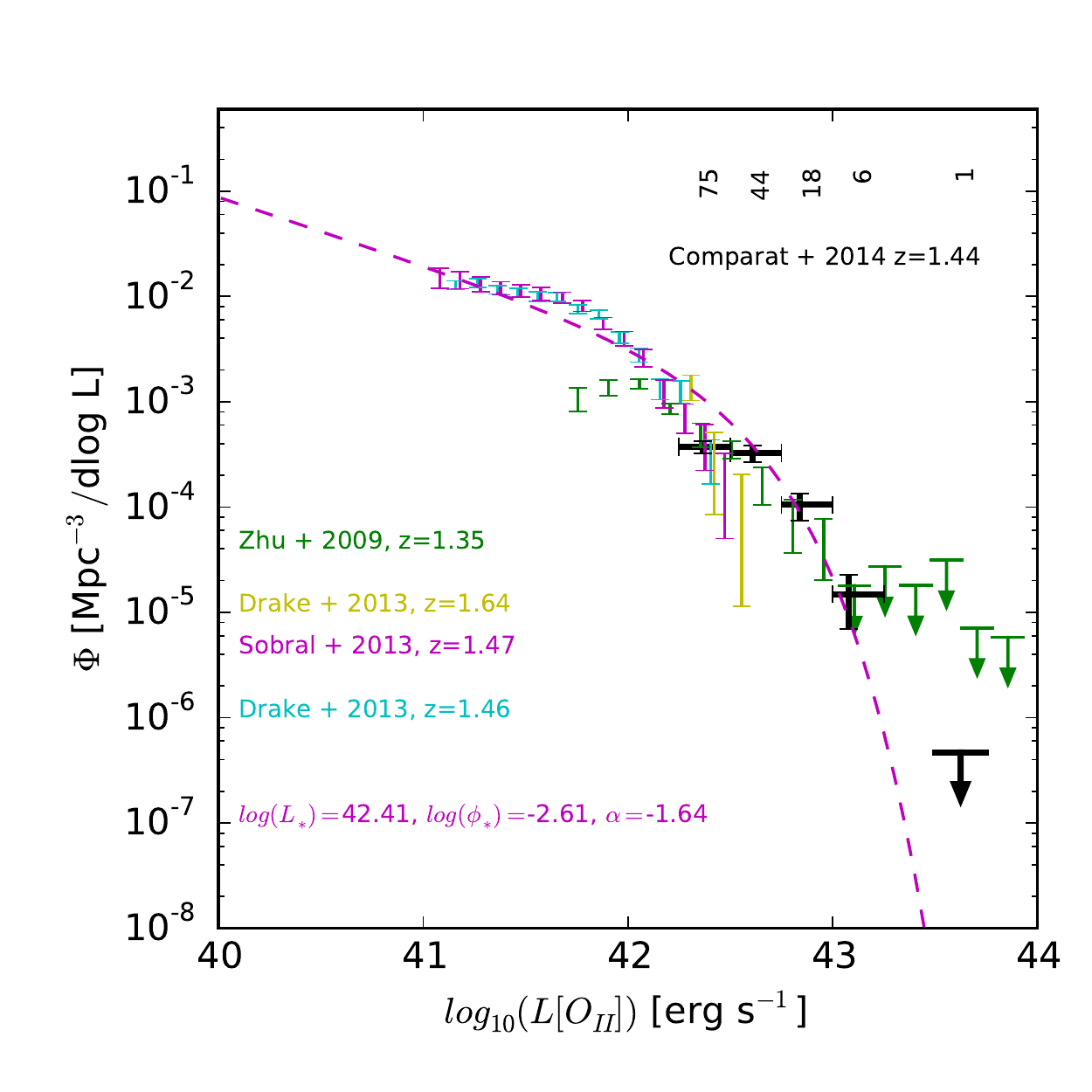}
\includegraphics[type=pdf,ext=.pdf,read=.pdf,width=6cm]{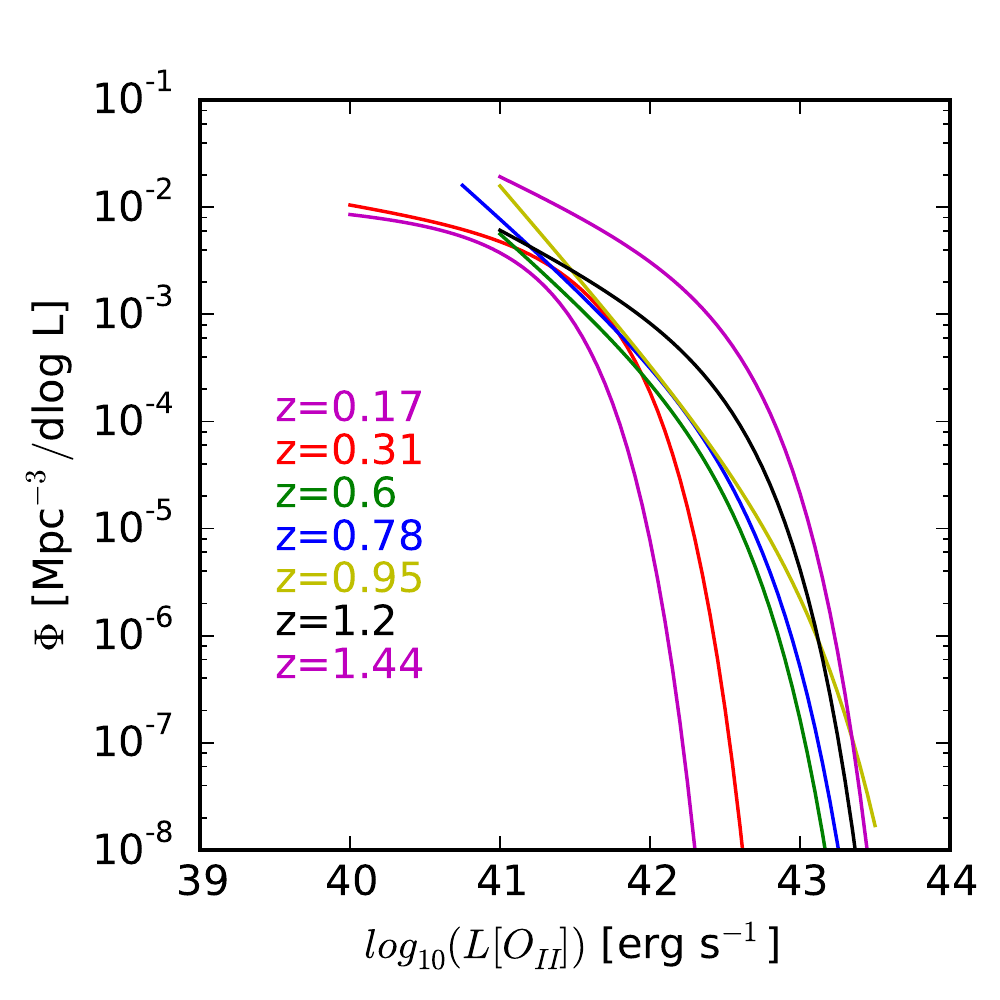}
\caption{\label{LF:measurement} Observed luminosity function compared to previous surveys estimates (in their closest redshift bin). The number on top of each LF point gives the exact number of galaxies used. The arrows going downwards correspond to measurements with an error consistent with 0. The Schechter functions fits are shown in magenta dashes. The last panel shows the evolution of the \OII LF from redshift 0.17 to 1.44 using the fits. The trend is that with redshift increasing there are more and more bright \OII emitters and the faint end slope gets steeper.}
\end{center}
\end{figure*}

We use the jackknife technique to estimate sample variance on the LF. We split the sample in 10 equal-in-number sub-samples and remeasure the LF on the subsamples. The standard deviation from the 10 estimations is our adopted sample variance error.
The LF measurements are shown in Fig.\ref{LF:measurement}. 
The error bars contain the error from the weight $(wErr)$ and the sample variance error computed with jackknife.
 
Thanks to the combination of the GAMA survey and the low redshift ELG observed in SEQUELS, we are able to estimate the \OII luminosity function at low redshift ($z<0.4$) and in particular measure accurately its bright end. The combined fit with the HETDEX measurement gives a good estimation of the parameters of the Schechter model.

In the redshift bin $0.4<z<0.5$, the completeness from GAMA drops and the other spectroscopic samples were selected to be at higher redshift. Therefore it is difficult to derive a clean LF in this bin and we exclude it from the analysis.

In the three redshift bins within $0.5<z<1.07$, we can compare to the \citet{2013MNRAS.433..796D} measurement in the first bin and to the \citet{2009ApJ...701...86Z} measurements in the following two bins. In the first bin the \citet{2013MNRAS.433..796D} measurement is at slightly lower redshift (0.53 compared to 0.6) and given the quick evolution of the LF at this epoch the discrepancy found is reasonable.
In the following two bins, given that the DEEP2 measurements are at a slightly higher redshift, they are brighter. In these bins, a \citet{1976ApJ...203..297S} model fits well the LF.
In particular, the faint completeness limit allows to fit well the faint end slope of the LF.

In the last two redshift bins, $1.07<z<1.65$, the LF measurement is in very good agreement with the previous DEEP2 measurements, although our data sets are limited to the bright end that correspond to a part only of the SFR density in these bins. To constrain the faint end slope of the Schechter fits in these bins, we use the measurements from \citet{2012MNRAS.420.1926S,2013MNRAS.433..796D}. We note that these measurement are in very good agreement in the overlapping region.

From an evolution point of view, our measurement shows clearly the evolution of the bright end of the \OII LF: as redshift increases from 0.165 to 1.44, there are more luminous \OII emitters. The last panel of Fig. \ref{LF:measurement} show the evolution using the fits. Thanks to the combination with the measurements of the faint end by \citet{2012MNRAS.420.1926S,2013MNRAS.433..796D} in the last two redshift bins and by \citet{2010MNRAS.405.2594G,2013ApJ...769...83C} in the first two redshift bins, we can also notice the steepening of the faint end slope from redshift 0.165 to redshift 1.44.

We computed directly (without using any fit) the integrated luminosity density; see Table \ref{luminosity:density:integrated}. Then using the latest \OII SFR calibration from \citet{2006ApJ...642..775M} we converted this luminosity density into a SFR density:
\begin{equation}
\frac{\rho SFR_{\left[\mathrm{O\textrm{\textsc{ii}}}\right]}}{\rm M_\odot \; yr^{-1} Mpc^{-3}}=\frac{2.18\times 10^{-41} \mathcal{L}(\left[\mathrm{O\textrm{\textsc{ii}}}\right])} {[{\rm erg \; s^{-1} Mpc^{-3}}]},
\label{sfr:kennicutt}
\end{equation}
Given that \citet{2006ApJ...642..775M} demonstrated that the \OII SFR is subject to uncertainties of $\sim$ 0.4 dex (a factor 2.5); the numbers in Table \ref{luminosity:density:integrated} should be considered with care. Note that, this result is not corrected from the extinction. Given that we integrate directly on the measurement and not on the fit, these values can only be considered as lower limits to the total SFR density.

The overall trend, that is independent of any fits, is the increase of the integrated luminosity density $\mathcal{L} (\left[\mathrm{O\textrm{\textsc{ii}}}\right]) (z=0.165)\sim39$ to $\mathcal{L}(\left[\mathrm{O\textrm{\textsc{ii}}}\right])(z=1.44)\sim40$, which confirms that there are more \OII emissions at $z>1$ than at $z<1$.

\subsection{Functional form of the luminosity function}
The number density of galaxies in the luminosity range $L+dL$, denoted $\Phi(L)dL$, is usually fitted with a \citet{1976ApJ...203..297S} function of the form
\begin{equation}
\Phi(L)dL=\phi_* \left( \frac{L}{L_*}\right)^\alpha \exp{\left(-\frac{L}{L_*}\right)} \, d\, \left(\frac{L}{L_*}\right)
\end{equation}
where the fitted parameters are $\alpha$ the faint end slope, $L_*$ [${\rm erg \cdot s^{-1}}$] the characteristic Schechter luminosity, and $\phi_*$ [Mpc$^{-3}$] the density of galaxies with $L>L_*$. The parameters fitted are usually highly correlated. 
\citet{2002ApJ...570L...1G} found a faint end slope of -1.2$\pm0.2$ with a Schechter model for ELGs in the local Universe. \citet{2010MNRAS.405.2594G} remeasured the \OII luminosity function in the local Universe, but found that a model with a double power-law and a faint end slope of $-1.6$ was a better representation of the data. \citet{2009ApJ...701...86Z} also found a double power law was a better description of the \OII LF. 

Our new LF measurement demonstrates that the Schechter model fits well the data. Based on the Schechter fits, we measure the evolution over 8 Gyr of $\log L_*$ from 42.41 at redshift 1.44 to 41.18 at redshift 0.165.

The parameter $\alpha$ is not well constrained in the literature, now, with this measurement, we have a better insight on its value and evolution. Beyond redshift $z>1.1$, the completeness limit of our sample is too bright to constraint $\alpha$, but combining with narrow band estimates of the LF enables the fit of $\alpha$. We measure the flattening of the LF from redshift 1.44 to redshift 0.165; see Table \ref{luminosity:density:integrated}. 

The results of the fits are summarized in Table \ref{luminosity:density:integrated} and are shown on the Fig. \ref{LF:measurement}.

\begin{landscape}  
\begin{table}
\caption{Fits on the luminosity function. Integrated luminosity density observed and corresponding SFR densities observed. %
The last column is computed with the relation provided by \citet{2014arXiv1403.0007M} eq. (15).}
\begin{center}
\begin{tabular}{c r c c c c c c c c c c c c}
\hline \hline
&\multicolumn{2}{c}{Validity}& \multicolumn{3}{c}{Schechter fit}& \\

&\multicolumn{2}{c}{$\log L$ range} &&  && && total  $\log \rho$SFR& percentage\\

mean& low&high &$\log L_*$ & $\log \Phi_*$  &$\alpha$&   $\log(\mathcal{L}^{\rm observed}(\left[\mathrm{O\textrm{\textsc{ii}}}\right]))$ &  $\log \rho SFR_{\left[\mathrm{O\textrm{\textsc{ii}}}\right]}$ & from UV and FIR& ratio\\

redshift & \multicolumn{2}{c}{[${\rm erg \cdot s^{-1}}$]}&[${\rm erg \cdot s^{-1}}$]&[Mpc$^{-3}$]&&[${\rm erg \cdot s^{-1}}$ Mpc$^{-3}$]& $M_\odot$ yr$ ^{-1}$ Mpc$^{-3}$& $M_\odot$ yr$ ^{-1}$ Mpc$^{-3}$&\%\\
\hline
0.165  &  40.0  &  42.5  & $ 41.18 \pm 0.08 $ & $ -2.16 \pm 0.4 $ & $ -1.1 \pm 0.2  $&  39.009  &  -1.652  &  -1.647  & $ 98.9 ^{+ 248.4 }_{- 39.4 }$ \\
0.307  &  40.0  &  42.7  & $ 41.52 \pm 0.04 $ & $ -2.31 \pm 0.55 $ & $ -1.2 \pm 0.2  $&  39.247  &  -1.414  &  -1.515  & $ 126.1 ^{+ 316.7 }_{- 50.2 }$ \\ 
0.604  &  41.0  &  43.5  & $ 42.31 \pm 0.1 $ & $ -3.8 \pm 0.3 $ & $ -2.2 \pm 0.1 $ &  39.045  &  -1.616  &  -1.285  & $ 46.7 ^{+ 117.3 }_{- 18.6 }$ \\ 
0.778  &  40.75  &  43.5  & $ 42.4 \pm 0.16 $ & $ -3.82 \pm 0.12 $ & $ -2.2 \pm 0.1$  &  39.273  &  -1.389  &  -1.176  & $ 61.3 ^{+ 154.0 }_{- 24.4 }$ \\ 
0.951  &  41.0  &  43.5  & $ 42.86 \pm 0.27 $ & $ -4.81 \pm -0.13 $ & $ -2.6 \pm 0.1 $ &  39.318  &  -1.344  &  -1.085  & $ 55.1 ^{+ 138.4 }_{- 21.9 }$ \\ 
1.195  &  41.0  &  43.5  & $ 42.39 \pm 0.1 $ & $ -3.18 \pm 0.35 $ & $ -1.7 \pm 0.1 $ &  39.447  &  -1.214  &  -0.985  & $ 59.0 ^{+ 148.2 }_{- 23.5 }$ \\ 
1.442  &  41.0  &  44.0  & $ 42.41 \pm 0.14 $ & $ -2.61 \pm 0.15 $ & $ -1.6 \pm 0.2 $ &  39.988  &  -0.673  &  -0.917  & $ 175.6 ^{+ 441.1 }_{- 69.9 }$ \\ 

\hline
\end{tabular}
\end{center}
\label{luminosity:density:integrated}
\end{table}%
\end{landscape}

\subsection{\OII flux limited redshift surveys, baryonic acoustic oscillation and emission line galaxy target selection}

Future large spectroscopic surveys that aim for a precise measurement of the baryonic acoustic oscillations (BAO) in the power spectrum of galaxies in the redshift range $0.7<z<1.6$, such as DESI\footnote{\url{http://desi.lbl.gov/}} or eBOSS\footnote{\url{http://www.sdss.org/sdss-surveys/eboss/}}, can be designed following three constraints. 

First, the measured power spectrum of the tracers surveyed must overcome the shot noise, which requires a high density of tracers. We can distinguish two regimes of selection below redshift $z<1.1$, and above.
At $z=0.7$, the power spectrum of the dark matter predicted by \textsc{CAMB} \citep{Lewis:1999bs} is $P(k=0.063 h$ Mpc$^{-1})=9.5\times 10^3 h^3$ Mpc$^{-3}$, thus a density of $\frac{3}{P(k)}\simeq3\times 10^{-4}$ Mpc$^{-3}$ is sufficient to overcome the shot noise by a factor three \citep{1986MNRAS.219..785K}. At $z=1.1$, $P(k=0.063 h$ Mpc$^{-1})=7.\times 10^3 h^3$ Mpc$^{-3}$, the density required is $4.2\times 10^{-4}$ Mpc$^{-3}$.
In the redshift range $0.7<z<1.1$, the massive $M>10^{11}M_\odot$ galaxy population consists of a mix of star-forming and quiescent galaxies, which number density is $\sim2\times 10^{-3}$ Mpc$^{-3}$ \citep{2013A&A...556A..55I}. The galaxy densities to overcome shot noise are therefore reachable either with the quiescent or the star-forming galaxies. 
At $z=1.6$, $P(0.063 h$ Mpc$^{-1})=5.\times 10^3 h^3$ Mpc$^{-3}$, the density required is $6\times 10^{-4}$ Mpc$^{-3}$.
From the galaxy evolution point of view, a significant change occurs in the redshift range $1.1<z<1.6$: the massive end ($M>10^{11}M_\odot$) of the mass function becomes dominated by star-forming galaxies. 
The density of massive star-forming galaxies is around $\sim 10^{-3}$ Mpc$^{-3}$, whereas the density of massive quiescent galaxies drops from $6\times 10^{-4}$ Mpc$^{-3}$ at $z=1.1$  to $10^{-4}$ Mpc$^{-3}$ at $z=1.6$. 
Therefore above redshift $z>1.1$, the density of massive quiescent galaxies decreases too rapidly to overcome the shot noise in the power spectrum of galaxies. However, the density of massive star-forming galaxies in the redshift range $0.7<z<1.6$ is sufficient to sample the BAO: this tracer covers consistently this redshift range. Therefore to overcome shot noise and measure the BAO in the power spectrum of galaxies, one must target star-forming galaxies.

Secondly, because of the large load of required data in BAO experiments, accurate spectroscopic redshifts must be acquired in an effective manner. Star-forming galaxies have strong emission lines in their spectrum, they are therefore good candidates: \citet{2013MNRAS.428.1498C} demonstrated that one can select efficiently star-forming galaxies to sample the BAO to redshift $z=1.2$.
The \OII luminosity function measurement presented here extends this measurement to redshift 1.65 and provides insight on the galaxy population considered by future BAO studies compared to the global galaxy population.

Thirdly, there is the need to survey massive galaxies that are well correlated to the whole matter field (luminous and dark) in order to obtain the highest possible signal-to-noise ratio in their power spectrum. \citet{2013MNRAS.433.1146C} demonstrated that the color-selected galaxies for BAO have a relatively high galaxy bias $b\sim1.8$, and their luminous matter - dark matter cross-correlation coefficient measured using weak-lensing is consistent with 1. But we are not considering this point in this article. Therefore one needs to select the most luminous galaxies of the redshift range to maximize the galaxy bias.

To detect the BAO at redshifts above $z>0.7$, with optical spectrographs, an \OII flux-limited sample is therefore the best way to cover the entire redshift range in a minimum amount of telescope time with a dense enough galaxy population. This selection is equivalent to making a SFR-selection plus a dust selection (selecting galaxies with the least amount of dust) so that lines emitted are well observed (not obscured).
This sample will neither be a mass-limited sample nor a SFR-limited sample.

Based on the catalog gathered to compute the LF, we can derive relations between the \OII flux observed and magnitudes to help the planning of theses surveys, in particular the target selection algorithms.

We investigate the eventual correlations between the observed \OII fluxes and the $ugriz$ broad-band magnitude (in the CFHT Megacam system). We set two redshift ranges : 
\begin{itemize}
\item $0.7<z<1.1$ corresponding to eBOSS-ELG and where the \OII data is complete to a luminosity of $10^{41}$ ${\rm erg \cdot s^{-1}}$, which corresponds to a flux $2.3\times 10^{-17} $\uflux$ $ at redshift 0.9.
\item $1.1<z<1.6$ corresponding to DESi-ELG and where the \OII data is complete to a luminosity of $10^{42}$ ${\rm erg \cdot s^{-1}}$, which corresponds to a flux $8.2\times 10^{-17} $\uflux$ $ at redshift 1.3.
\end{itemize}
The magnitude that correlates best with the \OII flux is the $g$ band in the redshift range $0.7<z<1.1$ and the $r$ band in the range $1.1<z<1.6$; see Fig. \ref{mag:oii:core}. These bands should thus be used to construct in the most efficient way an \OII flux limited sample. The correlation in the higher redshift bin might be biased because below the flux completeness limit the data is not representative of the complete population: $u$ or $g$ band could also be used for targeting at redshift $1.1<z<1.6$.

\begin{figure}
\begin{center}
\includegraphics[width=80mm]{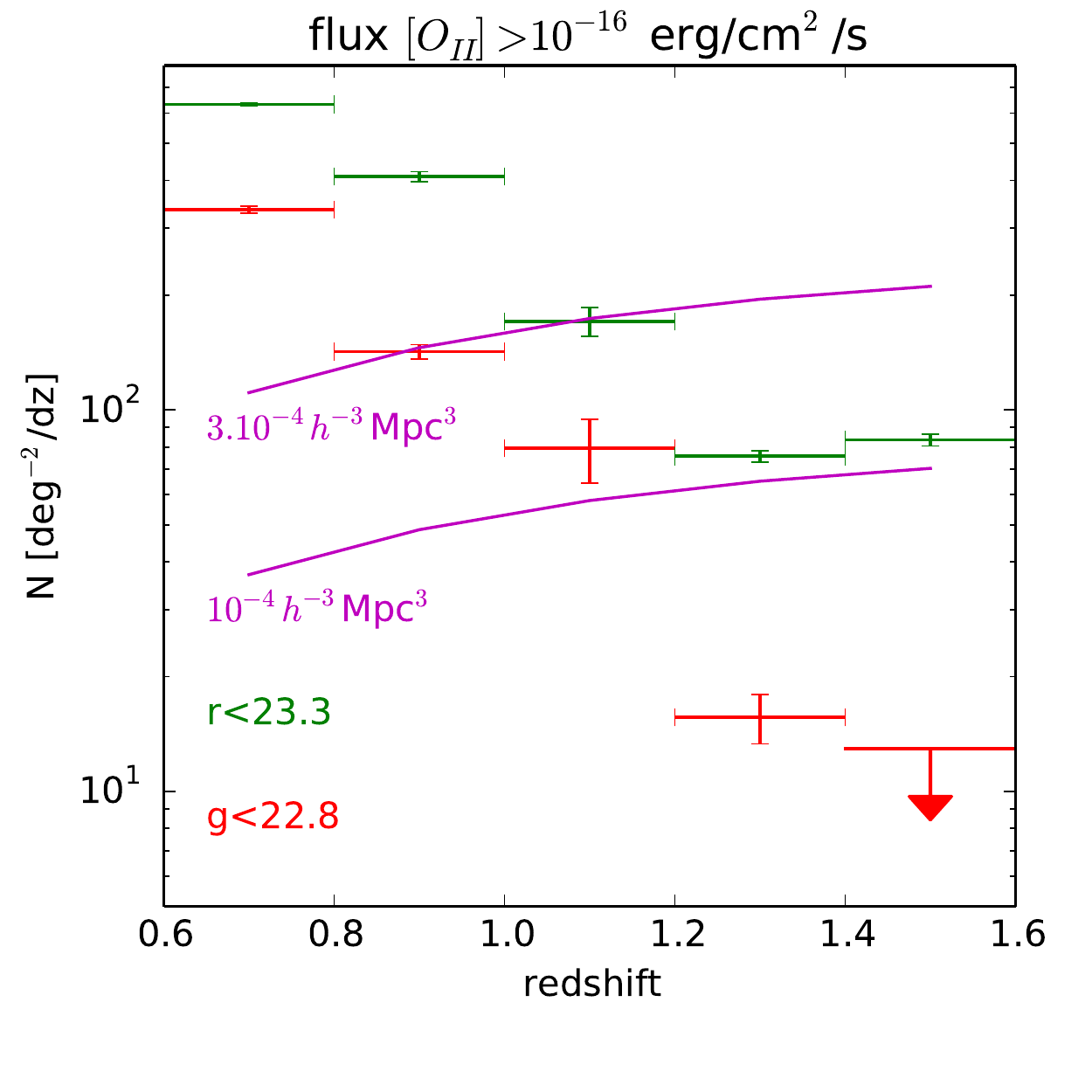}
\caption{\label{DESi:selection} Redshift distribution per square degree of galaxies with \OII flux greater than $10^{-16}$\uflux and magnitude $g$ brighter than $r<23.3$ (green) and $g<22.8$ (red) compared to constant density of $10^{-3}$ and $3\times 10^{-4}$ $h^3$Mpc$^{-3}$ (purple).}
\end{center}
\end{figure}

\begin{figure}
\begin{center}
\includegraphics[width=40mm]{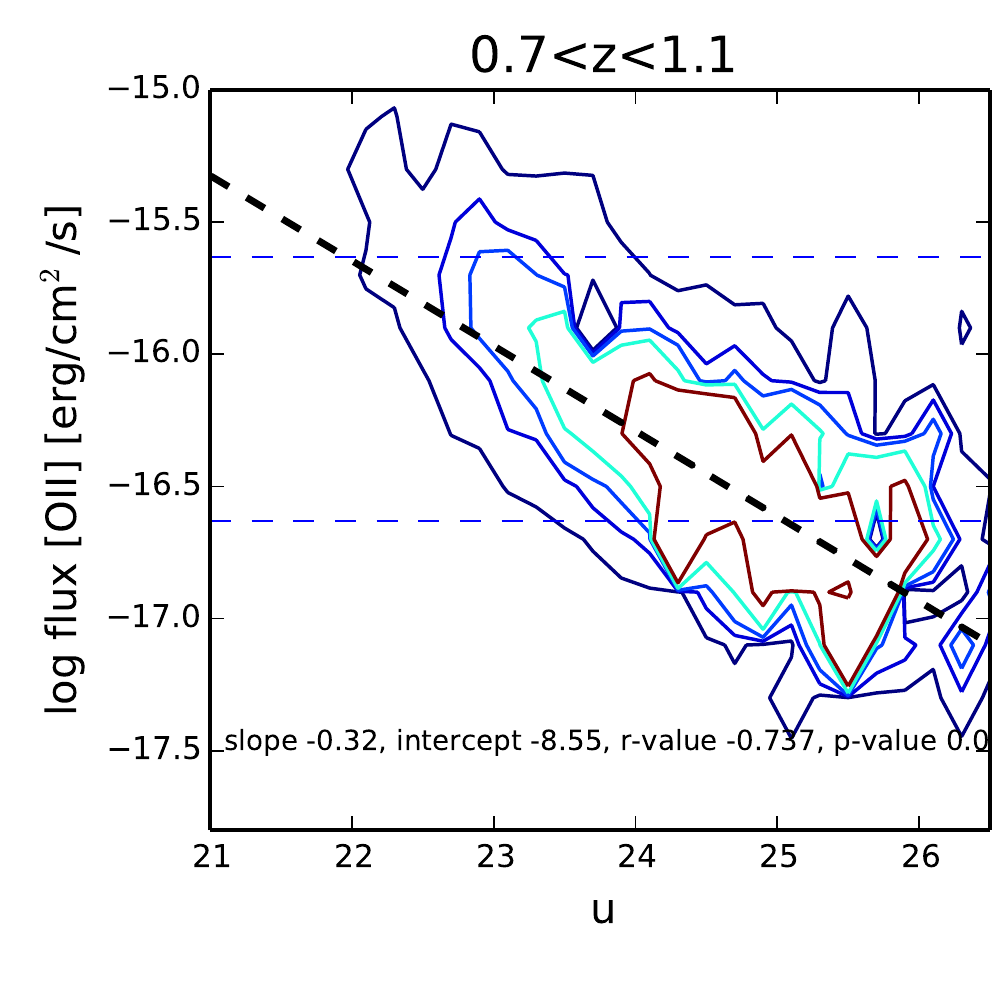}
\includegraphics[width=40mm]{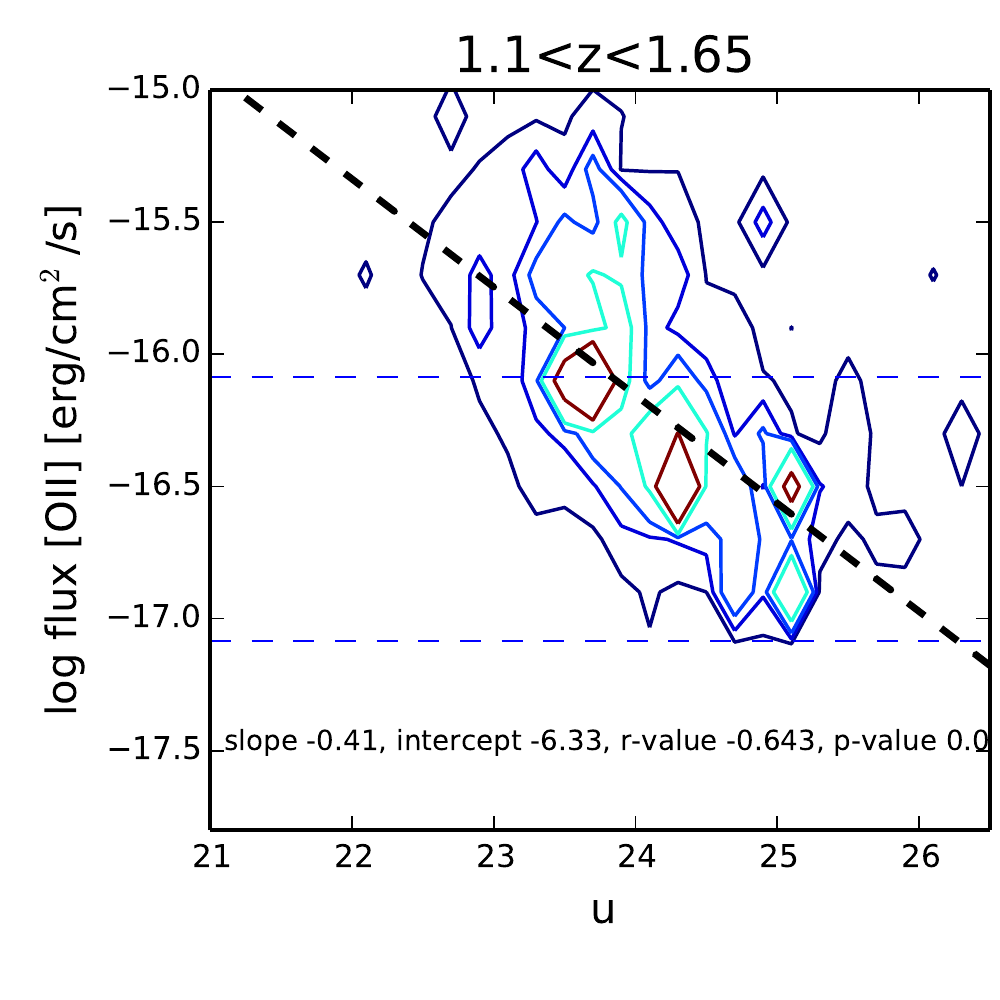}\\\includegraphics[width=40mm]{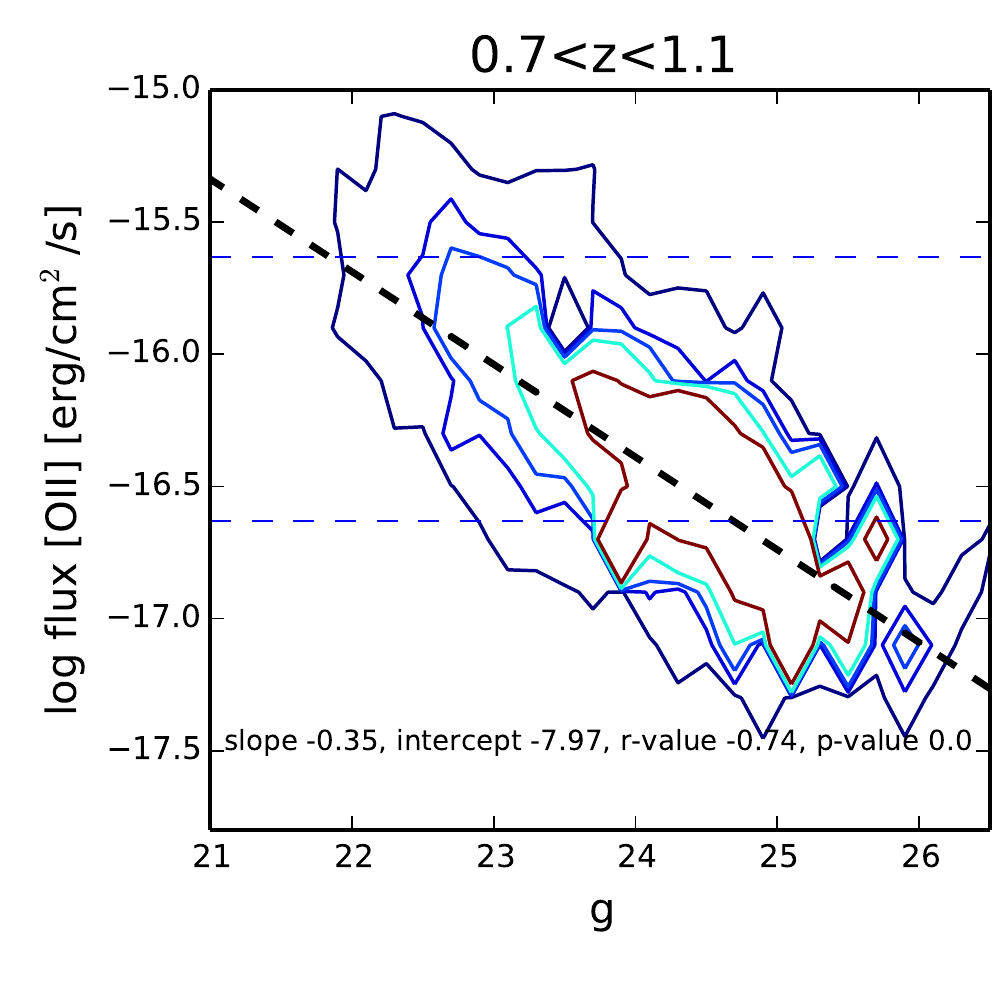}
\includegraphics[width=40mm]{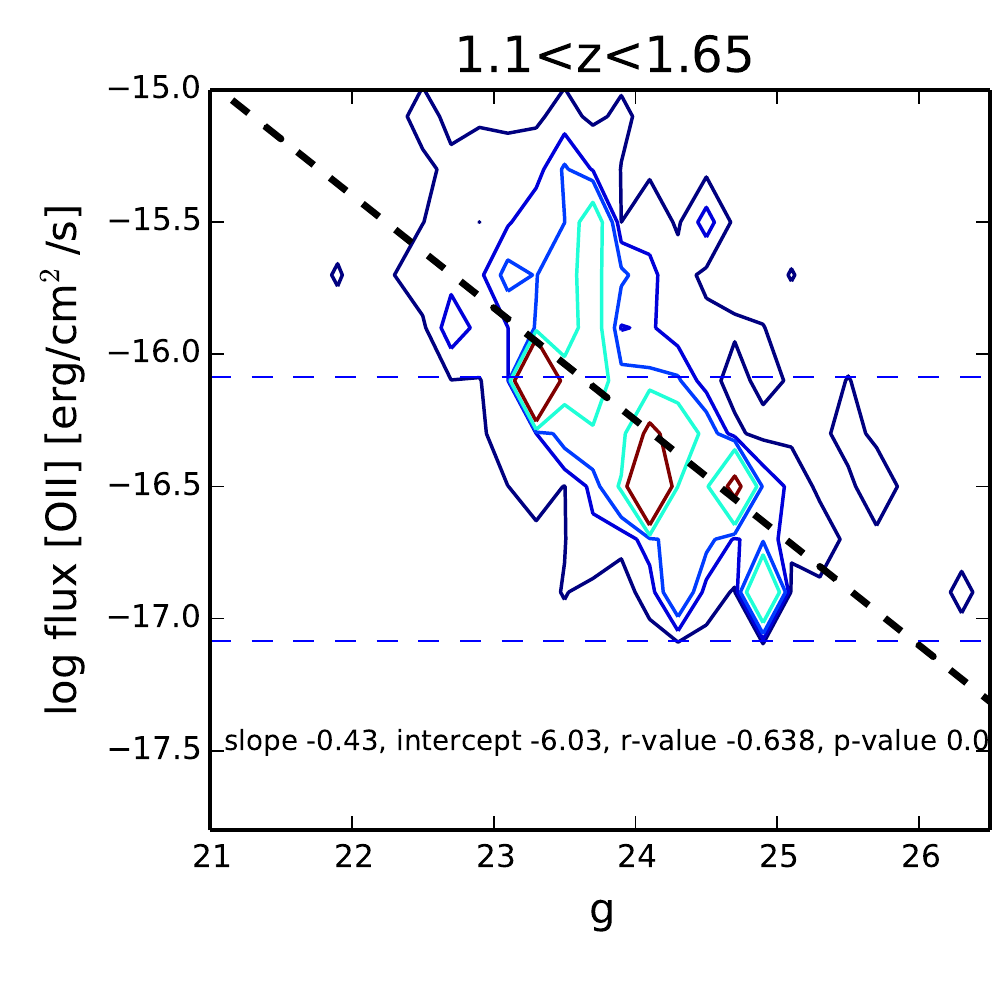}\\
\includegraphics[width=40mm]{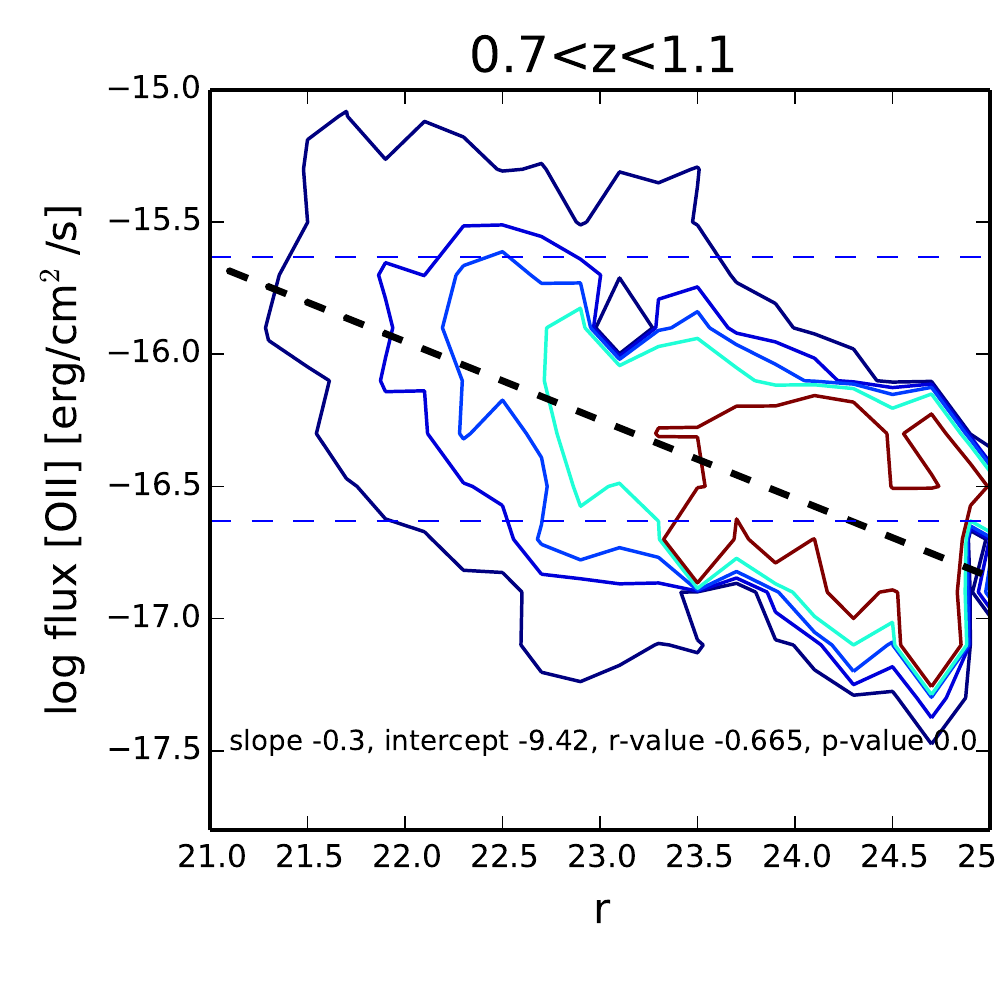}
\includegraphics[width=40mm]{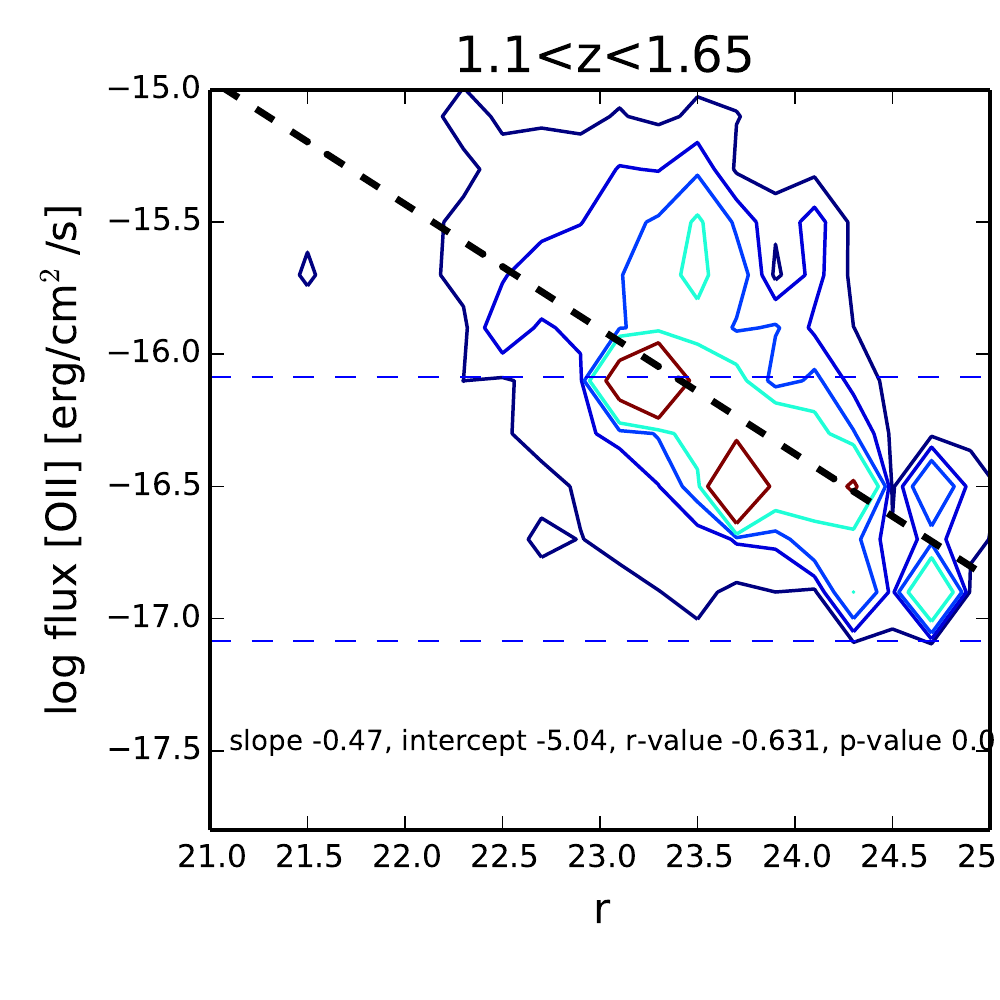}\\
\includegraphics[width=40mm]{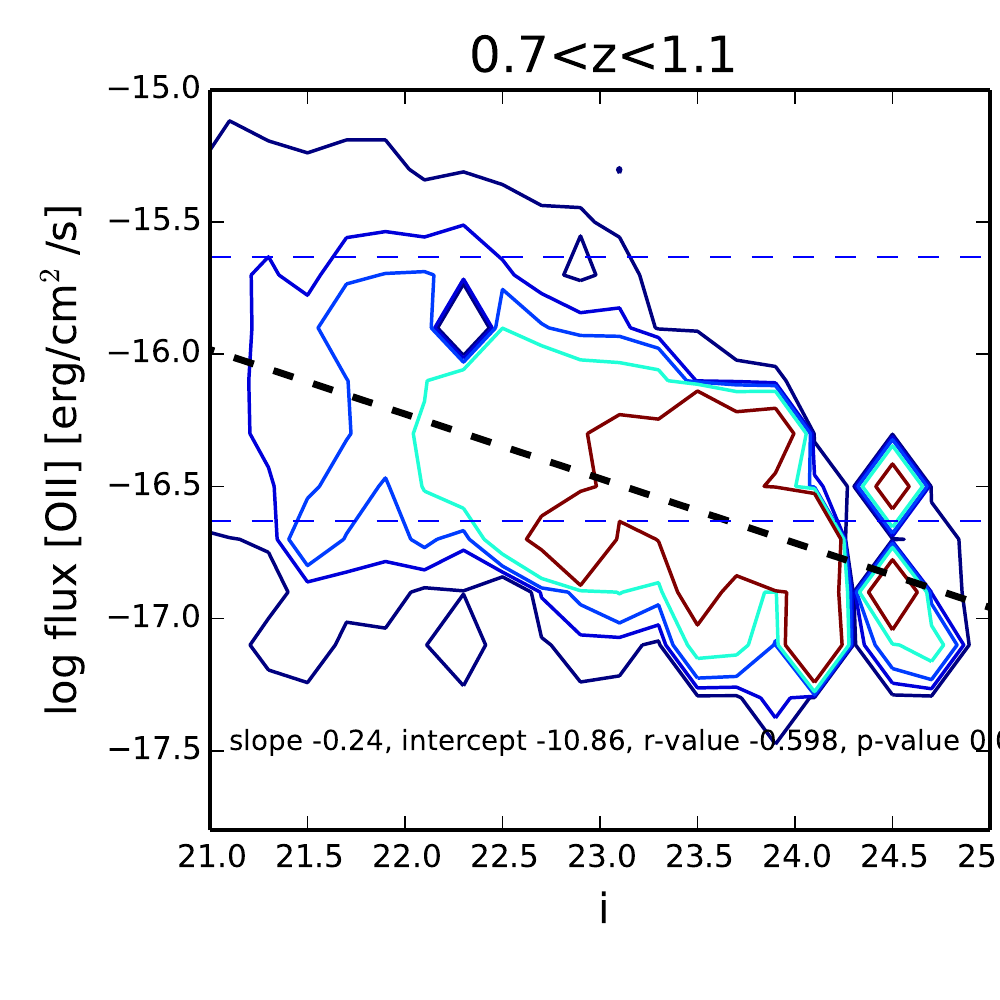}
\includegraphics[width=40mm]{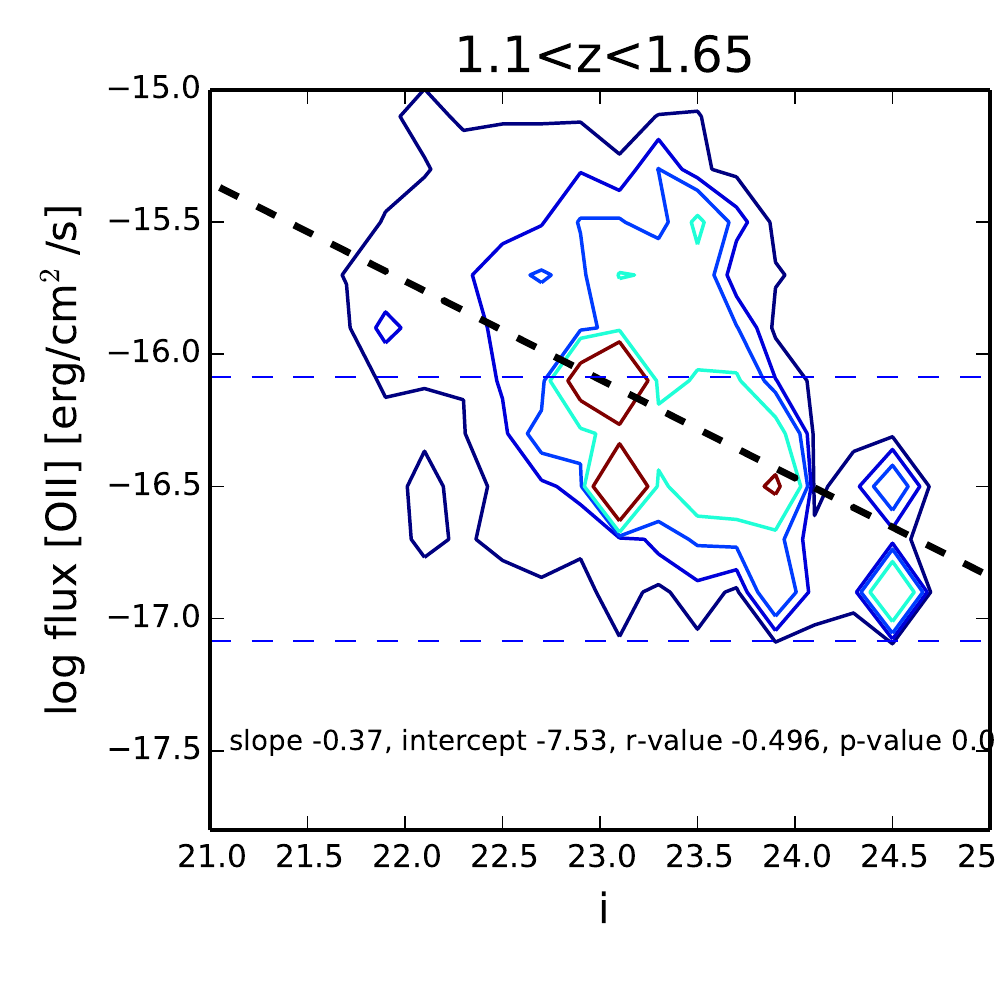}\\
\includegraphics[width=40mm]{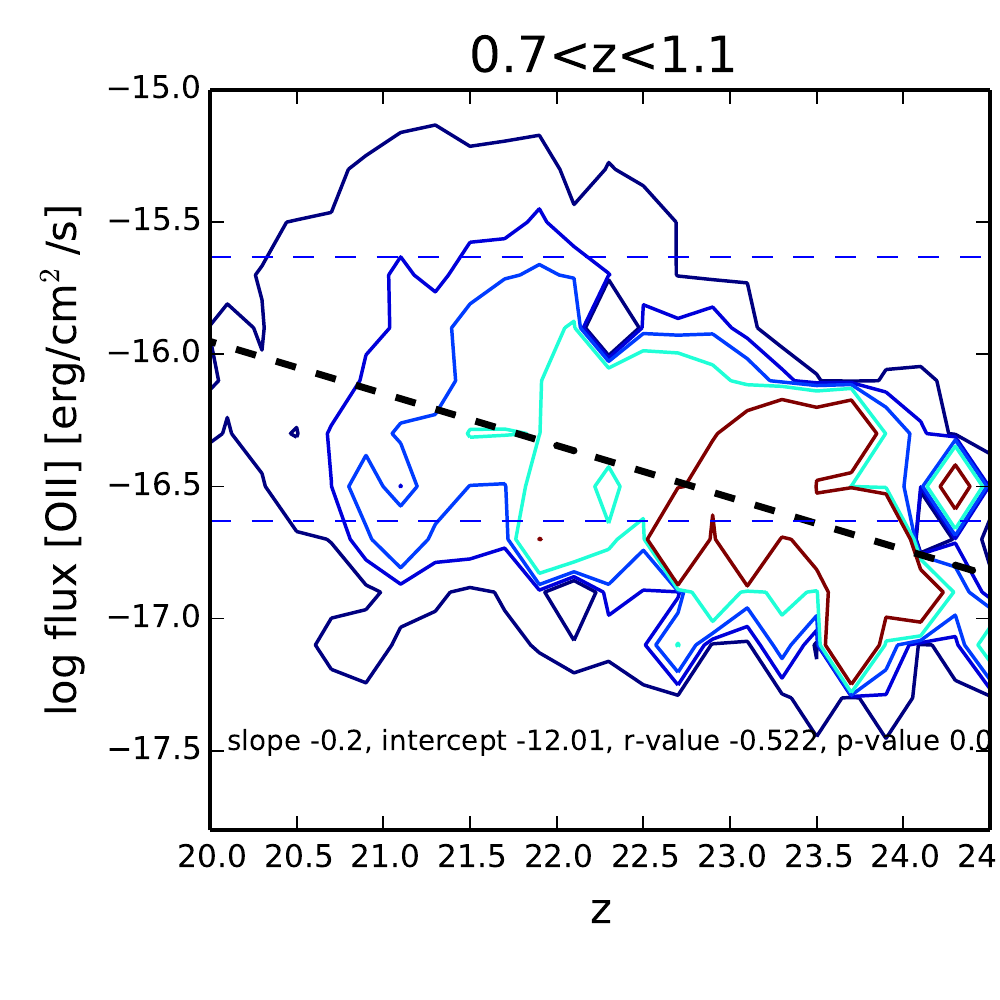}
\includegraphics[width=40mm]{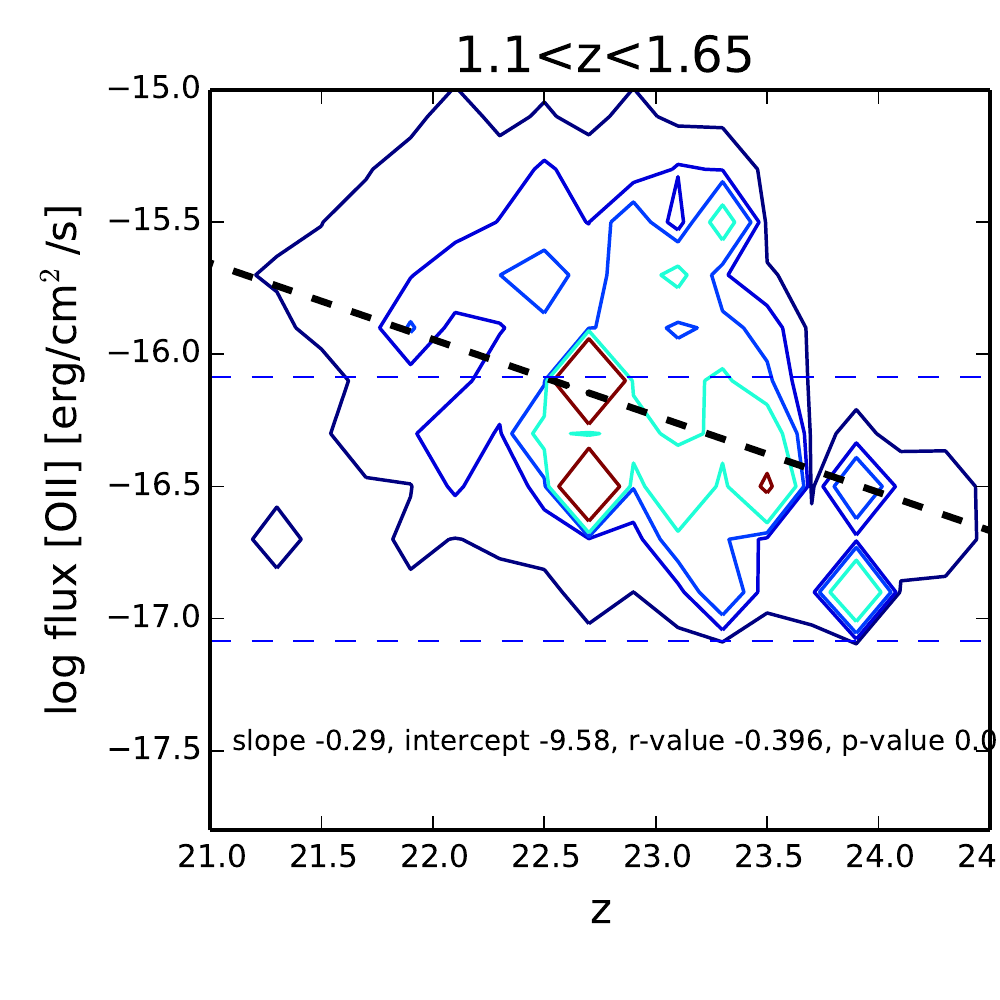}
\caption{\label{mag:oii:core} Correlations between the broad band magnitudes and the \OII flux. The contours represent the density of galaxies predicted by the weighted data (from dark blue to brown 10, 50, 100, 200, 500 galaxy deg$^{-2}$). The $g$ band correlates best in the redshift range $0.7<z<1.1$. The $r$ magnitude correlates best in the range $1.3<z<1.6$. The fluxes corresponding to a luminosity of $10^{41}$ and $10^{42}$ ${\rm erg \cdot s^{-1}}$ and the mean redshift (0.9 and 1.3) are represented in dashed blue.}
\end{center}
\end{figure}

We project the LF densities as a function of redshift to derive the brightest $g$ and $r$ limiting magnitude that will provide a sufficient density of the brightest \OII emitters (a flux limit of $10^{-16}$ \uflux) to sample the BAO.
We find that a survey with magnitude limit of $g<22.8$ and can target a tracer density greater than $10^{-4}$ galaxies $h^3$Mpc$^{-3}$ to $z\sim1.2$ ({\it e.g.} eBOSS). 
A survey with magnitude limit of $r<23.3$ can target a tracer density greater than $10^{-4}$ galaxies $h^3$Mpc$^{-3}$ to $z\sim1.6$ ({\it e.g.} DESI); see Fig. \ref{DESi:selection}. 
Further color selection is needed to sculpt the redshift distribution, in particular to remove lower redshift galaxies. We do not investigate color selections to separate \OII emitters in a given redshift range from the bulk of the galaxy populations at unwanted redshifts here. These color selection are dependent on the photometric survey used and should rather be discussed in each survey paper.

Comparing the magnitude -- \OII fluxes correlations and their densities with the current observational plans of surveys such as DESI or eBOSS broadly confirms their feasibility. 

\subsection*{\OII and stellar mass}
Beyond the relation between \OII flux and observed magnitudes, to plan future surveys and run N-body simulations with the adequate resolution, the stellar mass of the targeted ELG is of interest. We use the stellar mass catalog from \citep{2013A&A...556A..55I} to estimate the average stellar mass of the samples mentioned above. The mean stellar mass of the eBOSS-ELG sample is $10^{10.2\pm0.3}M_\odot$ and the mean stellar mass for the DESI-ELG sample is $10^{9.9\pm0.2}M_\odot$. This estimate confirms that ELG samples are not  mass-limited sample (complete in mass). This corroborates that an \OII-selected sample is likely to miss the dusty and star-forming galaxy population \citep{2013MNRAS.430.1042H} that lies in the massive end of the galaxy population \citep{2010MNRAS.409..421G}.

\subsection{\OII and star formation rate}

The oxygen \OII emission line is also an SFR indicator that is measurable in the optical wavelengths for galaxies with redshift $0<z<1.7$, thanks to its strength and its blue rest-frame location \citep{2004AJ....127.2002K}, although the SFR-\OII$ $ relation is not as direct as SFR-\Ha. The oxygen emission lines are not directly coupled to the ionizing continuum emitted by stars but are sensitive to metal abundance, excitation, stellar mass and dust-attenuation ({\it e.g.} \citealt{2006ApJ...642..775M}). The \OII lines are therefore more weakly correlated to the SFR due to a number of degeneracies \citep{2010MNRAS.409..421G}. In the past, the \OII luminosity functions have been derived and related to the SFR by \citet{2002ApJ...570L...1G,2007ApJ...657..738L,2009A&A...495..759A,2009ApJ...701...86Z,2010MNRAS.405.2594G}. In order to derive a clean estimation of the SFR density sampled, we would need to re-calibrated the \OII SFR relation in each redshift bin using a sample containing the \OII fluxes, the FUV and IR luminosities. We leave this work for a future study.


\section{Comparison to semi-analytical models}
\label{sec:SAMS}
In this section we compare our observations to the predictions from two semi-analytical models, {\sc galform} \citep{cole00} and {\sc sag} \citep{orsi14}, which are based on a $\Lambda$CDM 
universe with WMAP7 cosmology \citep{wmap7}. In order to make the comparison with these models, we have recomputed the observed LF for a WMAP7 cosmology.

Semi-analytical models use simple, physically motivated recipes and rules to follow the 
fate of baryons in a universe in which structure grows hierarchically 
through gravitational instability \citep[see][for an overview of 
hierarchical galaxy formation models]{baugh06,benson10}. 

Here, we compare our observations to predictions from both the \citet{gp14}
(thereafter GP14) flavour of the {\sc galform} model and the
\citet{orsi14} flavour of the {\sc sag} model (thereafter OR14). Both models follow the physical processes that shape the formation and evolution of galaxies, including: 
\begin{enumerate}
\item the collapse and merging of dark matter haloes;
\item the shock-heating and radiative cooling of gas inside dark matter haloes, leading to the formation of galaxy discs;
\item star formation bursts that can be triggered by either mergers or
  disk instabilities;
\item quiescent star formation in galaxy discs which in the OR14
  model is assumed to be proportional to the total amount of cold gas,
  while in the GP14 model it takes into
  account both the atomic and molecular components of the gas
  \citep{lagos11};
\item the growth of super massive black holes in galaxies;
\item feedback from supernovae, from active galactic nuclei and from photoionization of the intergalactic medium; 
\item chemical enrichment of the stars and gas;
\item galaxy mergers driven by dynamical friction within common dark matter haloes, leading to the formation of stellar spheroids.
\end{enumerate}
The end product of the calculations is a
prediction for the number and properties of galaxies that reside
within dark matter haloes of different masses. 

Although both the GP14 and OR14 models assume the same cosmology, they
used different N-body simulations for generating their respective dark
matter halo merger trees. The GP14 model uses the MS-W7 N-body
simulation (Lacey et al. in preparation), with a simulation box of 500
$h^{-1}$Mpc side. The OR14 model was run using an N-body
simulation of volume (150$h^{-1}$Mpc)$^3$. This volume is too small to
adequately model the properties of the brightest observed galaxies. Tests using the GP14 model showed
that a simulation box with side of at least 280$h^{-1}$Mpc is required
to study the bright end of the \OII luminosity function.

The free parameters in the GP14 model where chosen in order to reproduce the observed luminosity functions at z=0 in both b and K-band and to give a reasonable evolution of the rest-frame UV and V luminosity functions. For calibrating their free parameters, the OR14 model used the z=0 luminosity functions but also the z=1 UV luminosity function and SN Ia rates \citep{ruiz14}.

Both the GP14 and OR14 models reproduce reasonably well the evolution of the \Ha
LF \citep{lagos14,orsi14}. The \Ha is a recombination line and thus, its
unattenuated luminosity is directly proportional to the Lyman
continuum, which is a direct prediction of the semi-analytical
models. 

Below we briefly describe how the emission lines are claculated in both models.

\subsection{The {\sc galform} model}

In the {\sc galform} model the ratio between the \OII and the
Lyman continuum is calculated using the \HII region
models by \citet{sta90}. The
{\sc galform} model uses by default eight \HII region models spaning a
range of metallicities but with the same uniform density of 10 hydrogen particles per $cm^{-3}$ and one ionising star in the center of the region with an effective temperature of 45000 K. The ionising parameter\footnote{The ionising    parameter is defined here as a dimensionless quantity equal to the the ionizing photon flux  per unit area per hydrogen density, normalised by the speed of light  \citep[see][]{sta90}.} of these \HII region models is
  around $10^{-3}$, with exact values depending on their metallicity in a non
  trivial way. 
  These ionising parameters are typical within the
  grid of \HII regions provided by \citet{sta90}. 

In this way, the {\sc galform} model is assuming a nearly invariant ionization parameter. Such an assumption although reasonable for recombination lines, is likely too simplistic for other emission lines such as the \OII one \citep[e.g.][]{sanchez14}.
  
\subsection{The {\sc sag} model}

\citet{orsi14} combined the {\sc sag} semi-analytical model with a
photo-ionization code to predict emission line strengths
originated in \HII regions with different ionization parameters. In order to do this, this model assumes an ionization parameter that
depends on the cold gas metallicity of the galaxy. Such dependency is suggested by a number of observational studies \citep[e.g.][]{shim13,sanchez14}.

The dependency of the ionization parameter with metallicity
introduced two new free parameters in the OR14 model, an exponent and a
normalization, that where chosen in order to reproduce the observed BPT diagram
and the \OII and \OIII LF at different redshifts obtained by narrow
band surveys. 

\subsection{The predicted \OII luminosity function}

\begin{figure*}
\begin{center} 
\includegraphics[type=pdf,ext=.pdf,read=.pdf,width=7.5cm]{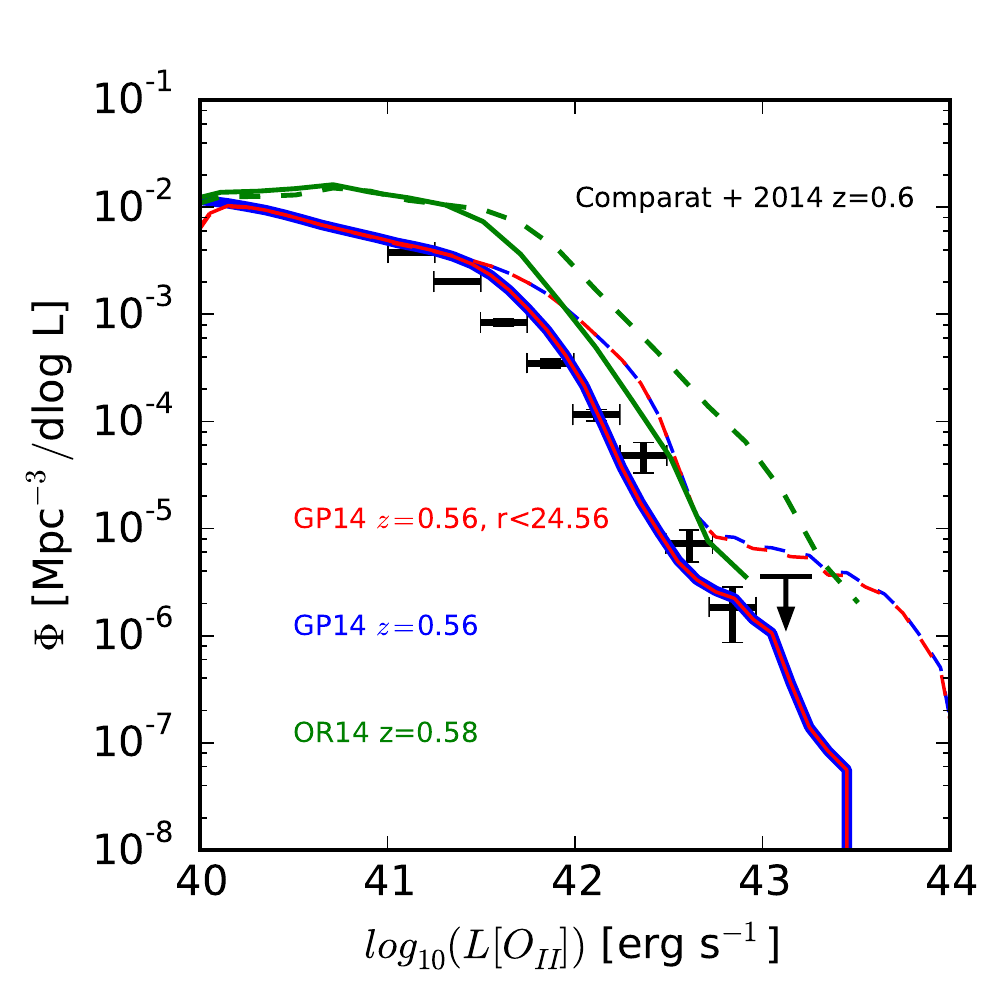}
\includegraphics[type=pdf,ext=.pdf,read=.pdf,width=7.5cm]{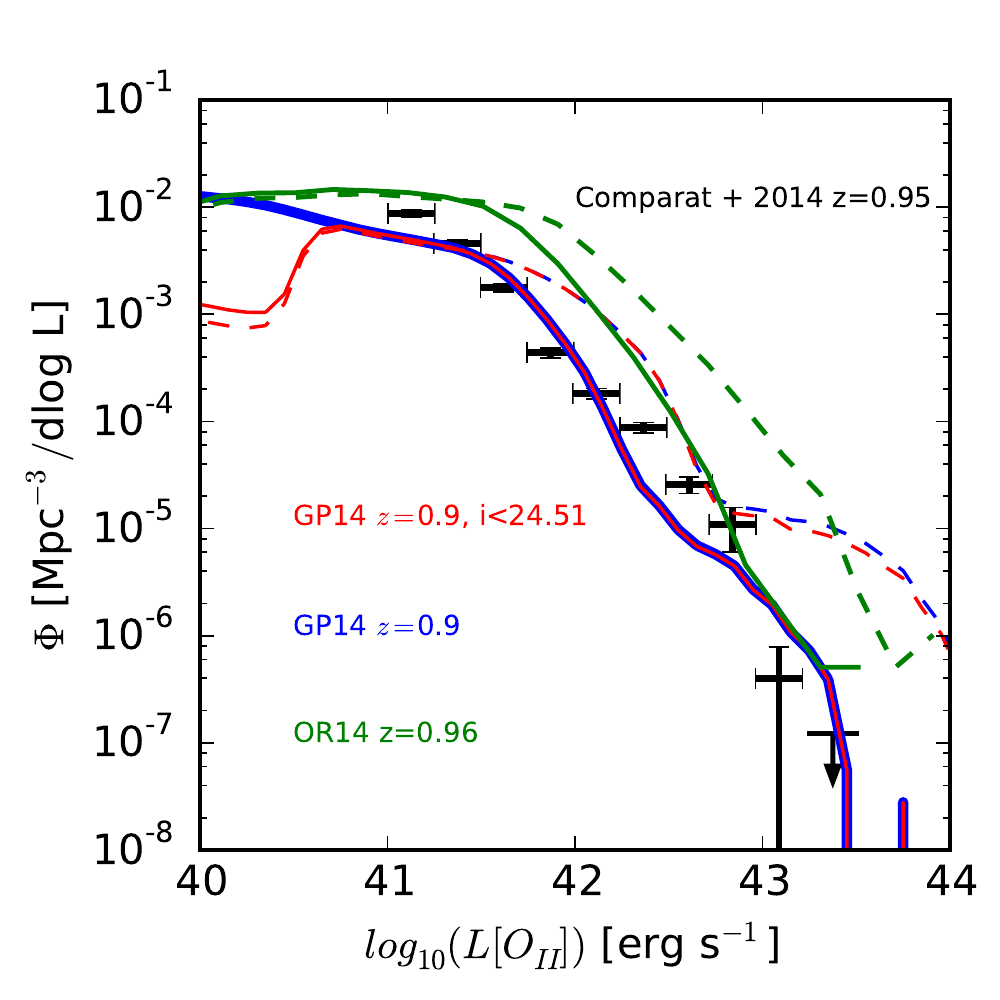} \\
\includegraphics[type=pdf,ext=.pdf,read=.pdf,width=7.5cm]{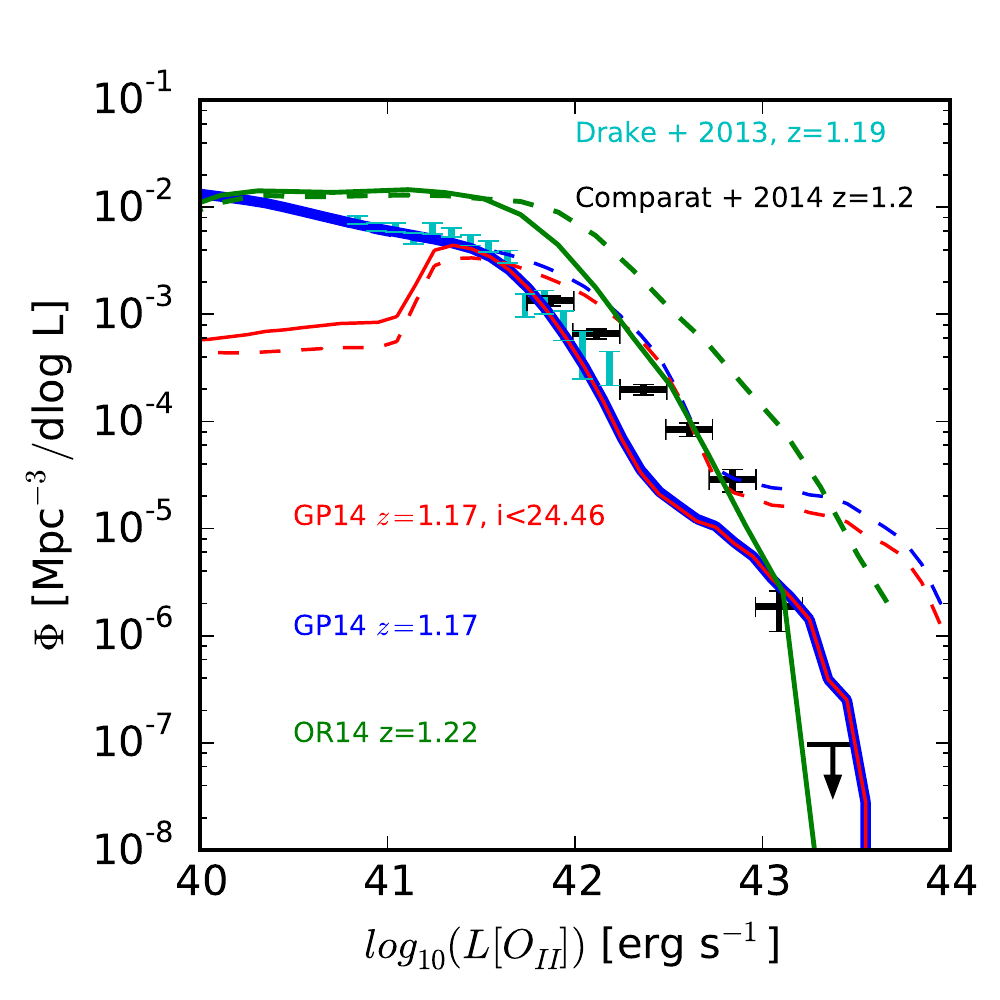}
\includegraphics[type=pdf,ext=.pdf,read=.pdf,width=7.5cm]{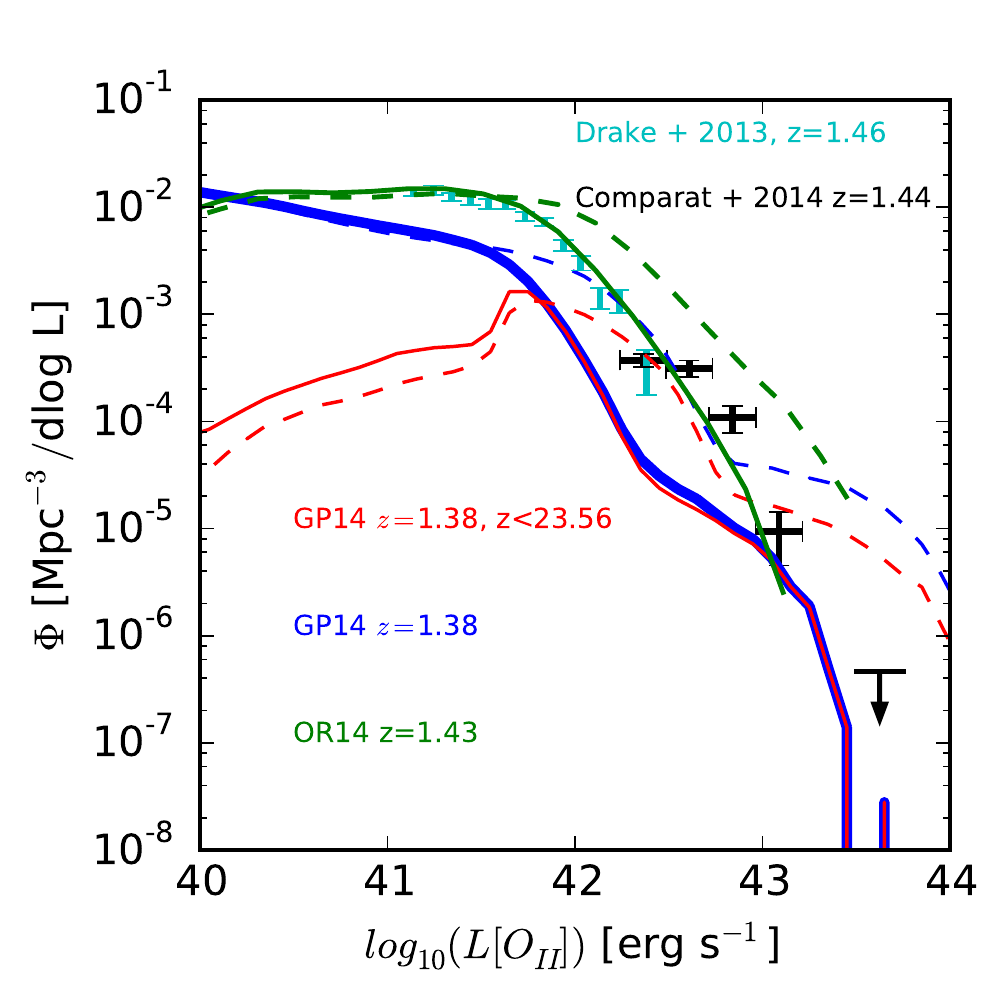}
\caption{\label{LF:sams:gal:form} Our observed LF (black symbols) compared to the predictions from the GP14 (solid blue lines) and the OR14 model (solid green lines). The solid red lines show the predictions from the GP14 model when an extra cut in magnitude is included, as indicated in each legend. The corresponding predictions without including the dust attenuation are shown as dashed lines of the corresponding color. We have also included for comparison the observational data from Drake et al. 2013 as cyan symbols.
}
\end{center}
\end{figure*}

Our observed \OII LF at z=0.6, 0.95, 1.2 and 1.44 are compared with the predictions from both the GP14 and the OR14 models in Fig. \ref{LF:sams:gal:form}. It is important to stress that the GP14 model was not callibrated to reproduce any observed \OII LF and that the OR14 was callibrated attempting to reproduce the \OII LFs of narrow-band observations, which do not suffer from the same selection effects than the ones derived here.

The OR14 model predicts an \OII LF with a bright end slope that agrees with our observations at all redshifts. As shown in Fig. \ref{LF:sams:gal:form}, the \OII LF predicted by the OR14 model at z=1.44 is in excelent agreement with both ours on others observations. However, at lower redshifts this model overpredicts the density of faint \OII emitters. Fig. \ref{LF:sams:gal:form} shows that the predictions of the OR14 are affected by the modelling of dust.

Fig. \ref{LF:sams:gal:form} shows that the observed \OII LFs at $z \simeq 1.2$ is reasonably reproduced by the prediction from the GP14 model. The GP14 model underpredicts the observed \OII LF at z=1.44, except for the brightest bins, which is dominated by the emission of central galaxies. The predicted \OII LF by the {\sc galform} model is sensitive to the assumed
ionization parameter. Using extreme values from \citet{sta90} grid of
\HII regions we obtain predicted \OII LFs bracketing those shown in
Fig. \ref{LF:sams:gal:form}. The default
characteristics of the \HII region model assumed in the {\sc galform}
model might not be adequate at the higher redshifts $z=1.44$. For the two highest redshift bins shown in Fig. \ref{LF:sams:gal:form}, the observations are actually closer to the predicted LF without dust attenuation, though is unlikely for these galaxies to be dust free. In particular if we take into account that the dust extinction applied to the \OII line is the same as experienced by the continuum at that wavelength; and thus the line could actually be more attenuated than predicted here. The main uncertainty for the \Ha line is the dust attenuation \citep[see][for a detailed description of the dust treatment in the models]{gp13}. 

Fig. \ref{LF:sams:gal:form} also shows the \OII LF predicted by the GP14 model imposing a cut in magnitude similar to that done observationally. Comparing with the model predictions, we expect our observations to be complete at the faintest end of the LF.

A detailed exploration of the source of the discrepancy between our observations and the predictions from both models is beyond the
scope of this paper.

\section{Conclusion}
In this work we have measured the \OII luminosity function every gigayear in the redshift range $0.1<z<1.65$ with unprecedented depth and accuracy. This has allowed us to witness the evolution of its bright end: we measure the decrease of $\log L_*$ from $42.4$ erg/s at redshift $1.44$ to $41.2$ at redshift $0.165$. Moreover, by combining our measurements with the fainter ones by \citet{2012MNRAS.420.1926S,2013MNRAS.433..796D} and \citet{2010MNRAS.405.2594G,2013ApJ...769...83C}, we measure the faint end slope flattening from redshift 1.44 to 0.165. 

Such a measurement has been possible by combining in a novel way observations from
 the FORS2 instrument at VLT on the COSMOS field  (released along with the paper);
 the SDSS-III/BOSS spectrograph ELG ancillary programs, 
and with public flux calibrated spectroscopy of \OII emitters. Indeed, we created a new weighting scheme that combines robustly different data sets, for observations which provide the measurement of the fluxes in the lines, the corresponding aperture correction and the parent photometry.

The measurement of the bright end of the LF demonstrates the feasibility of eBOSS and DESi emission line galaxy target selection, {\it i.e.} we have shown here that the density of galaxies with emission lines fluxes $>10^{-16}$\uflux$ $ is sufficient to sample the BAO to redshift 1.6.

We have compared our observed \OII LF to predictions from two
state-of-the-art semi-analytical models, finding a good agreement. This comparison is encouraging for the viability of
producing realistic mock catalogs of \OII flux limited surveys, though more work is needed to understand
the discrepancies found. 

\section*{Acknowledgements}
\vspace{0.2cm}

We thank Alvaro Orsi for sending us his model predictions and for the useful discussion. 

We thank the referee for constructive and insightful comments on the draft.

JC  acknowledges financial support from MINECO (Spain) under project number  AYA2012-31101.

JR acknowledges support from the ERC starting grant CALENDS.

JPK and TD acknowledge support from the LIDA ERC advanced grant.

Funding for SDSS-III has been provided by the Alfred P. Sloan Foundation, the Participating Institutions, the National Science Foundation, and the U.S. Department of Energy Office of Science. The SDSS-III web site is \url{http://www.sdss3.org}.

SDSS-III is managed by the Astrophysical Research Consortium for the Participating Institutions of the SDSS-III Collaboration including the University of Arizona, the Brazilian Participation Group, Brookhaven National Laboratory, University of Cambridge, Carnegie Mellon University, University of Florida, the French Participation Group, the German Participation Group, Harvard University, the Instituto de Astrofisica de Canarias, the Michigan State/Notre Dame/JINA Participation Group, Johns Hopkins University, Lawrence Berkeley National Laboratory, Max Planck Institute for Astrophysics, Max Planck Institute for Extraterrestrial Physics, New Mexico State University, New York University, Ohio State University, Pennsylvania State University, University of Portsmouth, Princeton University, the Spanish Participation Group, University of Tokyo, University of Utah, Vanderbilt University, University of Virginia, University of Washington, and Yale University. 

The BOSS French Participation Group is supported by Agence Nationale de la Recherche under grant ANR-08-BLAN-0222.

This work used the DiRAC Data Centric system at Durham University, operated by the Institute for Computational Cosmology on behalf of the STFC DiRAC HPC Facility (www.dirac.ac.uk). This equipment was funded by BIS National E-infrastructure capital grant ST/K00042X/1, STFC capital grant ST/H008519/1, and STFC DiRAC Operations grant ST/K003267/1 and Durham University. DiRAC is part of the National E-Infrastructure. VGP acknowledges support from a European Research
Council Starting Grant (DEGAS-259586) and the Science and Technology Facilities Council (grant number ST/F001166/1). 

Based on observations obtained with MegaPrime/Megacam, a joint project of CFHT and CEA/DAPNIA, at the Canada-France-Hawaii Telescope (CFHT) which is operated by the National Research Council (NRC) of Canada, the Institut National des Science de l'Univers of the Centre National de la Recherche Scientifique (CNRS) of France, and the University of Hawaii.

The SCUSS is funded by the Main Direction Program of Knowledge Innovation of Chinese Academy of Sciences (No. KJCX2-EW-T06). It is also an international cooperative project between National Astronomical Observatories, Chinese Academy of Sciences and Steward Observatory, University of Arizona, USA. Technical supports and observational assistances of the Bok telescope are provided by Steward Observatory. The project is managed by the National Astronomical Observatory of China and Shanghai Astronomical Observatory.

GAMA is a joint European-Australasian project based around a spectroscopic campaign using the Anglo-Australian Telescope. The GAMA input catalogue is based on data taken from the Sloan Digital Sky Survey and the UKIRT Infrared Deep Sky Survey. Complementary imaging of the GAMA regions is being obtained by a number of independent survey programs including GALEX MIS, VST KiDS, VISTA VIKING, WISE, Herschel-ATLAS, GMRT and ASKAP providing UV to radio coverage. GAMA is funded by the STFC (UK), the ARC (Australia), the AAO, and the participating institutions. The GAMA website is \url{http://www.gama-survey.org}.

\bibliographystyle{aa}
\bibliography{biblio.bib}

\appendix

\section{Weights}
\label{sec:weights}
This appendix describes the details of the weighting scheme.

\begin{figure*}
\begin{center}
\includegraphics[width=4.cm]{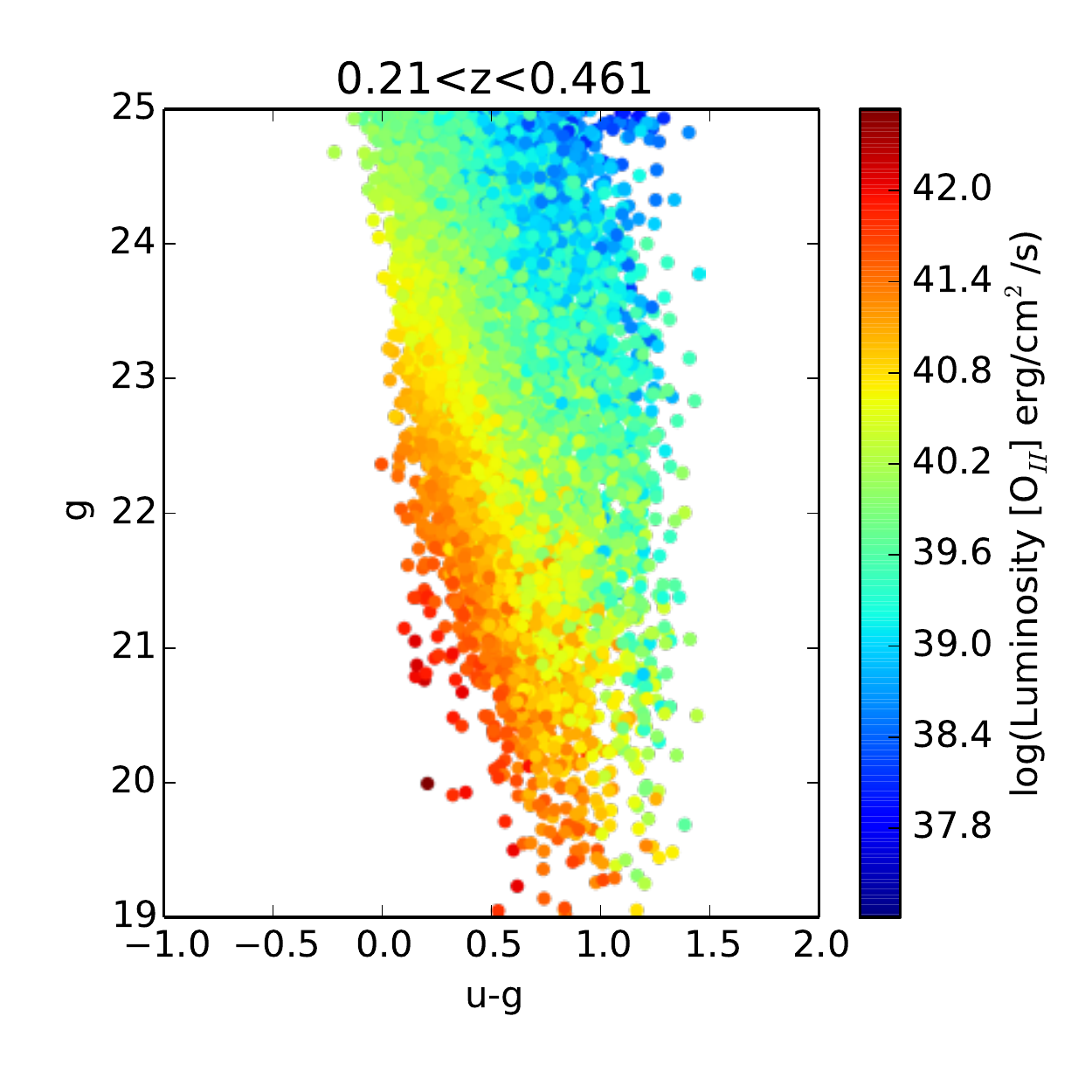}
\includegraphics[width=4.cm]{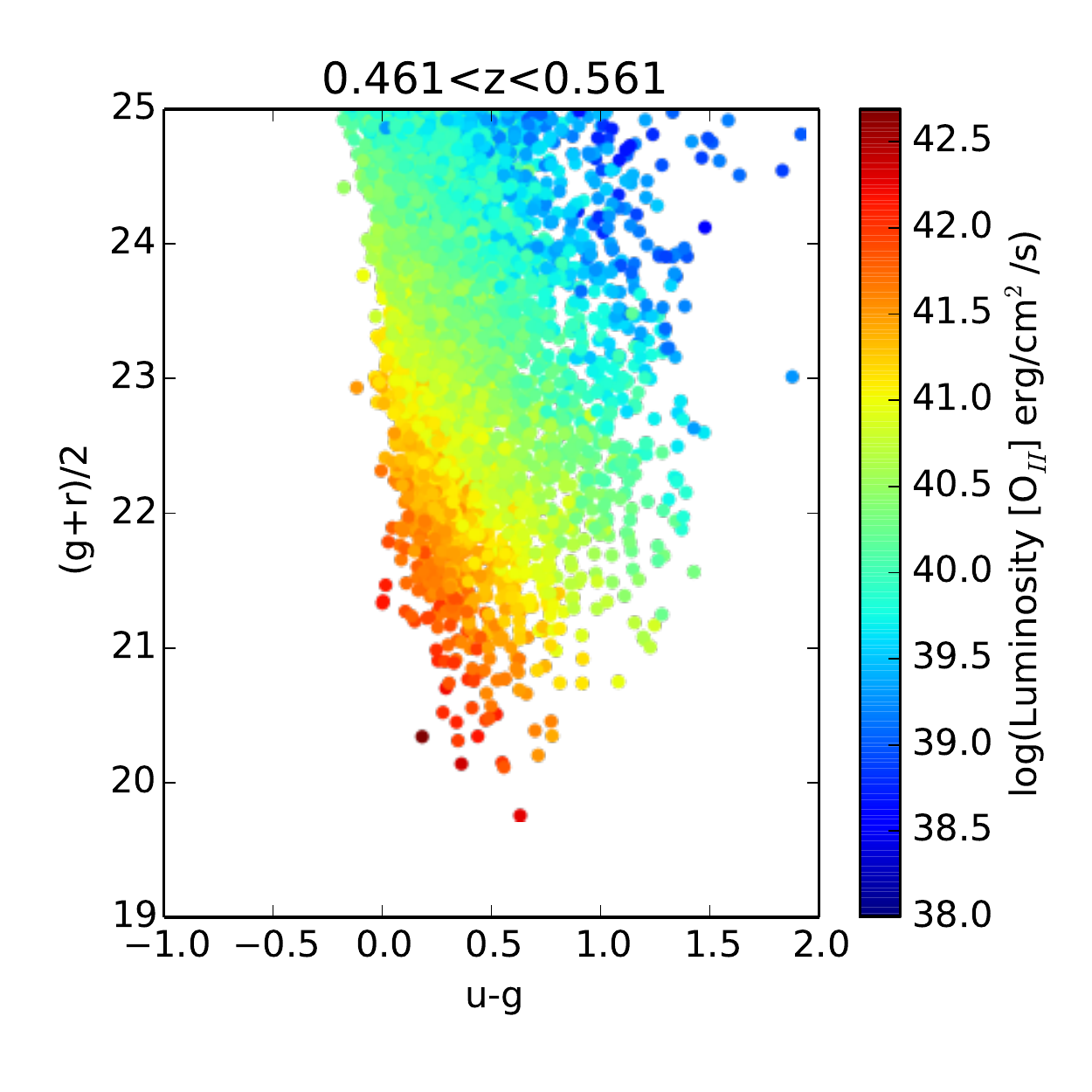}
\includegraphics[width=4.cm]{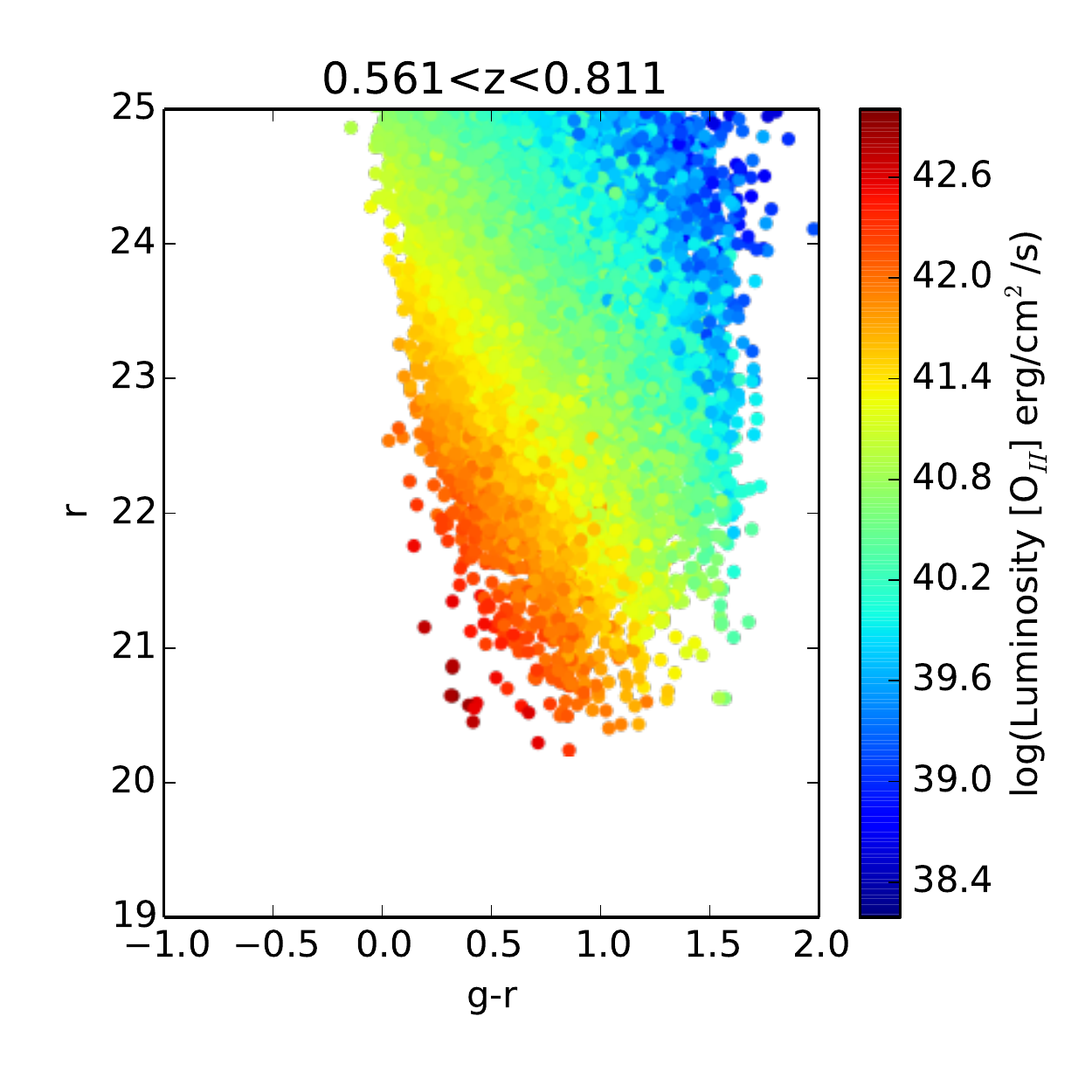}
\includegraphics[width=4.cm]{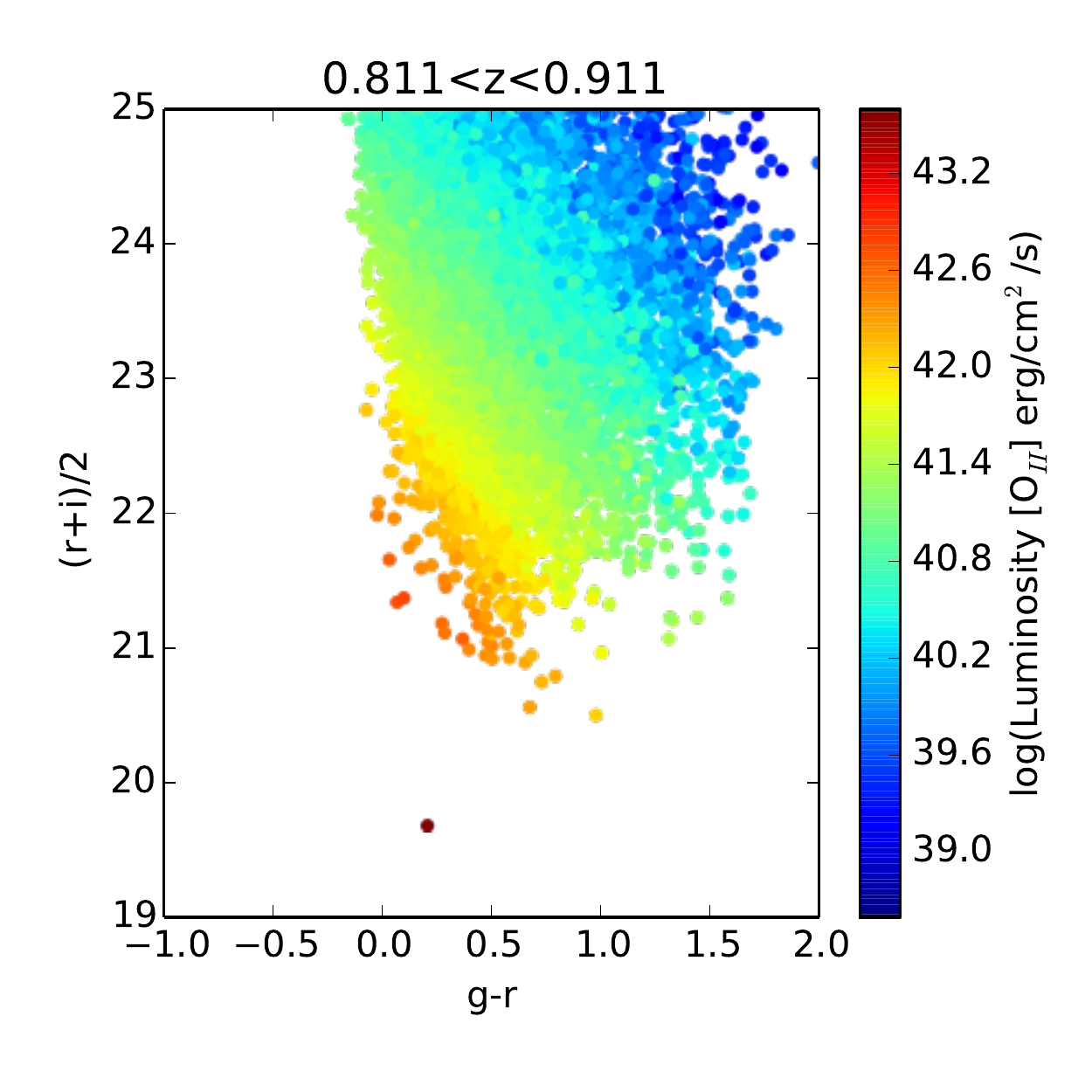}\\
\includegraphics[width=4.cm]{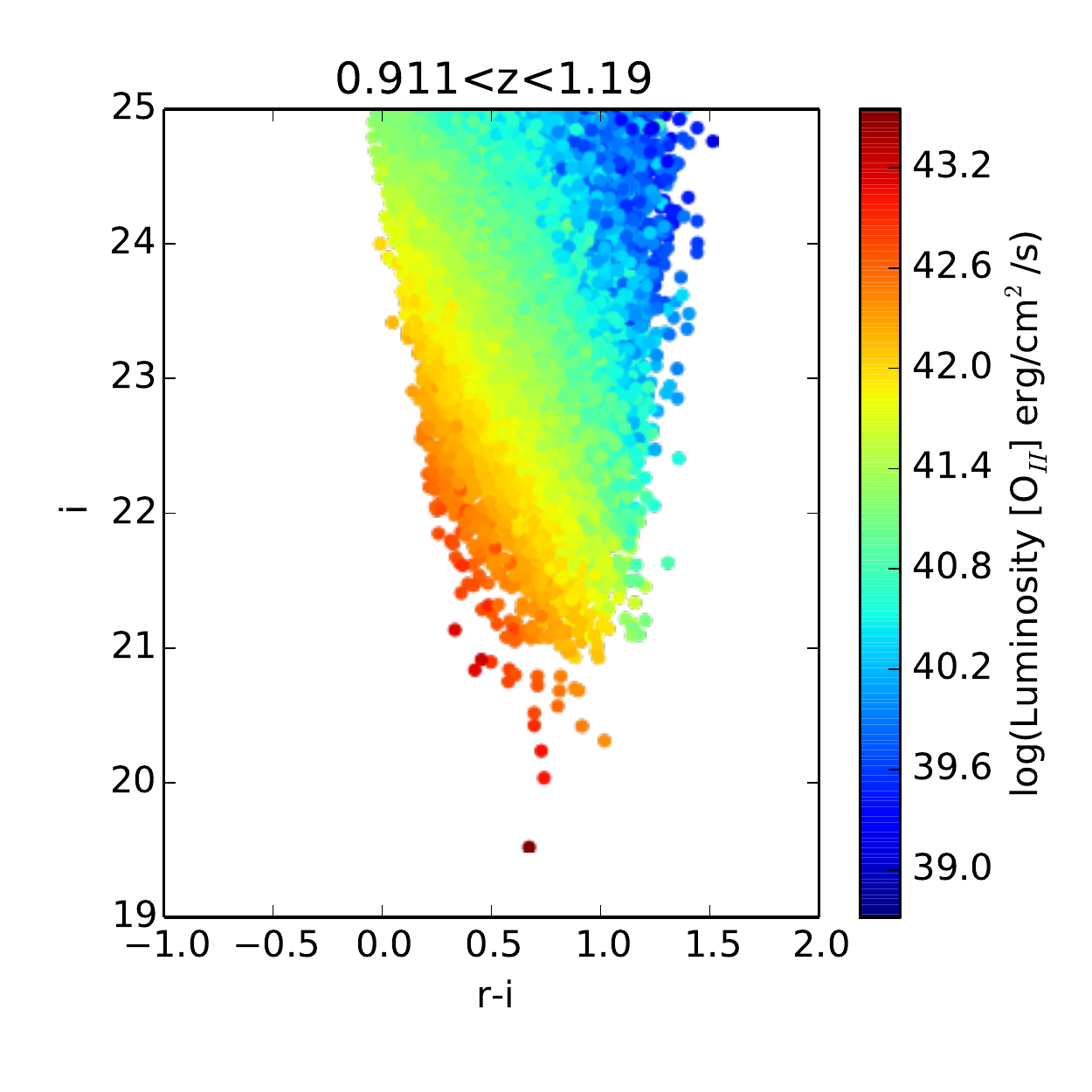}
\includegraphics[width=4.cm]{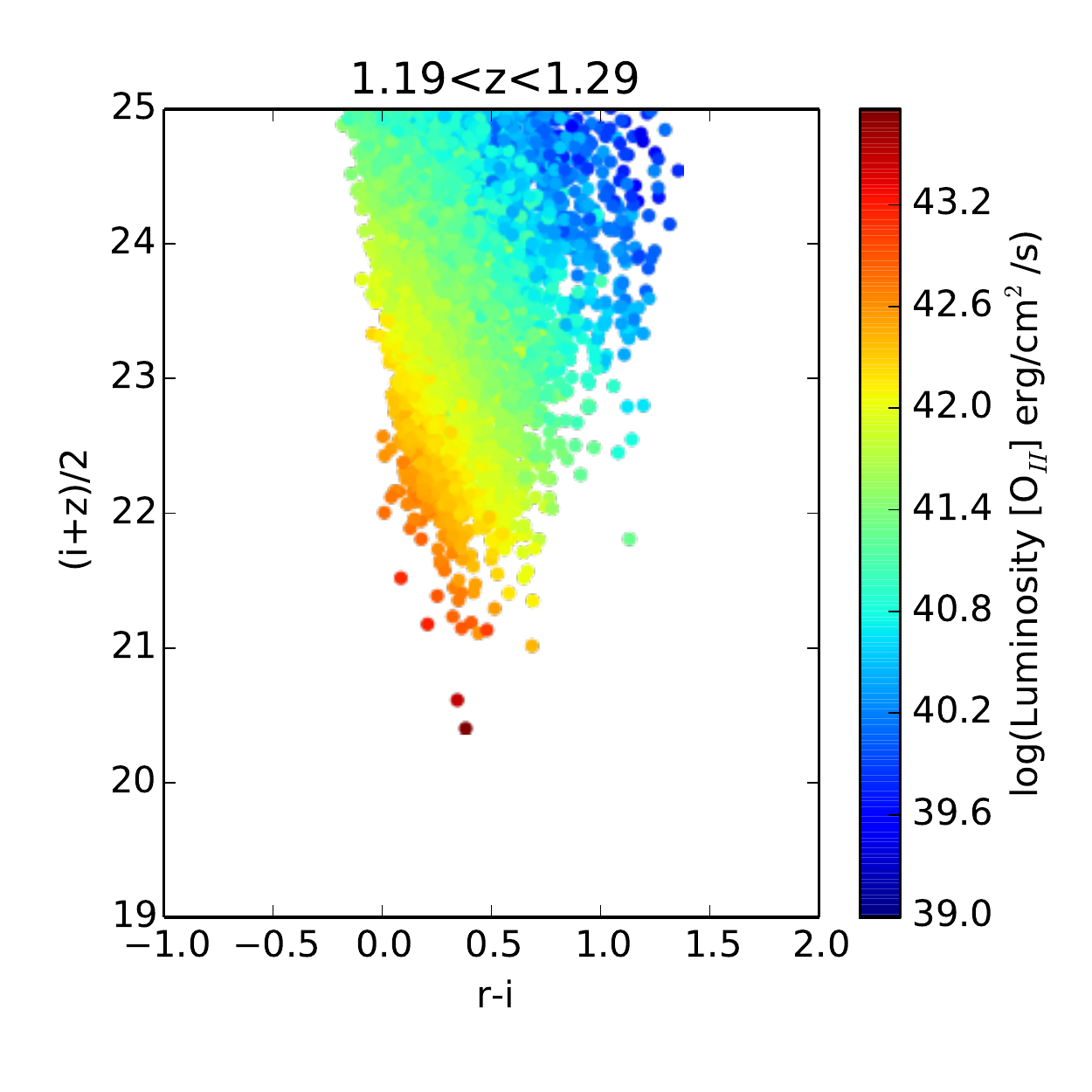}
\includegraphics[width=4.cm]{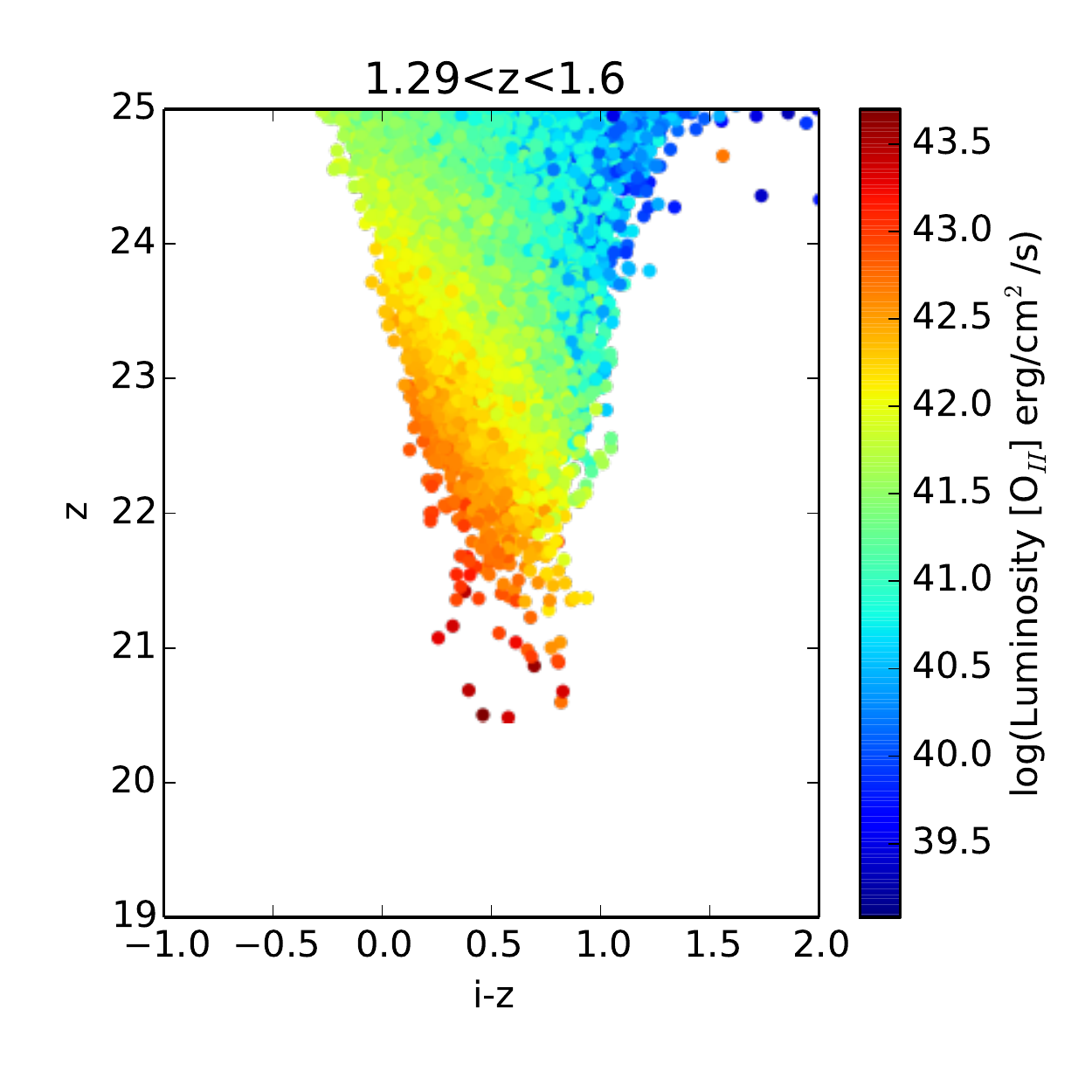}
\caption{Predictions from the CMC for magnitude vs. color (in the CFHT system) and observed \OII luminosity for the CFHT magnitude redshift bins described in Table \ref{color:weight:scheme:tab} and Fig. \ref{color:weight:scheme}.}
\label{weight:theory:1}
\end{center}
\end{figure*}

The theoretical relation between the magnitude containing \OII, the color before this magnitude, redshift, and the \OII luminosity, is shown using the Cosmo Mock Catalog \citep{2009A&A...504..359J} in Fig. \ref{weight:theory:1}. This representation does not take into account the dust present in the galaxies that will induce scatter in this figure. For a constant magnitude, the most luminous galaxies have a blue color. This simulation is based on Kennicutt laws, an extrapolation of the DEEP2 \OII LF, and ignores dust effects.  Therefore this test cannot be used at face value, even less to determine the completeness limit of our sample. This analysis provides an idea on the relation between the magnitude limit and the luminosity completeness limit we can reach with a sample.

In the text, we quote as best value for the tree search a distance of 0.15. This distance corresponds to a maximum distance in each direction of 0.088, and constrains the search for neighbors within about $\sim\pm0.5$mag around the magnitude, about $\sim\pm0.25$mag around the colors and about $\sim\pm0.15$ around the redshift. These values approximately correspond to the area a given galaxy population occupies; see Fig. \ref{weight:theory:1}.

\begin{figure*}
\begin{center}
\includegraphics[width=4.5cm]{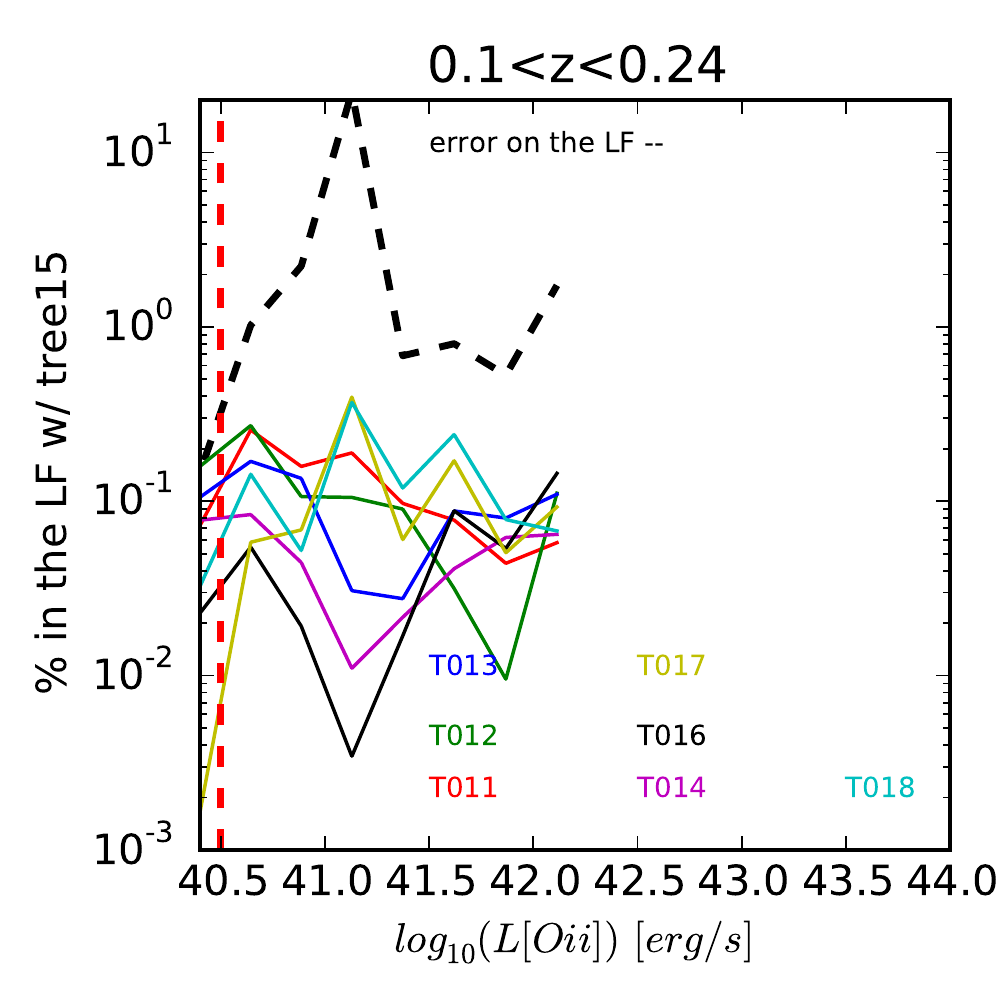}
\includegraphics[width=4.5cm]{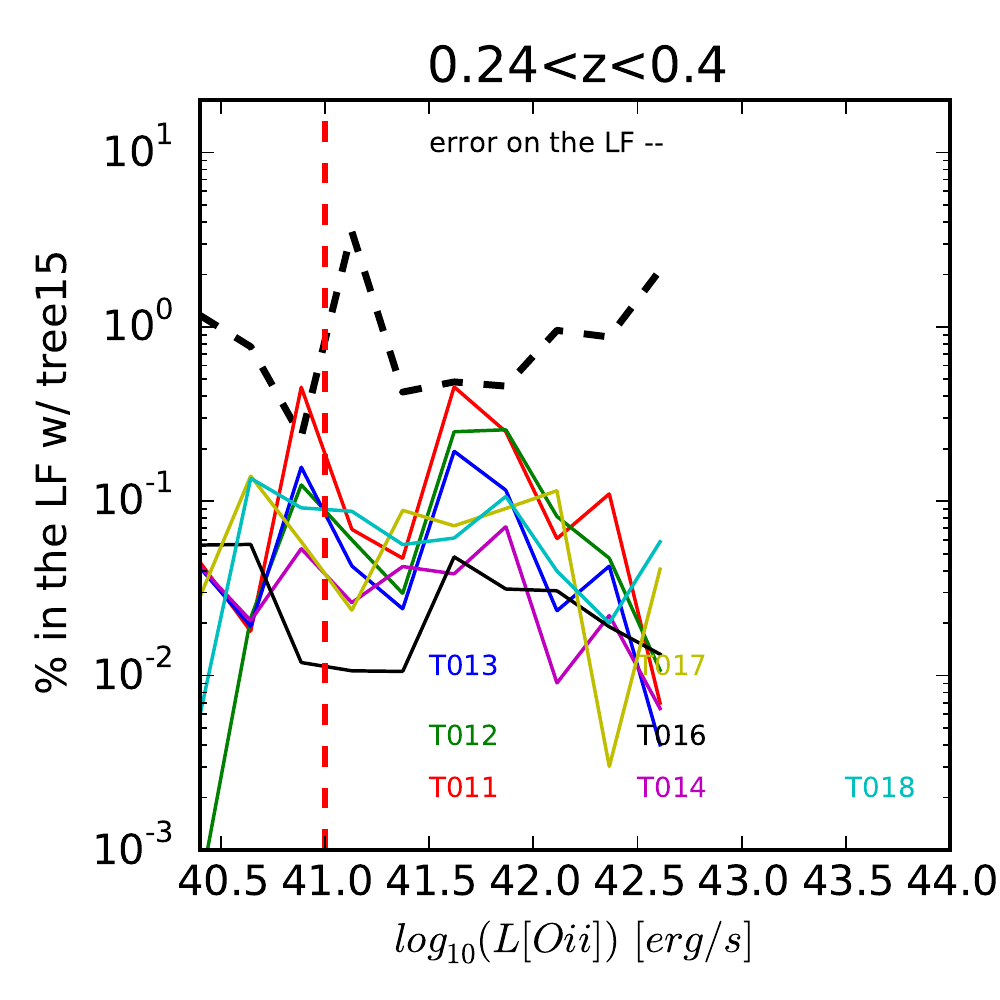}
\includegraphics[width=4.5cm]{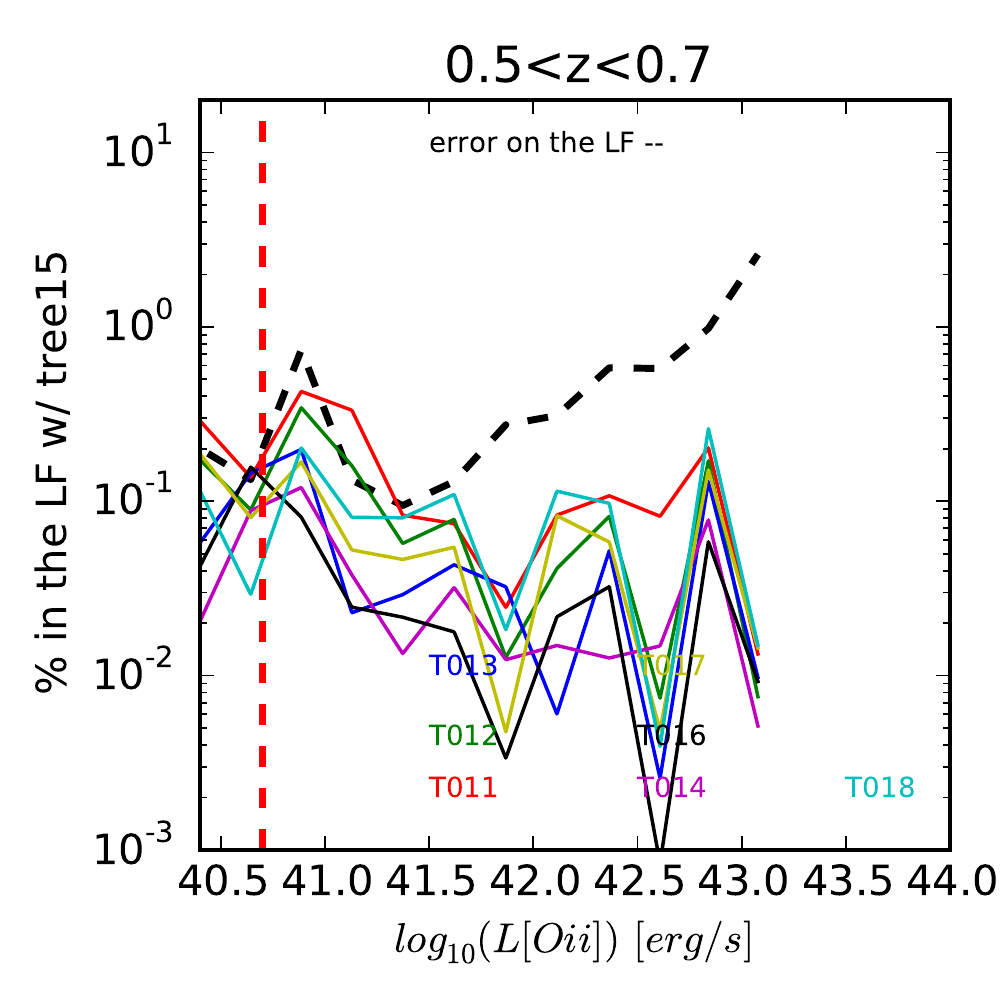}
\includegraphics[width=4.5cm]{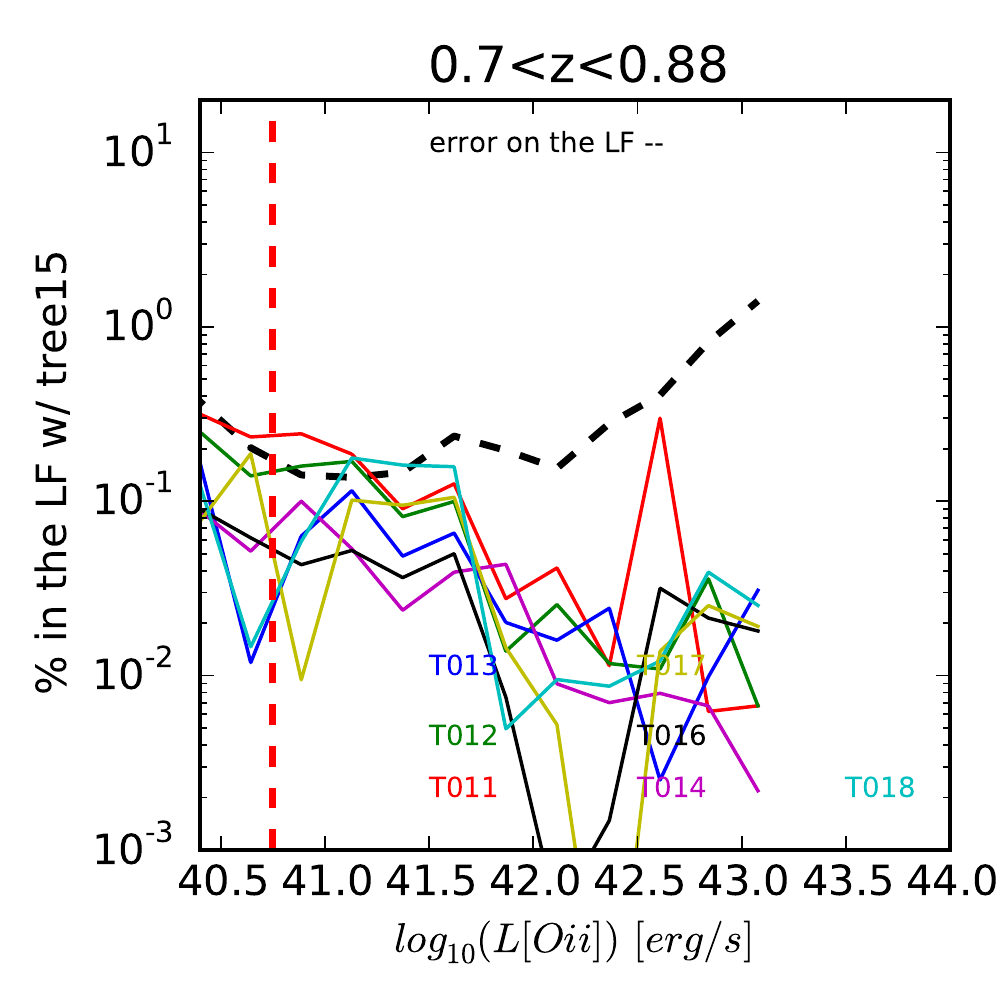}
\includegraphics[width=4.5cm]{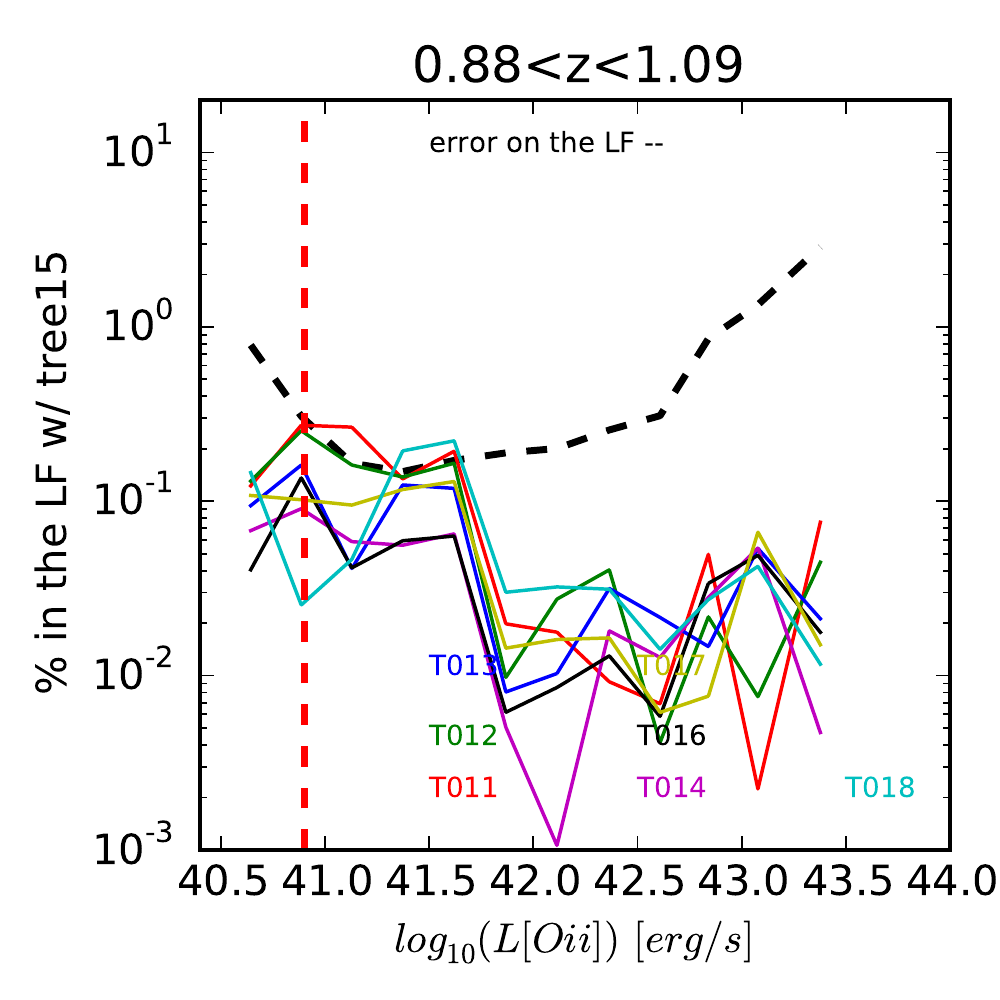}
\includegraphics[width=4.5cm]{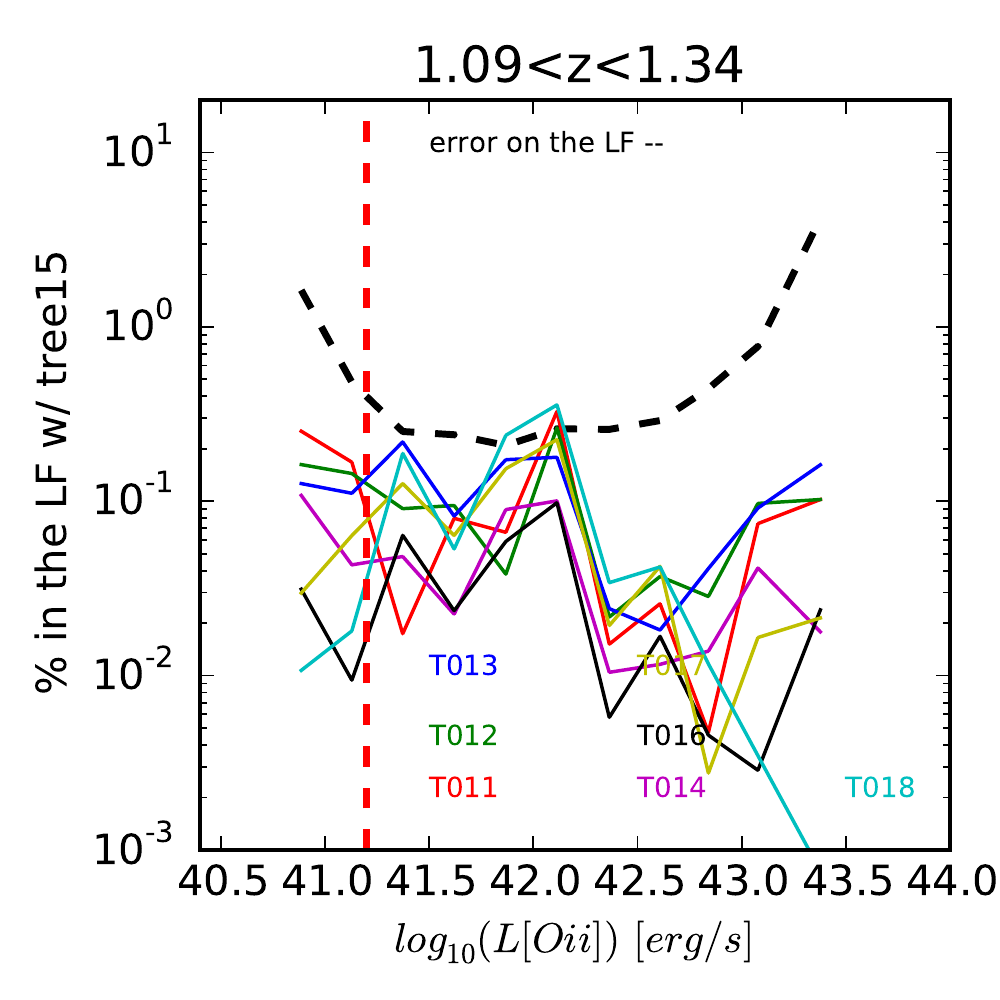}
\includegraphics[width=4.5cm]{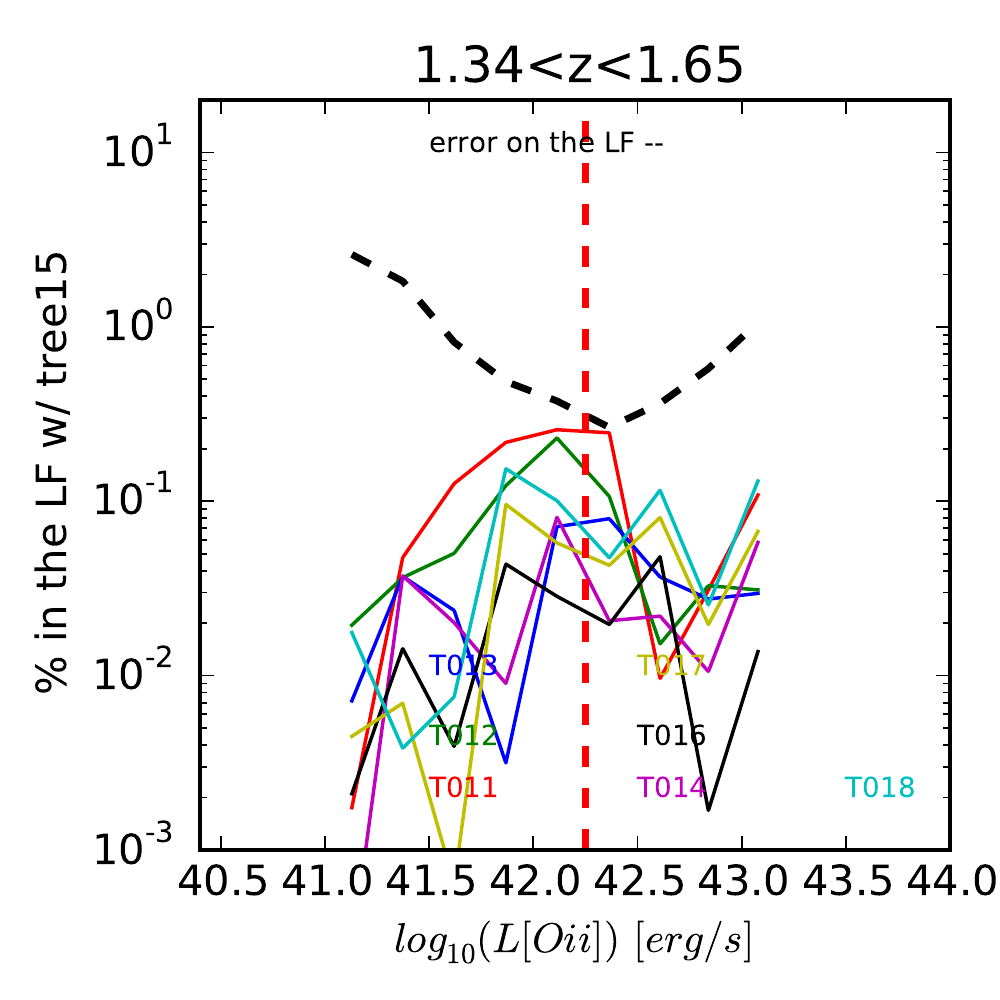}
\caption{LF(d)/LF(0.15) ratio for d=0.11, 0.12, 0.13, 0.14, 0.16, 0.17, 0.18 divided by the LF determined with 0.15. The vertical red line is the luminosity completeness limit. The error on the LF is shown in black dashes. The LFs with radius 0.14 and 0.16 stay well within the uncertainty on the LF, while larger or smaller radii approach to the limit of the uncertainty of the LF.}
\label{weight:distanceSearch}
\end{center}
\end{figure*}

We tested the LF estimation for different distance values and found that a limit at $0.15\pm0.01$ was stable and variations in the measurement of the LF would be smaller than the uncertainty on the LF. Fig. \ref{weight:distanceSearch} displays the variation in the LF compared to the LF estimate using the tree search distance 0.15. For tree searches that are too wide, $>$0.17, the  weighting scheme begins to fail {\it i.e.}, the LF is inconsistent at 1$\sigma$ with the fiducial LF. For tree searches too narrow, $<$0.13, the weights become inaccurate, the weight error increases and the LF is less accurate.

\end{document}